\documentclass[fleqn,usenatbib]{mnras}

\usepackage{framed}
\usepackage{graphicx,epsfig,psfig,multicol}
\usepackage[dvipsnames]{xcolor}
\usepackage[]{inputenc,amssymb}
\usepackage{amsmath}
\usepackage{amssymb}
\usepackage{mathtools}
\usepackage{color}
\usepackage{epsfig}
\usepackage{epstopdf}
\usepackage{textcomp}
\usepackage{array}
\usepackage{relsize}
\usepackage{longtable}
\usepackage{siunitx}
\usepackage{pdfpages}
\usepackage{xcolor,cancel}
\usepackage[dvipsnames]{xcolor}
\usepackage{comment}
\usepackage{subcaption}
\usepackage{accents}
\newcommand\thickbar[1]{\accentset{\rule{.25em}{.6 pt}}{#1}}

\usepackage{tikz}
\usetikzlibrary{shapes.geometric, arrows,calc,positioning}
\definecolor{blue1}{RGB}{61,243,255}
\definecolor{blue2}{RGB}{43,114,255}

\usepackage[normalem]{ulem}%strikeout text
\newcommand{\dd}[1]{#1} 
\newcommand{\ddd}[1]{\textcolor{Magenta}{\sout{#1}}} %delete text
\newcommand{\rdd}[2]{{\textcolor{Magenta}{\sout{#1}}}{\textcolor{Green}{[{\bf DO}: #2]}}} %delete arg#1 and replace with arg#2
\usepackage[export]{adjustbox}% this is for telling the positioning of the figure eg. center ..
\usepackage{adjustbox}
\usepackage{placeins}% This is for using the float barrier ...

\usepackage{float}
\usepackage{morefloats}

\usepackage{lipsum}
\usepackage{mwe}
\usepackage{arydshln} %for dashed line
\usepackage{stmaryrd}

\usepackage[toc,page]{appendix}

\usepackage[T1]{fontenc}

\DeclareRobustCommand{\VAN}[3]{#2}
\let\VANthebibliography\thebibliography
\def\thebibliography{\DeclareRobustCommand{\VAN}[3]{##3}\VANthebibliography}

\title[Fragmentation in Collapsar Disks]%{Fragmentation in Collapsar Disks}
{Fragmentation in Collapsar Disks: Migration, Growth, and Emission}

\author[Y. Lerner, N.C. Stone \& D.D. Ofengeim]{
Yonatan Lerner$^1$,\thanks{E-mail: yonatan.lerner@mail.huji.ac.il}
Nicholas C. Stone$^{1, 2}$,
Dmitry D. Ofengeim$^1$
\\
$^1$Racah Institute of Physics, The Hebrew University, 91904, Jerusalem, Israel\\
$^2$Department of Astronomy, University of Wisconsin, Madison, WI 53706, USA
}

\date{Accepted XXX. Received YYY; in original form ZZZ}

\pubyear{2025}

\usepackage{mathrsfs}
\begin{document}

\label{firstpage}
\pagerange{\pageref{firstpage}--\pageref{lastpage}}
\maketitle

\begin{abstract}
We present a parameter survey of fragmentation in collapsar disks, using a revised version of the Chen \& Beloborodov (2007) model that determines the structure of steady state hyperaccretion disks in a general relativistic and neutrino cooled framework. We map out the range of disk conditions leading to gravitational instability alongside an exploration of the dimensionless cooling time $\beta$, which together determine 
whether fragmentation is likely to occur. We estimate the initial mass and density of fragments, finding that they occupy a unique region in the space of self-gravitating compact objects, with masses $m_{\rm f,i} \sim 10^{-3} M_\odot -10^{-1} M_\odot$ and densities $\rho_{\rm f}\sim 10^8-10^{11}~{\rm g~cm}^{-3}$. We then calculate their migration and mass growth (via Bondi-Hoyle accretion) as they inspiral through the collapsar disk. During a fragment's migration to the central black hole, it can grow its mass up to a range $m_{\rm f}\sim 10^{-1} M_\odot - 1  M_\odot$.  In most cases, the final fragment mass is larger than the minimum cold stable neutron star mass but much smaller than any observed neutron star. The fragment briefly achieves peak accretion rates comparable to (or even larger than) that of the central engine. We propose that these bound fragments may give rise to observable astrophysical phenomena, and we approximately model two of these: (i) gamma ray burst variability produced by a secondary, fragment-launched jet; (ii) the generation of non-vacuum gravitational waveforms accompanied by electromagnetic counterparts.
\begin{comment}
    things to discuss:
- We have used an advanced GR, neutrino, microphysical framework. with an improved model.
- We made a big parameter survey of gravitational instability, while simultaneously making a parameter survey of the cooling time parameter $\beta$ which is also crucial for fragmentation.
- We then made an estimation of fragment initial mass and density while calculating how the fragment migrates and accretes mass through Bondi Hoyle accretion.
- We hypothesize that this effect might be the origin of  GRB variability, non-vacuum GWs with electromagnetic counterparts, ejections of single fragments or even binary fragments from the system that would merge in vacuum, exotic low-mass NSs with no other astrophysical origin and a novel astrophysical site for r-process nucleosynthesis.
\end{comment}
\end{abstract}

% Select between one and six entries from the list of approved keywords.
% Don't make up new ones.

%\fontsize{12}{15}
%\selectfont

\twocolumn

%\begin{comment}  

\section{Introduction}
\label{sec:intro}

The core-collapse supernovae that terminate the lives of massive stars can sometimes produce collapsar, or hyperaccretion, disks \citep{Paczynski98}.  These disks, which revolve around a central neutron star (NS) or black hole (BH), are thought to power the relativistic jets observed as long gamma-ray bursts, or GRBs \citep{Woosley1993}.  Collapsar disks also shape the demographics of stellar-mass BHs \citep{MacFadyenWoosley99, ShapiroShibata02, Janiuk+08, JacqueminIde+24}, and may even serve as primary astrophysical sites for r-process nucleosynthesis \citep{Siegel+19, Siegel+22,Issa+24}. Collapsar disks may even generate detectable gravitational waves (GWs), either through the energy flux from a relativistic jet \citep{Sago+04, BirnholtzPiran13, LeiderschneiderPiran21} or due to non-axisymmetric instabilities emerging in the system \citep{KobayashiMeszaros03, Siegel+22, Gottlieb+24}.

Collapsar disks are extreme astrophysical environments, with accretion rates that can exceed $10^{14}$ times the naive (photon) Eddington limit \citep{DiMatteo2002}.  The resulting densities in these disks lie in between the densities of white dwarfs and those of NSs, but the underlying microphysics differs from that of compact objects due to very high temperatures, which partially lift electron degeneracy and also power copious neutrino emission \citep{Beloborodov2003}.  At the highest accretion rates ($\dot M \gtrsim 1 M_\odot~{\rm s}^{-1})$, collapsar disks can develop the Toomre \citep{Toomre64} self-gravitational instability \citep{CB2007}.  

The outcomes of Toomre instability in collapsar disks are not well understood, in part because the large dynamic range and multi-scale physics of the problem has prevented 3D simulations from resolving the relevant spatial scales.  %\footnote{Toomre instability does not lead to fragmentation in 1D or axisymmetric 2D geometry.}.  
In other astrophysical contexts, Toomre instability can produce self-gravitating fragments, such as gas giants in protoplanetary disks \citep{Cameron78}, stars in active galactic nuclei \citep{LevinBeloborodov03}, or giant molecular clouds in galactic gas disks \citep{MartinKennicutt01}.  Some past work has examined the possible formation of analogous fragments in collapsar disks, beginning with \citet{Perna2006, PiroPfahl07}.  If gravitationally unstable regions of collapsar disks are able to cool efficiently \citep{Gammie2001edit2}, then bound objects may form.

The implications of a population of self-gravitating fragments in collapsar disks are manifold.  Such fragments can migrate inwards \citep{PiroPfahl07} until they tidally disrupt, modulating the accretion rate onto the central engine and creating variability in the GRB prompt emission \citep{Perna2006, DallOsso+17, ShahamatAbbassi20}.  This migration may produce a detectable GW signal on its own, observable with ground-based laser interferometers \citep{PiroPfahl07, Shahamat2021}.  Finally, the mass range of self-gravitating fragments may involve low-mass, high-density objects that cannot be produced in standard core collapse supernova explosions, but which nevertheless can reach hydrostatic equilibrium under a cold neutron star equation of state.  Toomre instability in collapsars may be the only way to form such low mass NSs in the Universe \citep{PiroPfahl07, Metzger+24}.  Recently, it was even suggested that interactions between these low mass NSs could have significant implications for r-process nucleosynthesis and associated electromagnetic transients \citep{Metzger+24}.

Most existing investigations of fragmentation in collapsar disks work with either order-of-magnitude estimates for disk conditions \citep{PiroPfahl07, Metzger+24} or, if they use an explicit disk model, consider only a small number of possible parameters \citep{ShahamatAbbassi20, Shahamat2021}.  Since the central engines of long GRBs likely cover a wide range of accretion rate and system mass, it is desirable to explore the parameter space of this problem more fully.  Likewise, past work has also not widely explored interactions between fragments and the surrounding collapsar disk, such as migration and fragment growth through Bondi-Hoyle accretion.  Ideally, one would explore collapsar fragmentation through multidimensional magneto-hydrodynamic simulations, but this is challenging in practice: existing numerical work has generally examined disks that are insufficiently massive to fragment \citep{Siegel+19, Gottlieb+24}.%, or has simulated massive, potentially unstable disks in 2D, where axisymmetry may fundamentally change the nature  \citep{Fujibayashi+20, Shibata+24}.

With these motivations, we perform a parameter survey of 1D, steady state general relativistic models for neutrino-cooled collapsar disks.  Our work also explores subsequent interactions (migration and mass growth) between bound fragments and the disk they formed from.  Our approach, which is based on the model of \citealt{CB2007}, is much more approximate than a self-consistent magneto-hydrodynamic simulation, but is also far cheaper computationally, and allows us to explore the large parameter space of central masses, accretion rates, and disk microphysics.  In \S \ref{sec:model}, we introduce the underlying physics of the disk model (though details are elaborated in Appendices \ref{app: Clarification to the CB07 model}-\ref{app:numerics}).  In \S \ref{sec:results}, we present the primary results of our survey, focusing on which regions of parameter space exhibit sufficiently strong self-gravity and sufficiently short cooling times to form self-bound fragments.  We also explore the evolution of these fragments due to orbital migration and mass growth through accretion.  In \S \ref{sec:observables}, we make tentative estimates concerning the contribution of these fragments to GRB variability and GW production in collapsar disks, which are further discussed in \S \ref{sec:discussion}.  We conclude in \S \ref{sec:conclusions}.  Additional results that more fully explore the parameter space of our model are presented in the Supplementary Material (Appendix \ref{app: Results}).

% Methods
\section{Disk Model}
\label{sec:model}

% The model itself, including differences from C&B
\subsection{Hyperaccretion Disk Structure}
\label{sec:disk}
Our model is based on the earlier work of \citealt{CB2007} (CB07), which built a steady-state, one dimensional, general relativistic, axisymmetric model for collapsar accretion disks. The model accounts for the high density microphysics of a neutron-proton-electron-$\alpha$ plasma, where some regions are efficiently cooled by neutrino losses and become geometrically thin, while others are thick and advection-cooled. We have modified the original model to make improvements and corrections, though these do not change the final results at leading order.

Our approach involves numerically solving three master equations, consisting of one local, algebraic equation, and two ordinary differential equations. The coupled solution of these equations gives one-dimensional profiles of disk quantities (e.g. mid-plane density $\rho$) as functions of cylindrical radius $r$ and global parameters such as the central mass $M$ and steady state accretion rate $\dot{M}$. While the CB07 model could in principle be applied to collapsar disks around either a central NS or a central BH, we specialize in this paper to disks with central BHs (as we shall see, larger values of $M$ are generally more interesting for our scenario).  The first master equation is pressure balance:
\begin{equation}
P=P_{\rm \gamma}+P_{\rm b}+P_{\rm e^{-}}+P_{\rm e^{+}}+P_{\rm \nu}+P_{\rm \thickbar{\nu}}=\left(\frac{GM\dot{M}\mathcal{S}}{6\pi r^{3}}\frac{\mathcal{J}}{\alpha}\right)^{2/3}\rho.
\label{eq: pressure equality}
\end{equation}
Here $P_{\rm \gamma}$, $P_{\rm b}$, $P_{\rm e^{-}}$, $P_{\rm e^{+}}$, $P_{\rm \nu}$ and $P_{\rm \thickbar{\nu}}$ represent pressure contributions from photons, baryons, partially degenerate electrons and positrons, neutrinos and antineutrinos, respectively. For each contribution described in the pressure sum, we calculate a respective energy density $U_{\rm i}$ which 
%\rdd{sums}{sum} 
sums to a total energy density $U$ (for exact calculations of all disk structure elements, except for a few corrections listed in appendix \ref{app: Clarification to the CB07 model}, we refer to CB07). The right hand side of the equation is the pressure $P$ needed to support vertical hydrostatic equilibrium, where $G$ is the gravitational constant. $\mathcal{S}$ and $\mathcal{J}$ are dimensionless general relativistic prefactors that go to unity as $r$ increases (see appendix \ref{app:GR}). $\alpha$ is the dimensionless Shakura-Sunyaev (\citealt{ShakuraSunyaev73}) free parameter that relates the effective viscosity $\nu$
to the scale height $H$ and the sound speed $c_{\rm s}$, by the ansatz $\nu =\alpha c_{\rm s} H$\label{eq: SS alpha anzats}.

The second master equation is energy balance:
\begin{equation}
F^{+}-F^{-}=u^{\rm r}\left(\frac{{\rm d}(UH)}{{\rm d}r}-\frac{(U+P)}{\rho}\frac{{\rm d}(\rho H)}{{\rm d}r}\right).\label{eq: collapsar energy relation}
\end{equation}
This equation relates the difference between local viscous heating (denoted by $F^+$, as described in CB07) and local cooling processes $F^-$ (representing the combined effects of neutrino-antineutrino emission and the photodisintegration of $\alpha$-particles, see also CB07), to heat transfer due to advection (right-hand side of Eq. \ref{eq: collapsar energy relation}). Here $u^r$ is the radial component of the mean 4-velocity for the fluid flow, and is related to the steady state accretion rate $\dot{M}=-4\pi u^r H \rho$\label{eq: baryon conservation} (from baryon conservation). All heating and cooling rates $F^{\pm}$ are per-area and defined for a single face of the disk.

The third master equation is the lepton number conservation equation:
\begin{equation}
\dot{S}_{\rm \nu}-\dot{S}_{\rm \thickbar{\nu}}=-\frac{u^{\rm r}}{ r}\frac{{\rm d}}{{\rm d}r}\left(r H \left(Y_{\rm e}\frac{\rho}{m_{\rm p}}+(n_{\rm \nu}-n_{\rm \thickbar{\nu}})\right)\right).\label{eq: lepton number advection equation}\end{equation}
This is an advection equation of the form $\dot{S} +\Vec{v}\cdot \nabla S=0$ where $S$ is the lepton number surface density (here we implicitly assume that $\nabla\cdot\Vec{v}$ is negligible), expressed in cylindrical coordinates. The left hand side is the local difference between the number fluxes of neutrino species $\dot{S}_{\rm \nu}$ and $ \dot{S}_{\rm \thickbar{\nu}}$. %(denoted in CB07 as $\dot{N}_{\rm \nu}$ and $ \dot{N}_{\rm \thickbar{\nu}}$ respectively). 
The right hand side represents advection of lepton number; inside the radial derivative are two terms, the first of which is the number density of protons (we assume charge neutrality, so this accounts for the lepton number density from electrons and positrons). The second term is the difference between neutrino and antineutrino number densities ($n_{\rm \nu}$ and $n_{\rm \thickbar{\nu}}$ respectively).  Here $m_{\rm p}$ is the proton mass and $Y_{\rm e} = n_{\rm p}/(n_{\rm n} + n_{\rm p})$ is the electron fraction\footnote{In regions of the disk with non-zero positron densities, this is not literally the electron fraction, but we use the standard terminology for $Y_{\rm e}$.}, expressed in terms of neutron and proton number densities ($n_{\rm n}$ and $n_{\rm p}$, respectively).

%should we disscuse that Andrei advised us to do this?
CB07 in their paper used a different advection equation for the lepton number:
\begin{equation}
\dot{S}_{\rm \nu}-\dot{S}_{\rm \thickbar{\nu}}=- u^{\rm r} H\left(\frac{\rho}{m_{\rm p}}\frac{{\rm d}Y_{\rm e}}{dr}+\frac{{\rm d}}{{\rm d}r}(n_{\rm \nu}-n_{\rm \thickbar{\nu}})\right).\label{eq: lepton number advection equation CB}\end{equation}
This is the same as Eq. \ref{eq: lepton number advection equation} but it neglects three different radial derivatives: of $r$, $H$ and $\rho$. Because $r$, $H$ and $\rho$ can change considerably with respect to distance $r$ we retain the more complete advective derivative\footnote{We thank Andrei Beloborodov for suggesting this revision to the advection equation to us.}. We now have three equations in three unknowns; we follow CB07 in parametrizing the unknown variables as electron fraction $Y_{\rm e}$, dimensionless temperature $\theta = k T / m_{\rm e} c^2$, and a dimensionless electron degeneracy parameter $\eta = \mu_{\rm e} / k T$.  Here $T$ is the mid-plane temperature, $\mu_{\rm e}$ is the electron chemical potential, $k$ is Boltzmann's constant, $c$ is the speed of light, and $m_{\rm e}$ is the electron mass.

Because of these changes we also need to change our numerical method of solution from the one suggested by CB07. The new lepton number equation forces us to perform a 3D root-finding, where we need to find the correct combination of  $\eta_{\rm e}(r_{\rm i}),$ $\theta(r_{\rm i})$ and $Y_{\rm e}(r_{\rm i})$ for each radius $r_{\rm i}$, that satisfy our three master equations. This is in contrast to the method of CB07 that assumes a $Y_{\rm e}(r_{\rm i})$ from the lepton equation, which allowed CB07 to only search for $\eta_{\rm e}(r_{\rm i})$ and $\theta(r_{\rm i})$ in a 2D root-finding procedure. The modified method is described in the flow chart of appendix Fig. \ref{app: Flow chart algo}.  We follow CB07 in setting our two outer boundary conditions at an outer radius $r=1000 r_{\rm s}$ (where $r_{\rm s} = 2 GM/c^2$ is the Schwarzschild radius): at this outer radius, we set $Y_{\rm e}=0.5$ (i.e. assuming an $\alpha$-particle dominated baryon population\footnote{As our model neglects the formation of nuclei heavier than $^4$He, the assumption that $Y_{\rm e}=0.5$ will break down at the largest radii, but (i) these are not important parts of our solutions, and (ii) we will later perform post-hoc checks to identify radii where heavy elements begin to form.}) and set $U/\rho = GM/r$ (i.e. a virialized, advective disk).  The second of these boundary conditions is clearly approximate but the inwards solution is insensitive to the exact choice for the prefactor on $U/\rho$ at the outer edge.

\subsection{Disk Instability and Fragmentation}
\label{sec:instability_theory}
Despite the laminar nature of any 1D idealization of hydrodynamics, real accretion disks are turbulent flows (indeed it is magnetized turbulence that sources the angular momentum transport approximated by the $\alpha$-viscosity ansatz we use; \citealt{Balbus1991}). 
 Any turbulent flow experiences transient fluctuations in density; for a massive disk, these perturbations could experience runaway growth due to self-gravity, or decrease back to the steady state, 1D configuration due to restoring forces. In the next section we shall discuss the onset and outcomes of gravitational instability.
\subsubsection{Toomre instability ($Q$)}
Toomre instability to gravitational collapse occurs when the gravitational force within the disk exceeds the restoring forces provided by both pressure gradients and the disk's differential rotation \citep{Toomre64}.
If we consider a rotating thin disk, with surface density $\Sigma$ and angular frequency $\Omega$ (here we use the Kerr metric angular frequency for circular equatorial orbits; see Appendix \ref{app:GR}), linear stability analysis shows that the disk is unstable to gravitational collapse if the following condition is satisfied:
\begin{equation}\label{eq: Tommre critarion}
    \begin{aligned}
     Q=\frac{c_{s} \Omega}{ \pi G \Sigma} < 1.
    \end{aligned}
\end{equation}
%\subsection*{Initial fragment mass
%}
where $Q$ is the dimensionless Toomre stability parameter.
We can make a rough estimate for the time it takes a fragment to form from the linear perturbation analysis used to derive the Toomre criterion \citep{BinneyTremaine08}. The fastest growing mode %corresponds to the following wave number:
%\begin{equation}
%k_{\rm fast}=\frac{ \pi G \Sigma}{c_{\mathrm{s}}^{2}}.
%\end{equation}
has a wavenumber of $k_{\rm fast} = \pi G \Sigma / c_{\rm s}^2$, and thus a wavelength of \citep{PiroPfahl07}:
\begin{equation}
\lambda_{\rm fast}=\frac{c_{\rm s}^{2}}{ \pi G \Sigma}=Q\frac{c_{\rm s}}{\Omega}.
\end{equation}
From the Toomre dispersion relation %$\omega_{\rm fast}^{2} \lesssim-(\Omega/Q)^2$, 
$\omega^2=c_{\rm s}^2 k^2 +4\Omega^2 - 2\pi G \Sigma |k|$, we find an estimated fragmentation time:
\begin{equation}
t_{\rm frag}\approx Q/\Omega .\label{eq: Duration of fragmentation approximation}
\end{equation}
Since fragmentation can only occur in regions with $Q\le 1$, the initial fragmentation timescale is by definition shorter than the local dynamical time, making it effectively the shortest timescale (locally) in disk evolution.  While the non-linear evolution of self-gravitating fragments is complex and may feature later bottlenecks, their initial formation occurs rapidly.  
%If the fragmentation timescale turns out to be small compared to both the duration of the entire collapse event (the average time for LGRBs is in tens of seconds, \citealt{Kouveliotou1993,Kumar2015,Bissaldi2017}) and the fragment migration times, we hypothesize that fragments might collide and scatter because of constant stochastic creation and rapid migration at different radii. For example if a fragment formed at an outer radius and starts migrating inwards it could overtake fragments that just finished forming at lower radii, potentially producing GWs and/or ejections of a few of those fragments (single fragments or even binaries), through few-body scatterings.

Assuming that only half of the wavelength contributes to an overdensity, we can estimate the fragment initial radius as $R_{\rm f}=\lambda_{\rm fast}/4$. Crudely assuming spherical symmetry of the density fluctuation, we estimate the initial fragment mass to be:
\begin{equation}
\begin{split}
&m_{\rm f,i}\approx\frac{4 \pi R_{\rm f}^3 \rho_{\rm i}}{3}= \frac{4 \pi \rho_{\rm i}}{3} \left(\frac{Q c_{\rm s}}{4 \Omega}\right)^3\\&\approx 5\cdot10^{-4} M_{\odot}Q^3\left(\frac{\Sigma}{5\cdot10^{18} {\rm g/cm}^{2}}\right)\left(\frac{H/r}{0.4}\right)^2\left(\frac{r}{50r_{\rm s}}\right)^2 \label{eq: initial fragment mass}.
\end{split}
\end{equation}
We note that this estimate for the fragment mass can differ dramatically from one computed using a fraction of the Jeans length as the initial fragment radius \citep{Shahamat2021}.  Had we used the Jeans length, it would have increased the initial fragment mass estimate by factor of $\sim10^2-10^4$. 
%In the earlier work \citep{ShahamatAbbassi20} pap
\begin{comment}
    \begin{equation}
\{m_{\rm f},\dot{m}_{\rm f},L_{\rm f},R_{\rm BH},h_{\rm GW},\Omega\}%\boldsymbol{\boldsymbol{\mathlarger{\mathlarger{\Uparrow}}}}
\end{equation}
\begin{equation}
\{r_{\rm f},R_{\rm f}\}
\end{equation}
\begin{equation}
L_{\rm f}\qquad \dot{m}_{\rm f}\gtrsim\dot{M}
\end{equation}
\end{comment}
We also note that our model, like CB07, neglects the role of dynamically important magnetic fields in determining disk structure.  While very strong magnetic fields are sometimes invoked for collapsar disks via the ``magnetically arrested disk'' (MAD) paradigm for jet launching \citep{Narayan+03, Tchekhovskoy+11}, strong but substantially sub-MAD fields can dominate the total pressure budget (in Eq. \ref{eq: pressure equality}) and create a ``magnetically elevated'' disk \citep{BegelmanPringle07}.  Magnetically elevated disks will often be more stable against fragmentation than their unmagnetized equivalents, both because of direct changes to $Q$ (i.e. a dominant magnetic pressure term will increase $c_{\rm s}$ and decrease $\Sigma$) and also because magnetic tension serves as an additional counter to self-gravity.  An investigation of fragmentation in strongly magnetized collapsar disks is beyond the scope of this work.

An implicit assumption in our discussion of Toomre instability is that the collapsar disk actually exists at large enough radii for $Q$ to fall below 1.  
In order for disk formation to occur, the angular momentum of the infalling matter needs to exceed $j_{\rm ISCO} \approx 2 GM/c$ \citep{Woosley1993}. Models of evolved stars, while uncertain, suggest that the circularization radius of the infalling gas, $r_{\rm circ}= j^2 / 2GM$, increases with time, while the accretion rate decreases with time \citep{WoosleyHeger2006, Gottlieb+22a}.  As fragmentation requires high accretion rates, the Toomre-unstable zone ($Q<1$) needs to be within $r_{\rm circ}$, and there will be a competition between the growth rates of $r_{\rm circ}$ and $r_{Q}$ (the smallest radius where $Q\le 1$).  Because the time evolution of $r_{\rm circ}$ ultimately depends on the details of the progenitor star structure, we do not treat it further here, but note that it may be an important limitation for parts of parameter space where $r_Q$ is large.   %As seen in Figs. 2,4,5 the fragmentation zone for the high accretion rates is within \sim 5-20 r_s (add citations on possible circularization radii range for comparison), and as the accretion rate decreases the fragmentation zone is pushed outwards. For fragmentation to occur at time t, the following condition needs to occur:
%r_{\rm circ}(t) >= fragmentation_radius(Mdot(t)).
%As  fragmentation_radius(Mdot(t)) is small for the initial high accretion rates and both radii increase with time it might be the case that this inequality holds. As the estimation of r_{\rm circ} is beyond the scope of this work we differ the answer to this question for future work.

\subsubsection{Cooling time ($\beta$)}\label{cooling time methodology}
Gravitational instabilities, which occur when the Toomre $Q<1$, can cause stochastic fluctuations in the density field to undergo gravitational collapse. In order for such overdensities to actually become self-bound fragments, they need to cool quickly enough so that the collapse happens before transient overdensities stochastically fluctuate back to the mean density. The classic work of \citet{Gammie2001edit2} conducted two-dimensional simulations using a simple cooling model, where the cooling time 
was set
inversely proportional to the angular frequency $\Omega$ and the efficiency of the cooling process was represented by the free parameter $\beta$, giving $t_{\rm cool} = \beta \Omega^{-1}$. We define the cooling time as the internal energy divided by the local cooling rate (i.e. neglecting advective cooling), letting us write:
\begin{equation}\label{eq: beta critarion}
    \begin{aligned}
     t_{\rm cool} = \frac{U_{\rm int}H}{F^{-}}=\frac{(U_{\rm \gamma}+U_{\rm b}+U_{\rm \nu}+U_{\rm \thickbar{\nu}})H}{F^{-}}.
    \end{aligned}
\end{equation}
Here $U_{\rm int}$ is the energy density without the zero-temperature contributions from electron and positron degeneracy. Redefining $t_{\rm cool}$ as a dimensionless $\beta$ parameter, we write:
\begin{equation}\label{eq: beta equation}
    \begin{aligned}
     \beta = \frac{(U_{\rm \gamma}+U_{\rm b}+U_{\rm \nu}+U_{\rm \thickbar{\nu}})H\Omega}{F^{-}}.
    \end{aligned}
\end{equation}
\citet{Gammie2001edit2} argued that a critical value of this dimensionless cooling time, $\beta_{\rm c}$, exists, so that if $\beta>\beta_{\rm c}$, inefficient cooling halts fragmentation and leads instead to a gravito-turbulent steady state \citep{Rafikov09}.  Conversely, if $\beta<\beta_{\rm c}$, then the disk undergoes fragmentation. With idealized two-dimensional simulations, \citet{Gammie2001edit2} found $\beta_{\rm c}\approx 3$.  We will later discuss more recent work that has found larger values for $\beta_{\rm c}$ or challenged the notion of a critical $\beta$ altogether \citep{Rice2005,Paardekooper2012,LinKratter2016, Hennebelle2021, Chen+23}. 

\dd{If the larger radii in our disks were to reach a gravitoturbulent state, they would likely self-regulate to the point of marginal stability with $Q\approx 1$.  A similar type of self-regulation would occur if accretion feedback onto fragments heats the disk \citep{SirkoGoodman03, Thompson2005}; in this scenario, the disk could even be over-stabilized to a $Q>1$ state \citep{GilbaumStone22}.  We do not consider the implications of self-regulation in this paper, deferring these questions for future work, but note that the impact of self-regulation will be minimized at the smallest radii in our Toomre-unstable zones.}

\subsection{Fragment migration and mass accumulation}
At radii $r$ with sufficiently low values of $Q$ and $\beta$, self-bound fragments will form in the disk.  Their internal evolution may involve non-trivial stages of dynamical collapse and slower, Kelvin-Helmholtz contraction.  We will qualitatively discuss this internal evolution later in \S \ref{sec:discussion} (though we defer quantitative modeling to future work).  Regardless of a fragment's internal properties, however, it will interact with the gaseous environment of the collapsar disk that spawned it, migrating radially through hydro-gravitational interactions and gaining mass through accretion.  We quantify these processes in this sub-section.
\label{sec: Fragment migration and mass accumulation}
\subsubsection{Reduced Bondi-Hoyle accretion
}
The standard Bondi-Hoyle formalism
(\citealt{BondiHoyle1944}, \citealt{Bondi1952} and \citealt{Shima1985},%\dd{Add Stone et al. 2017;Dittmannet al. 2021, Gilbaum and Stone 2022})
describes the accretion of matter onto a massive object from surrounding gas.  In this process, the gas in the accretion disk is pulled towards the massive object by its gravity.  %, and as it gets closer to the object, it picks up speed and forms a shock wave. 
The gas is decelerated in shocks and forms a smaller accretion disk around the embedded object, from which material can eventually accrete onto it. The accretion rate of gas onto the fragment's mini-disk is given by:
\begin{equation}
    \frac{{\rm d}m_{\rm f}}{{\rm d}t}=\sigma_{\rm f}(c_{{\rm s}}^{2}+V_{{\rm rel}}^{2})^{1/2}\rho.\label{eq: Bondi-Hoyle accretion rate}
\end{equation}
Where $\sigma_{\rm f}$ is an effective cross section for the accreted matter onto the fragment and $V_{\rm rel}$ is the magnitude of the relative velocity between the fragment (of semimajor axis $r$) and the ambient matter, and it is given by:
\begin{subequations}
    \begin{align}
    V_{\rm rel}&=\sqrt{V_{\rm \phi,rel}^2+V_{\rm r,rel}^2}, \\
    V_{{\rm \phi,rel}}&\approx c_{{\rm s}}\frac{\nabla_{P}}{2}\frac{H}{r}, \label{eq: subKeplerian} \\
    V_{\rm r,rel}&=u^r-\left(\frac{{\rm d}r}{{\rm d}t}\right)_{{\rm tot}}.
    \end{align}
\end{subequations}
Here, $V_{\rm \phi,rel}$ and $V_{\rm r,rel}$ are relative velocities in the azimuthal and radial directions, respectively.  $V_{\rm \phi,rel}$ originates from the sub-Keplerian motion of the gas, and Eq. \ref{eq: subKeplerian} defines this with the notation $\nabla_{x}=-{\rm d}\ln(x)/{{\rm d}\ln(r)}$.  To calculate $V_{\rm r,rel}$ we subtract the mean fluid radial velocity $u^r$ (from Eq. \ref{eq: baryon conservation}) off the total migration torque (given later on, in Eq. \ref{eq: total torque}). 

In standard Bondi-Hoyle formalism, gas in the disk will become gravitationally bound to the moving fragment if its impact parameter is smaller than the Bondi-Hoyle–Lyttleton radius, which is:
\begin{equation}
R_{{\rm BH}}=\frac{2 G m_{\rm f}}{c_{{\rm s}}^2+V_{\rm rel}^2}\label{eq: Bondi-Hoyle radius}.
\end{equation}
and defines the effective cross section as $\pi R_{{\rm BH}}^2$. In our case, the gas reservoir is the surrounding collapsar disk, so the accretion flow can be limited by the geometrical properties of the disk (i.e. the scale hight $H$) and the gravitational pull of the BH. %This is only true if $R_{{\rm BH}}$ is smaller than the scale hight $H$ and if we neglect the gravitational pull of the BH. 
In order to take these two limiting effects into account, we use the reduced Bondi-Hoyle formalism described in \cite{Stone2017, Dittmann+21, GilbaumStone22}, where an elliptical cross section is given by:
\begin{equation}
    \sigma_{\rm f}=\pi\cdot{\rm min}(R_{{\rm BH}}, \sqrt{R_{{\rm BH}}R_{{\rm H}}}) \cdot{\rm min}(R_{{\rm BH}}, \sqrt{R_{{\rm BH}}R_{{\rm H}}}, H).
\end{equation}
Where $\sqrt{R_{{\rm BH}}R_{{\rm H}}}$ is the maximum impact parameter within a Hill sphere of radius $R_{{\rm H}}=r\left(\frac{m_{\rm f}}{3 M}\right)^{1/3}$ \citep{MurrayAndDermott1999}.  \dd{These types of geometrical reductions in the naive Bondi-Hoyle cross-section have been tested and validated against numerical hydrodynamics simulations \citep{Choksi+23, Li+23}.}
%we get the following term:
%\begin{equation}
%\left(\frac{{\rm d}m_{{\rm f}}}{{\rm d}t}\right)_{{\rm BH}}\approx\frac{4\pi G{}^{2}\rho}{\left(c_{{\rm s}}^{2}(1+\frac{\nabla_{P}}{2}\frac{H}{a})^{2}+V_{{\rm a,rel}}^{2}\right)^{3/2}}m_{{\rm f}}^{2}\label{eq: Bondi-Hoyle accretion rate with cs}
%\end{equation}
%When $V_{{\rm a,rel}}{}^{2}/c_{{\rm s}}^{2}(1+\frac{\nabla_{P}}{2}\frac{H}{a})^{2}<0.1$ we neglect $V_{\rm a,rel}$, which simplifies Eq. \ref{eq: Bondi-Hoyle accretion rate with cs} to be:

\begin{comment}
The rate of mass accumulation into the mini-disk is given by the following formula:
\begin{equation}
\left(\frac{{\rm d}m_{\rm f}}{{\rm d}t}\right)_{{\rm BH}}=\frac{4\pi G^{2}\rho}{(c_{{\rm s}}^{2}+V_{{\rm rel}}^{2})^{3/2}}m_{\rm f}^{2} .\label{eq: Bondi-Hoyle accretion rate}
\end{equation}
\end{comment}

Inputting our approximation for $V_{\rm \phi,rel}$ into the Bondi-Hoyle accretion rate from Eq. \ref{eq: Bondi-Hoyle accretion rate}, and neglecting $V_{\rm r,rel}$ (an approximation we will check {\it post hoc} later), we get that the reduced accretion rate is:
\begin{equation}
\left(\frac{{\rm d}m_{{\rm f}}}{{\rm d}t}\right)_{{\rm RBH}}\approx \sigma_{\rm f} c_{{\rm s}}\rho\sqrt{1+\left(\frac{\nabla_{P}}{2}\frac{H}{r}\right)^2}\label{eq: reduced Bondi-Hoyle accretion rate with approx} .
\end{equation}

\subsubsection{Fragment migration}\label{sec: Fragment migration}
After formation, the fragment will experience torques from interactions with the surrounding gas and the BH. In this sub-section we will quantify the relevant torques that determine the migratory behavior of the fragment.  

The first torque we will consider is the ``type I'' torque arising from gravitational back-reaction: the gravity of the fragment linearly perturbs the collapsar disk, creating linear overdensities and underdensities, which in turn exert gravitational torques on the perturbing fragment \citep{GoldreichTremaine80}. The torque exerted by the disk on the orbiting body usually causes it to lose energy and momentum\footnote{\dd{In this paper, we neglect the role of disk turbulence in generating stochasticity in the migration torques that we employ.  Our treatment of migration is deterministic, but strong turbulence (from e.g. the gravitoturbulent $\beta \gg 1$ regime) can in some contexts lead to important stochasticity in migration \citep{Johnson+06, BaruteauLin10, Yinhao2024}.}}, causing it to migrate inward. We write the type I torque as \citep{Paardekooper2010}: 
\begin{equation}
\Gamma_{\rm I}=-C_{{\rm I}}\left(\frac{2m_{\rm f}}{M}\right)^{2}\left(\frac{H}{r}\right)^{-2}\Sigma r^{4}\Omega^{2} .
\end{equation}
%Originally in \cite{Paardekooper2010} the $C_{{\rm I}}$ was a factor, dependant only on the temperature ($T(a)\propto a^{-\nabla_{\rm T}(a)}$) and surface density ($\Sigma(a)\propto a^{-\nabla_{\rm \Sigma}(a)}$) power law behaviour. Later work by *** add two citations *** has made a different relation that fitted simulations better, which integrated the effect of Lindblad torque and a scaling regarding the adiabatic index $\gamma\equiv\nabla_{\rm P}/{\nabla_{\rm \rho}}$ (** add citation and reference to appendix \ref{app: fragmentation calculations}).
The dimensionless prefactor $C_{\rm I}$ has been calibrated from numerical simulations by many authors, and depends primarily on the local logarithmic slopes of disk temperature ($\nabla_T$) and surface density ($\nabla_\Sigma$).  We use the recent calibration of \citet{JimenezMasset17}, which is given explicitly in Appendix \ref{app: migration calculations}.

If the fragment is massive enough to form a gap in the disk (see appendix \ref{app: gap formation}) then Bondi-Hoyle mass accumulation is reduced and type I torques switch to ``type II'' ones. Type II migration %is caused by how interactions between the fragment and the walls of the gap cause a net gravitational pull on the fragment. 
refers to gravitational interactions with the non-linearly perturbed gas in and around the gap region \citep{Ward97}. The ratio between the surface densities with and without a gap is given by the proportionality factor \citep{Kanagawa+18}:
\begin{equation}
    \frac{\Sigma_{\rm gap}}{\Sigma}=\frac{1}{1+0.04\kappa_{{\rm gap}}}.
\end{equation}\label{eq: Sigma gap ratio}
Where $\kappa_{{\rm gap}}$ is given in appendix \ref{app: gap formation} and $\kappa_{{\rm gap}}\geq20$ is the gap opening criteria, which sets the transition between type I and type II torques and accretion rates. 

The resulting type II torque is approximately proportional to the type I torque, scaled by a proportionality factor \citep{GinzburgSari18}. Assuming that the scale height remains unchanged during gap formation, the ratio of surface densities is equivalent to the ratio of densities. This allows us to express the following relationships at gap formation:
\begin{subequations}
    \begin{align}
    \Gamma_{{\rm II}}&=\frac{1}{1+0.04\kappa_{{\rm gap}}}\Gamma_{{\rm {\rm I}}}\label{eq: Torque gap ratio}\\
    \left(\frac{{\rm d}m_{{\rm f}}}{{\rm d}t}\right)_{{\rm RBHG}}&=\frac{1}{1+0.04\kappa_{{\rm gap}}}\left(\frac{{\rm d}m_{{\rm f}}}{{\rm d}t}\right)_{{\rm RBH}}.
    \end{align}    
\end{subequations}
Where `RBHG' stands for reduced Bondi-Hoyle at gap. For simplicity we use the following notation:
\begin{subequations}
    \begin{align}
    \Gamma_{{\rm I/II}}&=\begin{cases}
\displaystyle\Gamma_{{\rm I}} & \kappa_{{\rm gap}}<20\\[2ex]
\displaystyle\Gamma_{{\rm II}} & \kappa_{{\rm gap}}\geq20
\end{cases},\\
    \Dot{m}_{{\rm f}}&=\begin{cases}
\displaystyle\left(\frac{{\rm d}m_{{\rm f}}}{{\rm d}t}\right)_{{\rm RBH}} & \kappa_{{\rm gap}}<20\\[2ex]
\displaystyle\left(\frac{{\rm d}m_{{\rm f}}}{{\rm d}t}\right)_{{\rm RBHG}} & \kappa_{{\rm gap}}\geq20.
\end{cases}\label{eq: total accretion rate transition}
    \end{align}
\end{subequations}
\dd{We note that this reduced Bondi-Hoyle accretion rate is self-limited by the gap-opening process. Once the fragment exceeds a critical gap-opening mass, the depleted surface density in its annular region will greatly reduce its ability to grow further.  Prior to gap opening, the reduced Bondi-Hoyle accretion rate can become quite large, in some cases even exceeding $\dot{M}$ by a factor of a few (Fig. \ref{fig:AccretionRateRatios_3_Panel}), but fragment growth is quickly saturated by gap opening and $\dot{m}_{\rm f}$ can only exceed $\dot{M}$ for a few milliseconds (see Fig. \ref{app:DurationOfHighAccretion_alpha002M20_a0}).} 

\dd{Our estimate for the accretion rate assumes that annular gaps can only open due to nonlinear angular momentum exchange with surrounding gas via Lindblad resonances (i.e. the classic Type II picture, in which a massive perturber pumps angular momentum into nearby gas faster than it can be viscously refreshed).  The unusual conditions in collapsar disks may motivate a second form of gap opening, one in which (reduced) Bondi-Hoyle accretion depletes surrounding gas faster than the viscous refresh time.  While a full examination of these ``depletion gaps'' is beyond the scope of this work, we remark that it may be an important topic for future examination, as the emergence of such gaps would diminish final fragment masses.}

Gas in accretion disks moves on quasi-circular orbits where attraction from the BH is balanced primarily by centrifugal force, but with a weak contribution from radial pressure gradients.  The radial pressure gradient causes azimuthal gas velocities to be generically sub-Keplerian, creating another source of migratory torque. As the fragment accretes mass with sub-Keplerian specific angular momentum, it decreases the fragment's specific angular momentum, leading to inward migration. The total torque exerted on the fragment from the gas\footnote{Due to the uncertainties of fragment accretion and the collapsar environment, we neglect ``thermal torques'' arising from accretion feedback from the fragment on the collapsar disk \citep{Masset17, Grishin+24}, though these can be important in other settings.} is thus the sum of the type ${\rm I/II}$ torque and the torque of the accumulated gas on the fragment:
\begin{subequations}
    \begin{align}
    \Gamma_{{\rm gas}}&=\Gamma_{{\rm I/II}}+\Gamma_{{\rm acc}}=\Gamma_{{\rm I/II}}+\Dot{m}_{{\rm f}}\mathcal{L}_{\rm gas}\label{eq: sum of gas torques}\\
    \mathcal{L}_{\rm gas}&\approx \mathcal{L}\left(1-\nabla_{P}\left(\frac{H}{r}\right)^{2}\right)\label{eq: specific gas torque}
    \end{align}
\end{subequations} 

Here, $\mathcal{L}$ is the specific angular momentum (see appendix \ref{app:GR}). The argument in parentheses in Eq. \ref{eq: specific gas torque} reduces the gas angular momentum to its sub-Keplerian value accounting for radial pressure gradients.
By expressing the torque as the rate of change of angular momentum we can make the following equations:
\begin{equation}
\Gamma_{\rm gas}=\frac{{\rm d}L}{{\rm d}t}=\frac{{\rm d}\left(m_{\rm f}\mathcal{L}\right)}{{\rm d}t}=\Dot{m}_{{\rm f}}\mathcal{L}+m_{\rm f}\left(\frac{{\rm d}r}{{\rm d}t}\right)_{{\rm gas}}\frac{{\rm d}\mathcal{L}}{{\rm d}r} .\label{eq: rate of change of angular momentum}
\end{equation}
%Where $\mathcal{L}$ is the specific angular momentum and it is found in the appendix 
%\dd{\ref{app:GR}} at Eq. \ref{specific angular momentum GR}.

Combining Eqs. \ref{eq: sum of gas torques}, \ref{eq: specific gas torque} and \ref{eq: rate of change of angular momentum} lets us write the following expression for the total migration rate due to gas interactions:
\begin{equation}
    \left(\frac{{\rm d}r}{{\rm d}t}\right)_{{\rm gas}}=\left(m_{{\rm f}}\left(\frac{{\rm d}\mathcal{L}}{{\rm d}r}\right)\right)^{-1}\left(\Gamma_{{\rm I/II}}-\Dot{m}_{{\rm f}}\mathcal{L}\nabla_{P}\left(\frac{H}{r}\right)^{2}\right) \label{eq: rate of migration from gas and type 1}.
\end{equation}

Migration due to gravitational wave emission can also become relevant as the fragment gains mass and shrinks its semimajor axis. In our calculation of the GW migration rate we neglect the contribution of mass enclosed in the disk, as GW torque dominates the migration in the inner region of the disk where the enclosed mass is small. Using the leading post-Newtonian gravitational wave emission rate \citep{peters1964}, and assuming quasi-circular motion, we write the gravitational wave migration rate as: 
\begin{equation}
    \left(\frac{{\rm d}r}{{\rm d}t}\right)_{{\rm GW}}=-\frac{64}{5}\frac{G^{3}M m_{\rm f}(M+m_{\rm f})}{c^{5}r^{3}} .\label{eq: type GW rate of migration}
\end{equation}
% \subsection*{Total migration time?
% }
% \begin{equation}
% \Delta t=\frac{L}{\Gamma_{I}+\Gamma_{{\rm GW}}} .
% \end{equation}
The total migration rate is thus:
\begin{equation}
\left(\frac{{\rm d}r}{{\rm d}t}\right)_{{\rm tot}}=\left(\frac{{\rm d}r}{{\rm d}t}\right)_{{\rm gas}}+\left(\frac{{\rm d}r}{{\rm d}t}\right)_{{\rm GW}}
\label{eq: total torque}
\end{equation}
Using Eqs. \ref{eq: total accretion rate transition} and \ref{eq: total torque} we can make the following differential relation:
\begin{equation}
\Dot{m}_{{\rm f}}\left[{\left(\frac{{\rm d}r}{{\rm d}t}\right)_{\rm tot}}\right]^{-1}=\frac{{\rm d}m_{\rm f}}{{\rm d}r}.\label{eq: accertion with radius 1}
\end{equation}
In order to numerically solve this differential equation we need to choose an initial condition, meaning an initial fragment at a specific initial mass and distance from the BH. In general, we choose such an initial condition from a given disk model (with fixed $M$, $\dot{M}$, $\alpha$) by examining all the possible fragmentation radii ($Q\le 1$, $\beta \le \beta_{\rm c}$) and picking the one that (i) has the maximum initial fragment mass and (ii) also has a viscous time $t_{\rm visc}=(\alpha\Omega(H/r)^{2})^{-1}$ below $5M/\dot{M}$ and (iii) an enclosed disk mass smaller than $M$. Disk regions that fail conditions (ii) and (iii) would be unlikely to be described by a steady state disk model\footnote{We pick a prefactor of $5$ because bulk re-adjustments of $\Sigma$ in simple $\alpha$-disk theory will often propagate on timescales up to an order of magnitude shorter than the naive viscous time, \citet{Pringle81}.} or a Kerr metric model, respectively.  In Fig. \ref{fig: Initial_Fragment_Mass_3_Panel}, we illustrate these selected radii in different disk models using pink points.  Solving the differential equation for ${\rm d}m_{\rm f}/{\rm d}r$ using this initial condition gives us the fragment mass evolution with radius: $m_{\rm f}(r)$. 

Finally, we can use Eq. \ref{eq: total torque} and the solution of $m_{\rm f}(r)$ to calculate the migration time of the fragment from its formation site until it reaches a smaller radius $r$:
\begin{equation}
\Delta t(r)=\int_{r_{{\rm init}}}^{r}{\left(\frac{{\rm d}r'}{{\rm d}t}\right)_{\rm tot}}^{-1}{\rm d}r' .\label{eq: final fragment migration time}
\end{equation}
Within the context of our model assumptions, we now have the full picture of the fragment's bulk evolution in mass, space, and time. 
%\subsection*{Fragmentation timescale ($Q/\Omega$)
%}

%\clearpage
% Results
\section{Results}
\label{sec:results}

Here we explore the general results from our disk modelling, exploring disk stability properties, as well as bulk fragment evolution.  

\subsection{Disk Stability}
\label{sec:instability_results}
In this section we present our parameter survey across a wide range of central masses $M$, accretion rates $\dot{M}$ and effective viscosities $\alpha$. We have chosen physically motivated values for these parameters: for example, we consider the full range of black hole masses that are likely to result from stellar evolution, $3 \lesssim M/M_\odot \lesssim 50$ \citep{Spera2017}, and a range of Shakura-Sunyaev viscosity coefficients, $0.01<\alpha < 0.3$, that is broadly consistent with magnetohydrodynamic (MHD) accretion simulations  \citep{Penna2013, Siegel+19, Jiang2019}. %In order to estimate the accretion rate, a good understanding of the radiative efficiency (add an explanation) of these processes is needed. It can be seen that this is still not well understood from how, a few different multi-D simulations of iron core collapse in rotating massive stars, 
The accretion rates in collapsar disks remain uncertain, and multi-dimensional simulations of iron core collapse in rotating, massive stars find a large range of peak $\dot{M}$, spanning three orders of magnitude, from $0.01$ to $10 M_\odot/{\rm s}$  \citep{Ott2011,Sekiguchi2011,Fujibayashi2022}. As we shall see, these high accretion rates do not need to be sustained for the duration of the collapsar event but only for relatively short periods of time (long enough for a quasi-steady state disk solution to be established).  We focus on higher accretion rates ($0.1 M_\odot~{\rm s}^{-1} \le \dot{M} \le 10 M_\odot~{\rm s}^{-1}$) because this is the portion of parameter space where gravitational instability can emerge (CB07).  For the results shown in the main body of the paper we set the BHs spin to be $a=0.95$, however we have also recalculated some of our findings for a non-fiducidal, non-spinning case in Appendix \ref{app: a0}.  As we will see, fragmentation almost always occurs at relatively large radii in collapsar disks where the effect of BH spin is relatively unimportant.  The final evolution of fragments can be substantially more sensitive to spin, as this will affect how far inwards fragments can migrate in the disk, but we will discuss this issue later in \S \ref{sec:GWs}.
\begin{comment}
As hyper-Eddington luminosities are measured from a few different highly energetic sources, such as GRBs, theorists and observers try to infer the associated accretion rates. In order to estimate the accretion rate a good understanding of the radiative efficiency of these processes, which is still not well understood, is needed (see Eq. \ref{eq: radiative efficency}). This uncertainty can be seen from how, a few different multi-D simulations of iron core collapse in rotating massive stars, find a large range of accretion rates that spans three orders of magnitude, from $0.1-10 M_\odot/s$ \citep{Ott2011,Sekiguchi2011,Fujibayashi2022}.
We focus our survey on the higher range of plausible accretion rates $0.2 M_\odot~{\rm s}^{-1} \le \dot{M} \le 10 M_\odot~{\rm s}^{-1}$ 
%(as discussed in \S \ref{accretion disk subsection}; 
\citealt{Ott2011,Sekiguchi2011,Fujibayashi2022}). We focus on higher accretion rates because this is the portion of parameter space where gravitational instability is expected to emerge (CB07).
\end{comment}
\subsubsection{Toomre instability: $Q$
}
\begin{figure}
    \centering
    \includegraphics[width=0.48\textwidth]{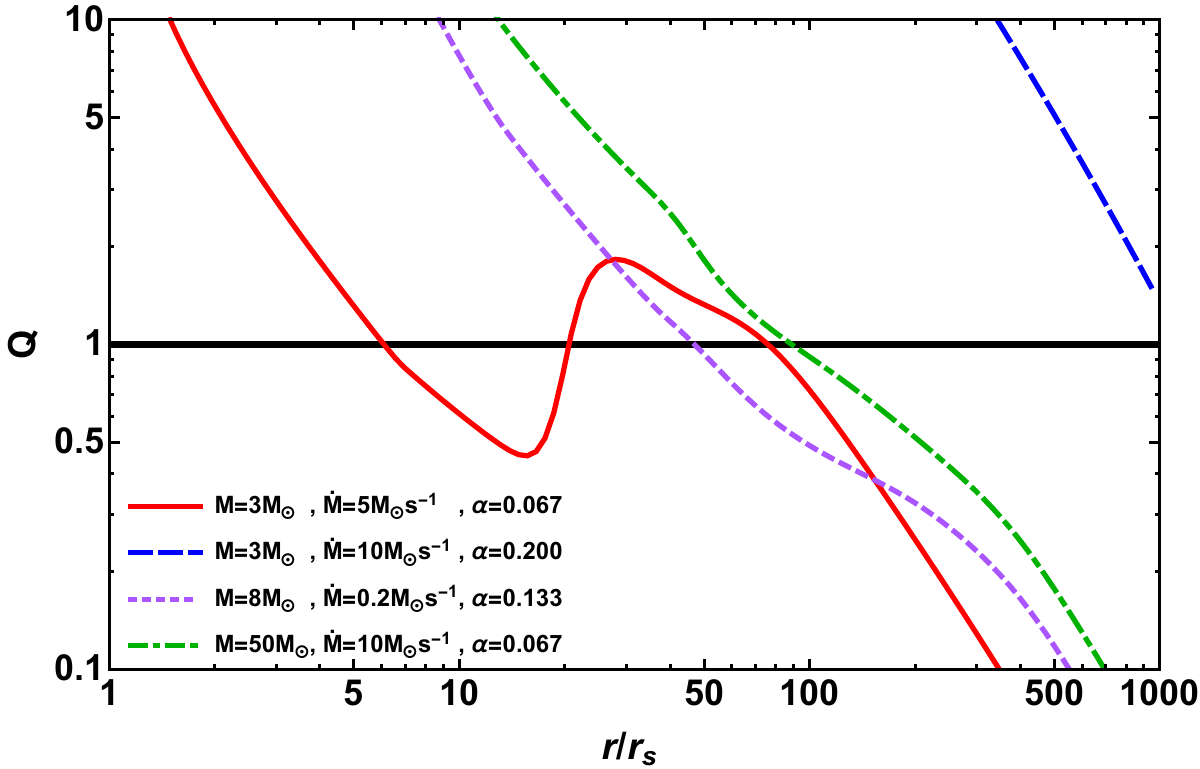}
    \caption{Toomre's $Q$ parameter as a function of radius (normalized by $r_{\rm s}$), for four 
    disk solutions with different
    mass $M$, accretion rate $\dot{M}$, and viscosity parameter $\alpha$. 
    The spin parameter $a=0.95$ ($r_{\rm ISCO}=0.97 r_{\rm s}$)
    for all solutions. The black dashed line represents $Q=1$, and regions under it are gravitationally unstable.  Both the purple solution (central mass of $3{M}_{\odot}$, accretion rate $0.4{M}_{\odot} {\rm s^{-1}}$ and $\alpha=0.010$) and the green solution (central mass of $3{M}_{\odot}$, accretion rate $10{M}_{\odot} {\rm s^{-1}}$ and $\alpha=0.200$) have one outer region of Toomre instability that extends until $r_{\rm out}=1000$ $r_{\rm s}$. These two solutions are quite similar, illustrating the tradeoff between $\dot{M}$ and $\alpha$ in setting $\Sigma$ and thus $Q$.  The blue solution, with a central mass of $8{M}_{\odot}$, accretion rate $0.2{M}_{\odot} {\rm s^{-1}}$ and $\alpha=0.133$ has no Toomre unstable regions inside $r_{\rm out}$. The red solution, with a central mass of $50{M}_{\odot}$, accretion rate $10{M}_{\odot} {\rm s^{-1}}$ and $\alpha=0.067$, has a bifurcation of the unstable region.  Collapsar disks become more prone to instability for higher $\dot{M}$ and lower $\alpha$; $M$ is less important.}
    \label{fig: Q examples}
\end{figure}

Every time we ran our model to calculate a different disk solution, varying $M$, $\dot{M}$,  $a$, and $\alpha$, we calculated the radial profile of the Toomre $Q$ parameter (see Eq. \ref{eq: Tommre critarion}) and searched for regions where $Q<1$. Fig. \ref{fig: Q examples} shows a few examples of the radial behavior of $Q$. We can classify the behavior of the Toomre $Q$ parameter in each of our disk models into three different categories: (1) $Q>1$ everywhere for $r<r_{\rm out}$; (2) $Q>1$ at small radii, continuously transitioning to a $Q<1$ zone that lasts until the outer radius $r_{\rm out}$; (3) Bifurcation of the unstable regions (where $Q<1$) into two distinct zones, an inner region at lower radii and an outer region that extends from some radius until $r_{\rm out}$, with two stable ($Q>1$) zones as well.  The inner unstable zone in the bifurcated solutions arises because of a bump in surface density $\Sigma$ at small radii.
% efficient neutrino cooling that reduces the inner disk aspect ratio $H/r$.
%a spike in surface density $\Sigma$ at small radii.
\begin{figure}
\includegraphics[width=0.48
\textwidth]{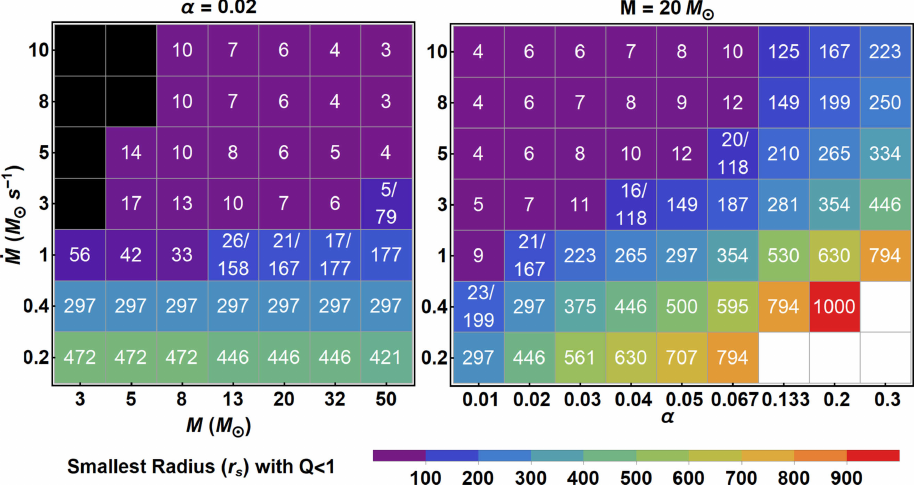}
\caption{The smallest radius (normalized by $r_{\rm s}$) with Toomre parameter $Q<1$, $r_Q$, shown on two array plots. {\it Left panel}: we take a constant viscosity parameter $\alpha=0.02$ and vary the central mass $M$ and the accretion rate $\dot{M}$. {\it Right panel}: we take a constant central mass $M=20{M}_{\odot}$ and vary the viscosity parameter $\alpha$ and the accretion rate $\dot{M}$. The white blank regions in the right panel have $Q>1$ for all $r < r_{\rm out}$. The black colored regions are points in parameter space for which our numerical model failed to reach a solution. In both panels we see examples of bifurcated solutions (where two numbers are separated by a slash). The top number in the bifurcated slot is the smallest radius of the inner $Q<1$ region and the bottom number is the smallest radius of the outer $Q<1$ region. Generally, varying $M$ does not significantly change the smallest (normalized) radius with $Q<1$, while variations in $\dot{M}$ and $\alpha$ have a bigger effect.}
\label{fig: Lowest_Q_Radius_2_panel}
\end{figure}

Fig. \ref{fig: Lowest_Q_Radius_2_panel} shows two cross sections in our parameter survey. The left panel is an array plot with constant viscosity parameter $\alpha=0.02$% varying the central mass from $3$ to $50{M}_{\odot}$ and the accretion rate from $0.2{M}_{\odot} {\rm s^{-1}}$ to $10{M}_{\odot} {\rm s^{-1}}$
; the right panel is an array plot with a constant central mass $M=20{M}_{\odot}$.  Each panel shows $r_{\rm Q}$, the radius at the inner edge of a Toomre-unstable region where $Q=1$ (in bifurcated solutions, we show both inner edges, and $r_{\rm Q}$ is defined by the dual condition that $Q=1$ and ${\rm d}Q/{\rm d}r < 0$).   % varying the viscosity parameter $\alpha$ from $0.01$ to $0.45$ and the accretion rate from $0.2{M}_{\odot} {\rm s^{-1}}$ to $10{M}_{\odot} {\rm s^{-1}}$. 
From Fig. \ref{fig: Lowest_Q_Radius_2_panel} we reach the following conclusions: (1) decreasing the $\alpha$ value causes the instability zones (i.e. $Q<1$) to appear at lower accretion rates, to extend to lower radii and to bifurcate (in some solutions) at smaller accretion rates; (2) changing the central mass $M$ has a very weak impact on the radius (normalized by $r_{\rm s}$) where $Q=1$; (3) as we increase the accretion rate $\dot{M}$, the unstable regions extend to smaller radii.  Conclusions (1) and (3) can be understood from the inverse dependence of $\Sigma$ on $\alpha$ and the proportionality between $\Sigma$ and $\dot{M}$ (i.e. $\dot{M}=-2\pi u^r \Sigma$).  A notable caveat to  conclusion (2) is that bifurcated instability zones only appear for relatively large $M$.
%NCS: Yonatan, see if you can come up with a simple physical explaation for point (2)

In collapsars with limited angular momentum budgets, fragmentation is only possible at times when $r_{\rm circ}(t) > r_Q$.  Fig. \ref{fig: Lowest_Q_Radius_2_panel} shows that for higher accretion rates, $r_Q \sim 5-20 r_{\rm s}$, so that the circularization radius is likely not restrictive, while for lower accretion rates (and lower $\alpha$ values), $r_Q \sim 100-1000 r_{\rm s}$, and inadequate disk angular momentum may limit the ability of fragmentation to occur.  

\subsubsection{Cooling time: $\beta$}\label{subsubsec: cooling time beta}
Although classic work on self-gravitating disks either finds \citep{Gammie2001edit2} or assumes \citep{Nayakshin+07} that $\beta < \beta_{\rm c}=3$ is a necessary condition for fragmentation, more recent studies have complicated this picture.
\citet{Paardekooper2012} first suggested that fragmentation is a stochastic process and can occur with some probability for every value of $\beta$. Alternatively, in \citet{LinKratter2016}, it is argued that there is actually no $\beta_{\rm c}$ for which a $Q<1$ zone is truly stable, but rather the linear growth rate of overdensities simply decreases with increasing cooling time (i.e. fragmentation can occur at any $\beta$ value, but it takes more dynamical times to develop for higher $\beta$). A recent 3D simulation of a $Q<1$ disk by \citet{Hennebelle2021} finds continued fragmentation for $\beta = 8$ and $\beta = 10$, and that for such large $\beta$ values, the emergence of self-bound fragments is stochastic. Their boundary does not converge fully with increasing resolution, leading \citet{Hennebelle2021} to conclude that a definite boundary between fragmenting and non-fragmenting $Q<1$ regimes may not fully exist, and that the transition between the two may be continuous. %Therefore, they recommend using a more statistical approach to further study this transition, a similar conclusion to Paardekooper. 
A recent local radiation-hydrodynamics simulation of $Q<1$ disks has also challenged the classic picture of $\beta_{\rm c}$, finding that as radiation pressure grows in importance (i.e. as the equation of state becomes softer), fragmentation will develop at higher values of $\beta$, in some cases as large as $\beta\approx 35$ \citep{Chen+23}.  This set of simulations shows that $\beta_{\rm c}$ can be enhanced dramatically as the gas adiabatic index $\gamma \to 4/3$, as is the case in large radial zones of a neutrino-cooled collapsar disk.  Interestingly, our disk models also feature zones with $\gamma < 4/3$, hinting that fragmentation may be able to proceed with even larger $\beta$ values.

\begin{figure}
    \centering    \includegraphics[width=0.48\textwidth]{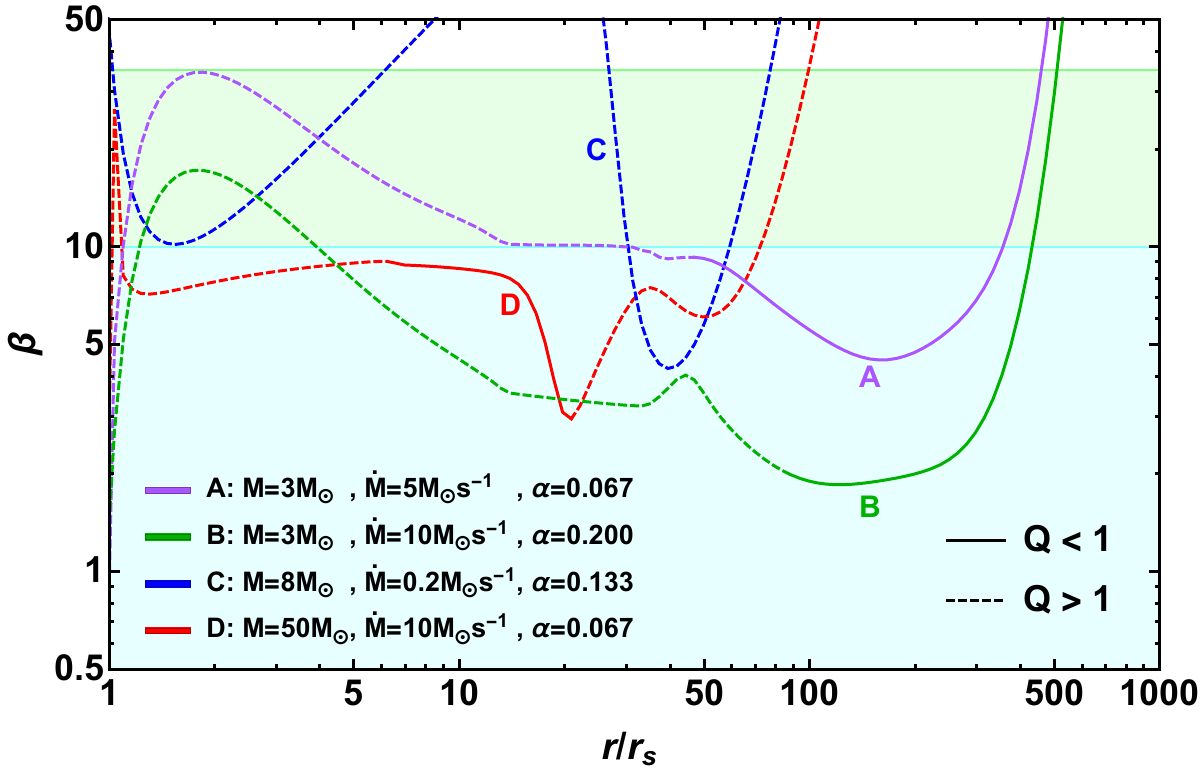}
    \caption{Gammie's $\beta$ parameter as a function of normalized radius, for the same four different disk solutions seen in Fig. \ref{fig: Q examples}. Lines are solid where $Q<1$ and dashed where $Q>1$. %The purple solution with central mass of $3{M}_{\odot}$, accretion rate $0.4{M}_{\odot} {\rm s^{-1}}$ and $\alpha=0.010$, reaches below $\beta = 28$ at the $Q<1$ regions. The green solution with central mass of $3{M}_{\odot}$, accretion rate $10{M}_{\odot} {\rm s^{-1}}$ and $\alpha=0.200$, reaches below the $\beta = 3$ 
    %\rdd{line}{level}, 
    %which is the original critical limit be Gammie. The blue solution, with a central mass of $8{M}_{\odot}$, accretion rate $0.2{M}_{\odot} {\rm s^{-1}}$ and $\alpha=0.133$ has $Q>1$ at all radii. The red solution, with a central mass of $50{M}_{\odot}$, accretion rate $10{M}_{\odot} {\rm s^{-1}}$ and $\alpha=0.067$, has a bifurcation of the $Q<1$ region and reaches the $\beta = 3$. 
    We shade two zones of $\beta$: in light-green, $10<\beta<35$, and in light-cyan $\beta<10$, reflecting upper limits for fragmentation found in the simulations of \citet{Chen+23} and \citet{Hennebelle2021}, respectively.  The solutions that achieve Toomre instability (purple, green, red) possess $\beta<10$ sections of their $Q<1$ zones, although it is only in relatively extreme cases (e.g. the green $M=3M_\odot$, $\dot{M}=10 M_\odot~{\rm s}^{-1}$) that $\beta$ falls below the classic $\beta_{\rm c}=3$ value \citep{Gammie2001edit2}.}
    \label{fig: beta examples}
\end{figure}

For each run of our disk models, we calculated the radial profile of the dimensionless cooling time $\beta$ (Eq. \ref{eq: beta equation}) and looked for regions where both $Q<1$ and $\beta<35$ (for the rest of the main text, we set $\beta_{\rm c}=35$ as our definition of the critical fragmentation boundary, motivated by \citealt{Chen+23}; in Appendix \ref{app: beta 10} we show more conservative results with $\beta_{\rm c}=10$). Fig. \ref{fig: beta examples} shows $\beta$ profiles for four different disk solutions. %In the figure, we mark some important values of $\beta$ (10 and 35) that we adopted from the models mentioned above. Later on we will be using these values to classify different $\beta$ ranges in our plots. In figure \ref{fig: beta examples} 
%Here we see a few examples of possible fragmentation regions, as seen by the red, magenta and green lines. These solutions exhibit $\beta$ values that go below the critical $\beta$ of 35 and even  below $\beta=2$ at the regions where the Toomre $Q$ is less than 1.
In the selected examples, every disk model with $Q<1$ regions contains extensive regions of Toomre instability with $\beta < 10$.  It is less common for $\beta$ to go below the classic $\beta_{\rm c}=3$ value \citep{Gammie2001edit2}, but in some of our models, particularly those with high accretion rates, it does (see e.g. the $M=3M_\odot$, $\dot{M}=10M_\odot~{\rm s}^{-1}$ example in Fig. \ref{fig: beta examples}).

\begin{figure}
\includegraphics[width=0.48\textwidth]{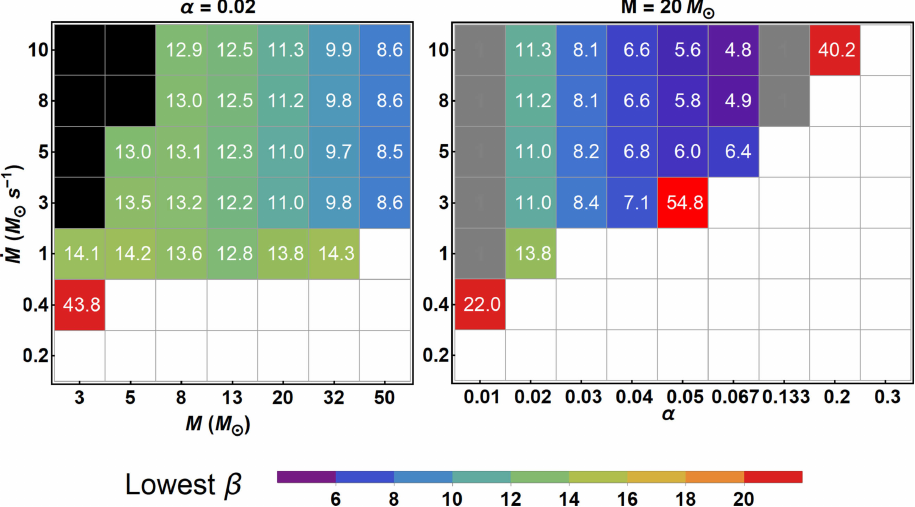}
\caption{The smallest cooling parameter $\beta$ achieved in radial zones with Toomre parameter $Q<1$. %The {\it left panel}: is an array plot with a constant viscosity parameter $\alpha=0.02$ varying the central mass $M$ and the accretion rate $\dot{M}$; {\it right panel}: is an array plot with a constant central mass $M=20{M}_{\odot}$ varying the viscosity parameter $\alpha$ and the accretion rate $\dot{M}$. 
{\it Left} and {\it right} panels have identical $\{M, \dot{M}, \alpha\}$ parameter choices to Fig. \ref{fig: Lowest_Q_Radius_2_panel}. White blank regions are disk models that lack radial zones inside $r_{\rm out}$ with both $Q<1$ and $\beta<100$. The black colored regions are points in parameter space where our numerical models failed to reach a solution. Grey indicates numerically converged solutions where the mass of the enclosed disk, within the smallest radius that satisfies both $\beta<35$ and $Q<1$, is larger than the central mass (i.e. the model assumptions are breaking down). %red colored regions are solution which have $\beta$ numbers above $22$. 
Colored regions with a $\beta$ number below 35 will likely exhibit fragmentation. Generally, varying $\dot{M}$ does not change the smallest $\beta$ significantly, while varying $M$ and especially $\alpha$ does.}
\label{fig: Lowest_beta_Radius_2_panel}
\end{figure}

We explore the parameter space of cooling times further in Fig. \ref{fig: Lowest_beta_Radius_2_panel}, which shows two cross sections of our parameter survey in $\{M, \dot{M}, \alpha \}$ (in the same format as Fig. \ref{fig: Lowest_Q_Radius_2_panel}). Here we show $\beta_{\rm min}$, the lowest $\beta$ value achieved in all radial zones with $Q<1$ (discarding solutions where $\beta_{\rm min} < 100$). %, the left panel is an array plot with a constant viscosity parameter $\alpha=0.02$ varying the central mass from $3$ to $50{M}_{\odot}$ and the accretion rate from $0.2{M}_{\odot} {\rm s^{-1}}$ to $10{M}_{\odot} {\rm s^{-1}}$; the right panel is an array plot with a constant central mass $M=20{M}_{\odot}$ varying the viscosity parameter $\alpha$ from $0.01$ to $0.45$ and the accretion rate from $0.2{M}_{\odot} {\rm s^{-1}}$ to $10{M}_{\odot} {\rm s^{-1}}$. 
Investigating Fig. \ref{fig: Lowest_beta_Radius_2_panel}, we see that there is a large parameter space for which $\beta_{\rm min} < 35$, and a smaller but still substantial parameter space where $\beta_{\rm min}<10$ (in Appendix \ref{app: beta 10}, we explore most of our main results using this more conservative critical value), suggesting favourable conditions for fragmentation in many high-$\dot{M}$ collapsar disks.  More specifically, we see that decreasing $\alpha$ causes the fragmentation zones to appear at lower accretion rates due to the smaller radii where $Q<1$ (see Fig. \ref{fig: Lowest_Q_Radius_2_panel}).  However, at fixed $\dot{M}$, decreasing $\alpha$ leads to a mild increase in $\beta_{\rm min}$, likely due to the higher $\Sigma$.  For each given $\alpha$ value, there seems to be a lower boundary on the range of $\dot{M}$ that supports fragmentation; below this $\dot{M}$, fragmentation cannot occur, but above it, further increasing the accretion rate has only a limited effect on $\beta_{\rm min}$.  Finally, we note that all else equal, increasing the central mass $M$ decreases cooling times, but the effect is realtively mild.

%NCS: add new figure showing array plot of betas at innermost Q<=1 for two panels: alpha vs M and MDot vs M
%\clearpage
%NCS: add fragment initial mass plots, 2 colors (beta < 10, beta < 35) and 2 shades (e- degenerate/non-degenerate).  Full-page figure, 3 panels, stacked vertically, one panel varies M, one panel varies MDot, one panel varies alpha

\subsection{Fragment Initial Mass}
\label{fragment initial mass}
As we have investigated the likelihood of fragmentation for various disk solutions in our $\{M, \dot{M}, \alpha\}$ parameter space, we now proceed to estimate the properties of fragments that would form in those unstable regions. Using Eq. \ref{eq: initial fragment mass}, we plot fragment masses as functions of the radius at which they were formed (so long as the formation conditions, $Q\le 1$ and $\beta \le 35$, are satisfied).

\begin{figure} % Use figure* to span both columns
    %\centering
    \begin{subfigure}{\textwidth}
    %\centering
        \includegraphics[width=0.47\textwidth]{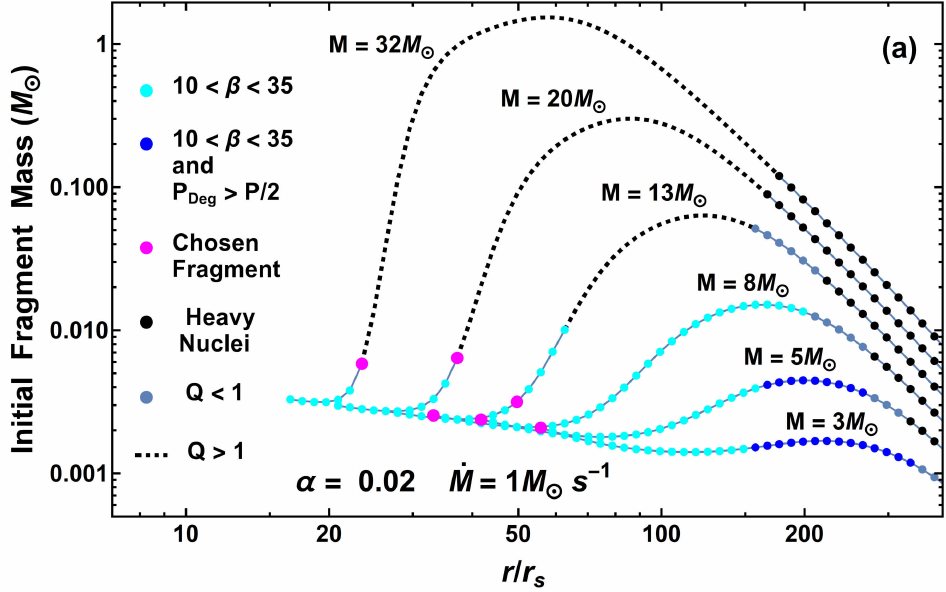}
        \label{fig: InitialFragmentMass_alpha002Mdot1}
    \end{subfigure}
    
    \begin{subfigure}{\textwidth}
    %\centering
        \includegraphics[width=0.47\textwidth]{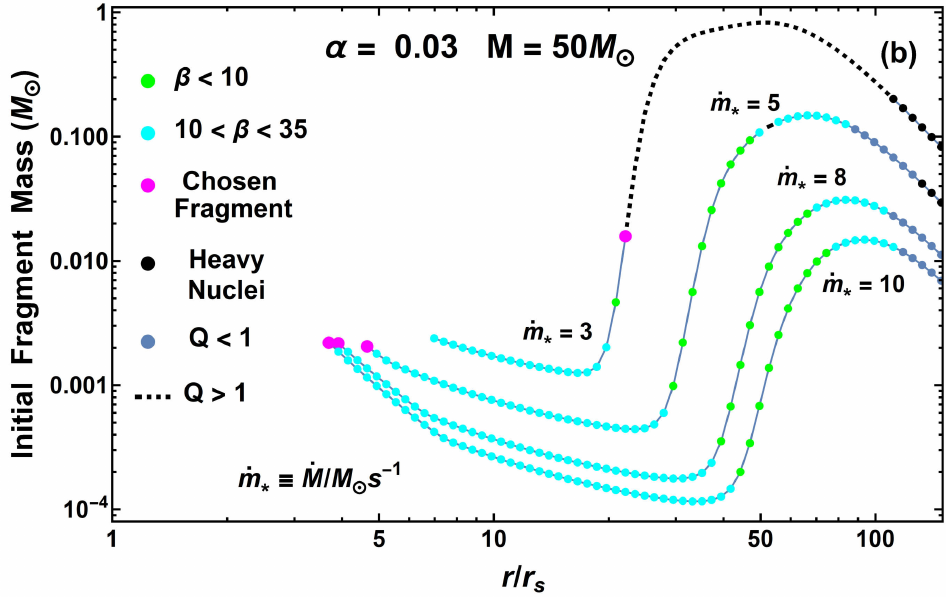}
        \label{fig: InitialFragmentMass_alpha003M50}
    \end{subfigure}

    \begin{subfigure}{\textwidth}
    %\centering
        \includegraphics[width=0.47\textwidth]{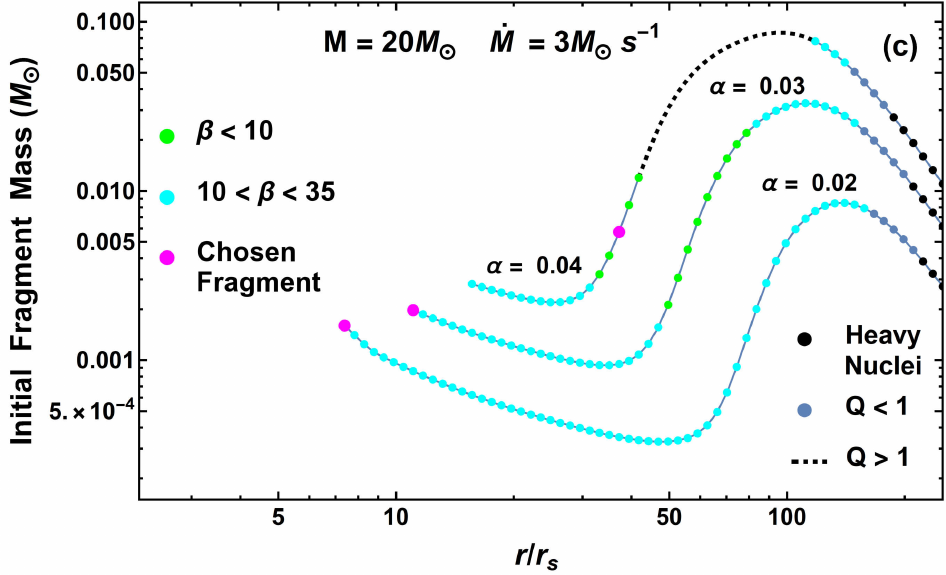}
        \label{fig: InitialFragmentMass_M20Mdot3}
    \end{subfigure}
    \caption{Initial fragment mass $m_{\rm f}$ plotted against dimensionless radius $r/r_{\rm s}$ for a wide variety of different disk parameters $M$, ${\dot{m}}_{\rm *} = \dot{M} / (M_\odot~{\rm s}^{-1}$) and $\alpha$.  Each string of points shows the regions of a given disk model where $Q \le 1$ {\it and} the cooling parameter $\beta$ is underneath a critical value $\beta_{\rm c}$.  Color coding indicates the value of $\beta_{\rm c}$ as well as the dominance (or not) of e$^{-}$ degeneracy pressure, as indicated in the legends. Grey dots indicate regions with $Q \le 1$ but inefficient cooling ($\beta > 35$) that likely prohibits fragmentation; black dots indicate regions where the disk models may
    fail due to heavy element production; magenta dots are placed at the innermost radii that can support fragmentation (we later choose these points to study fragment evolution).  For large $M$, the instability regions are sometimes bifurcated; $Q>1$ intermediate zones are shown with dashed black lines.  {\it Top panel}: results for fixed $\alpha=0.02$ and $\dot{M}=1 M_\odot~{\rm s}^{-1}$.  {\it Middle panel}: results for fixed $\alpha = 0.02$ and $M=20M_\odot$.  {\it Bottom panel}: results for fixed $M=20 M_\odot$ and $\dot{M}=3 M_\odot~{\rm s}^{-1}$.  Fragment masses generally fall in the range $10^{-4} \lesssim m_{\rm f}/M_\odot \lesssim 10^{-1}$. While holding two of the other disk variables constant, the initial masses of fragments tend (depending on the appearance of bifurcated $Q<1$ regions) to increase with increasing $M$ or $\alpha$, or with decreasing $\dot{M}$. 
    }
    \label{fig: Initial_Fragment_Mass_3_Panel}
\end{figure}

Fig. \ref{fig: Initial_Fragment_Mass_3_Panel} shows three panels containing multiple radial profiles of fragment masses, where in each panel we vary either $M$, $\dot{m}$ or $\alpha$. %The top panel shows results for fixed $\alpha=0.02$ and $\dot{M}=1 M_\odot~{\rm s}^{-1}$, the middle panel results for fixed $\alpha = 0.02$ and $M=20M_\odot$ and bottom panel results for fixed $M=20 M_\odot$ and $\dot{M}=3 M_\odot~{\rm s}^{-1}$. 
In the figure, we color code the points to indicate the range of $\beta$ as well as the possible dominance of e$^{-}$ degeneracy pressure. The dominance of degeneracy pressure is defined in the plots for when both $P_{\rm e^{-}}+P_{\rm e^{+}}> P/2$ and the dimensionless degeneracy parameter $\eta\equiv\mu_{\rm e}/k_{\rm b}T>1$.  Radial zones that are formally susceptible to fragmentation but which have gas dominated by degeneracy pressure may in reality be unable to form fragments, as even very efficient cooling ($\beta \ll \beta_{\rm c}$) would not permit fragments to radiate away their pressure support\footnote{On the other hand, if the relevant fermions are ultra-relativistic, it may not be necessary for fragments to radiate away internal energy in order for a dynamical collapse to ensue.}.  To the best of our knowledge, the evolution of $Q<1$ disks dominated by degenerate electron pressure has not yet been studied; we defer this subject for future work but label strongly degenerate regions in Fig. \ref{fig: Initial_Fragment_Mass_3_Panel} as a caution.

%Grey dots indicate regions with $Q \le 1$ but inefficient cooling that likely prohibits fragmentation. 
In Fig. \ref{fig: Initial_Fragment_Mass_3_Panel}, we also label fragment masses from the innermost radii that could support gravitational instability (usually this means $r=r_{Q}$, up to the limitations of numerical resolution); in later sections of this paper, we will choose these points for calculations of fragment migration and growth. We choose these specific points because: (1) out of all $Q\le 1$ zones, our disk model is most robust here, where the back-reaction of Toomre instability on disk structure (e.g. gravitoturbulence as in \citealt{Rafikov09}, or fragment feedback as in \citealt{SirkoGoodman03, GilbaumStone22}) is minimized; (2) at the $Q=1$ radius, 
the viscous time $t_{\rm visc} \lesssim M/\dot{M}$, the disk evolution time, meaning that our steady state assumption is more valid (we will return to this point in \S \ref{sec:discussion}); (3) the migration time from these radii is small compared to the evolution time $M/\dot{M}$, but this is not always the case at larger radii. 

Examining the different panels of Fig. \ref{fig: Initial_Fragment_Mass_3_Panel}, we see that fragment initial masses generally fall in the range $10^{-4} \lesssim m_{\rm f}/M_\odot \lesssim 10^{-1}$, significantly less than past estimates from more simplified disk models (which generally found $10^{-1} \lesssim m_{\rm f}/M_\odot \lesssim 1$; \citealt{PiroPfahl07, Metzger+24})\footnote{We note that fragment masses could be even lower if fragmentation is confined to global spiral arms in the disk, as is the case in some models of Toomre-unstable protoplanetary disks \citep{Boley+10}.}. Also it can be seen that while holding two of the other disk variables constant, the initial fragment mass range tends (depending on the appearance of $Q<1$ bifurcated regions) to increase with either increasing $\alpha$, or decreasing $\dot{M}$. The largest initial fragment masses, $m_{\rm f} \sim 0.1 M_\odot$, are achieved for higher $M$. 

An important caution should be made concerning the physical conditions in the disk. At larger radii, the density and temperature ranges that follow from our model often fall in the typical range for outer envelopes (i.e. the crust) of proto-NSs and remnants of NS mergers \cite[e.g.][]{HPY2007,Dehman+2024}. Typical equations of state of such matter predict that formation of heavy atomic nuclei with $Z\geqslant 3$ are energetically favorable. This option is ignored in the master Eqs. \eqref{eq: pressure equality}---\eqref{eq: lepton number advection equation}. However, to take heavy atomic nuclei into account, one has to consider a model of nuclear interactions, which is not a trivial task. Here we use the LS220 equation of state for hot dense matter \citep{LattimerSwesty1991}\footnote{provided by the CompOSE database \citep{Typel+2015} \texttt{https://compose.obspm.fr/}}. We use this model to explore qualitatively the possibility of heavy nuclei formation. Black dots in Fig. \ref{fig: Initial_Fragment_Mass_3_Panel} mean that at a given $r$, the physical conditions in the disk ($\rho$, $T$, $Y_e$) are such that LS220 predicts more than a 5\% mass fraction of heavy nuclei. This indicates that the present disk model may fail there. A comprehensive investigation of this problem is beyond the scope of the present work, but this post hoc check shows that in most cases, our neglect of heavy nuclei is only relevant at large radii well beyond $r_{\rm Q}$, and thus will not impact our estimates for fragment properties and evolution near the inner edge of the fragmentation zone.

\begin{figure}
\includegraphics[width=0.48\textwidth]{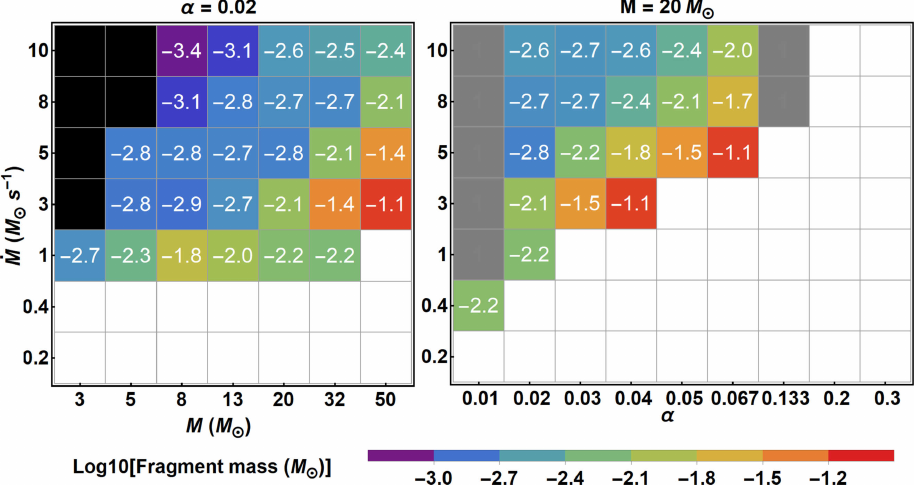}
\caption{Color-coded values for the largest initial fragment mass (represented as $\log_{10}(m_{\rm f,i}/M_\odot)$) in the fragmenting $Q<1$ region.  %Each array plot is a slice of our $\{M, \dot{M}, \alpha\}$ parameter survey in the same style as Figure \ref{fig: Lowest_Q_Radius_2_panel}. % shown for two array plots the rainbow color system represents the negative log base 10 value of the fragment initial mass. %: the {\it left panel}: is an array plot with a constant viscosity parameter $\alpha=0.02$ varying the central mass $M$ and the accretion rate $\dot{M}$; {\it right panel}: is an array plot with a constant central mass $M=20{M}_{\odot}$ varying the viscosity parameter $\alpha$ and the accretion rate $\dot{M}$; The white blank regions in the right panel don't have regions inside $r_{\rm out}$ with both $Q<1$ and $\beta<35$; The black colored regions are points in parameter space which our numerical model failed to reach a solution; Solutions where the mass of the enclosed disk, within the smallest radius that satisfies both $\beta<35$ and $Q<1$, exceeds the central mass are indicated in grey; 
{\it Left} and {\it right} panels represent the same $\{M, \dot{M}, \alpha\}$ parameter choices as in Fig. \ref{fig: Lowest_Q_Radius_2_panel}.  Black, white, and grey squares represent breakdowns in model assumptions or solutions as in Fig. \ref{fig: Lowest_beta_Radius_2_panel}.
We see that fragment initial masses span a wide range of $10^{-3} \lesssim m_{\rm f}/M_\odot \lesssim 10^{-1}$.  The initial masses tend to increase with declining accretion rate, or with increasing central mass and/or viscosity.}
\label{fig: InitialFragmentMass_2_panels}
\end{figure}

Fig. \ref{fig: InitialFragmentMass_2_panels} explores the $\{M, \dot{M}, \alpha\}$ parameter space to determine the largest possible initial fragment masses ($m_{\rm f}^{\rm max}$) in the fragmentation region, for two cross sections of our parameter survey. %The left panel is an array plot with a constant viscosity parameter $\alpha=0.02$ varying the central mass from $3$ to $50{M}_{\odot}$ and the accretion rate from $0.2{M}_{\odot} {\rm s^{-1}}$ to $10{M}_{\odot} {\rm s^{-1}}$; the right panel is an array plot with a constant central mass $M=20{M}_{\odot}$ varying the viscosity parameter $\alpha$ from $0.01$ to $0.45$ and the accretion rate from $0.2{M}_{\odot} {\rm s^{-1}}$ to $10{M}_{\odot} {\rm s^{-1}}$. 
This is a broader but more coarse-grained look at fragment masses than was presented in Fig. \ref{fig: Initial_Fragment_Mass_3_Panel}; for example, in the right panel of Fig. \ref{fig: InitialFragmentMass_2_panels}, there is a point corresponding to an accretion rate of $3{M}_{\odot} {\rm s^{-1}}$, a central mass $M=20{M}_{\odot}$, and a viscosity parameter $\alpha=0.04$.  Here the  maximum initial fragment mass is $10^{-1.1}{M}_{\odot}$ which can be seen as a maximum mass in panel (c) of Fig. \ref{fig: Initial_Fragment_Mass_3_Panel}.
Investigating Fig. \ref{fig: InitialFragmentMass_2_panels}, we see shows that $10^{-3} \lesssim m_{\rm f}^{\rm max}/M_\odot \lesssim 10^{-1}$, broadly consistent with Fig. \ref{fig: Initial_Fragment_Mass_3_Panel}. The maximum initial fragment mass tends to increase with either increasing $M$ or $\alpha$, or decreasing $\dot{M}$.  In almost all cases, $m_{\rm f}^{\rm max}$ is well below the minimum stable cold neutron star mass of $\approx 0.1 M_\odot$ 
\citep{Haensel2002,HPY2007}, and in every case it is below the minimum hot stable neutron star mass \citep{Gondek+97, StrobelWeigel01,Dehman+2024}.

%\clearpage
% All plots related to fragment mass, mass growth, migration etc go here
\subsection{Fragment Migration and Evolution}
After the fragment forms, its subsequent evolution can be greatly affected by interactions with the surrounding collapsar disk, as described in \S \ref{sec: Fragment migration and mass accumulation}. Bondi-Hoyle mass accretion causes the fragment to grow rapidly, but not indefinitely: the accumulation of sub-Keplerian gas, gravitational interactions with unbound disk gas, and eventually GW emission will all cause the fragment's orbit to inspiral. Here we study the combined accretion-migration evolution of a fragment forming at $r_{Q}$, the innermost radii of the fragmentation zone (for reasons specified in \ref{fragment initial mass}). 
\begin{figure*} % Use figure* to span both columns
    \centering
    \begin{subfigure}{\textwidth}
    \centering
        \includegraphics[width=0.99\textwidth]{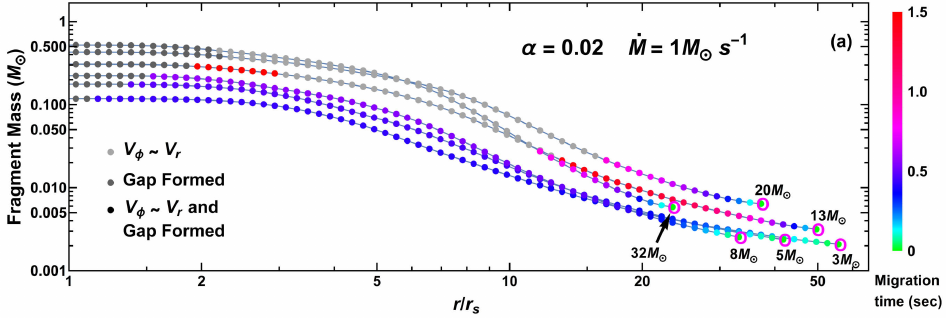}
         %\caption{}
        \label{fig: FragmentMassGrowth_alpha002Mdot1}
    \end{subfigure}
    
    \begin{subfigure}{\textwidth}
    \centering
        \includegraphics[width=0.99\textwidth]{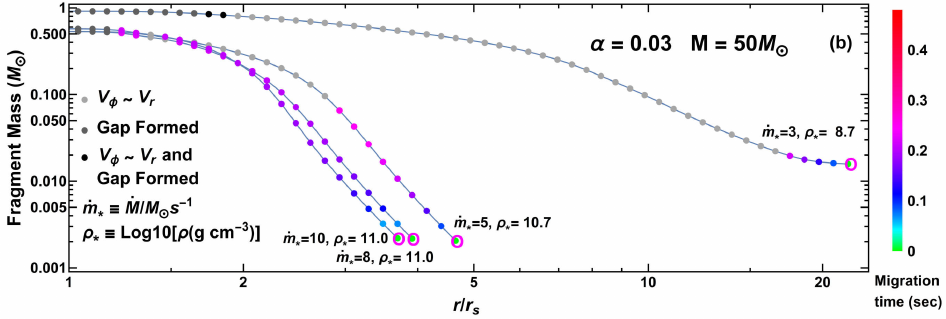}
        %\caption{}
        \label{fig: FragmentMassGrowth_alpha003M50}
    \end{subfigure}

    \begin{subfigure}{\textwidth}
    \centering
        \includegraphics[width=0.99\textwidth]{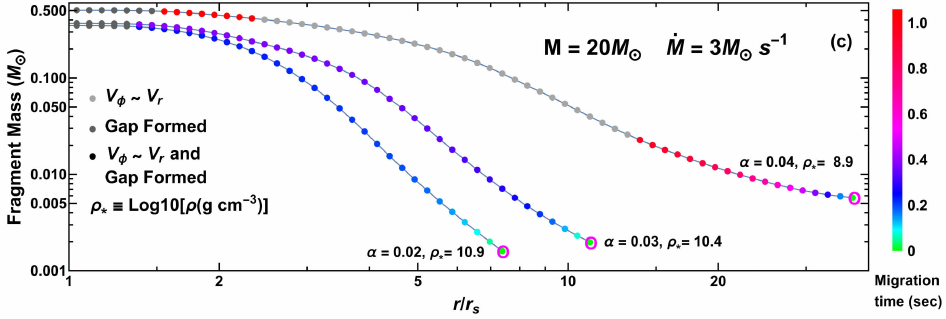}
        %\caption{}
        \label{fig: FragmentMassGrowth_M20Mdot3}
    \end{subfigure}
    \caption{Fragment mass growth and inward migration time, plotted for each dimensionless radius $r/r_{\rm s}$ that the fragment passes through. We explore fragment evolution for a wide variety of different disk parameters $M$, $\dot{m} = \dot{M} / (M_\odot~{\rm s}^{-1}$) and $\alpha$. {\it Top panel}: results for fixed $\alpha=0.02$ and $\dot{M}=1 M_\odot~{\rm s}^{-1}$, with multiple values of central mass $M$ shown as different evolutionary tracks.  {\it Middle panel}: results for fixed $\alpha = 0.02$ and $M=20M_\odot$, with multiple values of $\dot{m}$ shown as different evolutionary tracks.  Here we also label each track by $\rho_\star \equiv \log_{10}(\rho)$, where $\rho$ is the initial fragment density in cgs normalized by ${\rm g~cm}^{-3}$.  {\it Bottom panel}: results for fixed $M=20 M_\odot$ and $\dot{M}=3 M_\odot~{\rm s}^{-1}$, with multiple values of $\alpha$ shown as different evolutionary tracks (that again are also labeled by $\rho_\star$).  Color coding indicates the time it takes a fragment to reach a specific radius going inward from its initial formation radius on the right end of each track (denoted by a magenta ring, which corresponds to the magenta points in Fig. \ref{fig:  FragmentMassGrowth_3_Panel}). Grey-scale points indicates failed assumptions, where: (1) light grey represents where the relative radial velocity is no longer negligible compared to the azimuthal relative velocity; (2) grey represents gap formation; (3) black represents when both a gap is formed and the relative radial velocity isn't negligible. Fragments form with a typical density range of $10^{9}-10^{11}~{\rm g cm}^{-3}$ (above WD and below NS densities) and an initial mass of $\sim10^{-3}M_\odot$, and grow by a few orders of magnitude to $\sim1M_\odot$ where they form a gap in the disk and eventually migrate to the BH. }
    \label{fig:  FragmentMassGrowth_3_Panel}
\end{figure*}

Fig. \ref{fig:  FragmentMassGrowth_3_Panel} shows three panels containing many evolutionary profiles of fragments experiencing migration and mass accumulation, where in each panel we vary either $M$, $\dot{m}$ or $\alpha$. %The top panel shows results for fixed $\alpha=0.02$ and $\dot{M}=1 M_\odot~{\rm s}^{-1}$, the middle panel results for fixed $\alpha = 0.02$ and $M=20M_\odot$ and bottom panel results for fixed $M=20 M_\odot$ and $\dot{M}=3 M_\odot~{\rm s}^{-1}$. 
%In the figures we color-code the points to indicate the time it takes a fragment to reach a specific radius going inward from it's initial formation radius (noted by a magenta ring, which correspond to the magenta points in Fig. \ref{fig:  FragmentMassGrowth_3_Panel}). Grey scale indicates failed assumption, where: (1) light-grey represent where the relative radial velocity isn't negligible compered to the azimutal relative velocity (see Eq. \ref{}); (2) grey represents gap formation (see Eq. \ref{}); (3) black represents when both a gap is formed and relative radial velocity isn't negligible. In all three panels we mark if the fragment becomes bigger than 5\% of the mass of the BH, in order to show where our assumption of a small fragment becomes less valid (see \ref{}, in the top panel this is represented by a blue `X' and at the bottom two panels with a dashed line). It is shown 
We see that fragments form within a density range of $10^{9}-10^{11}~{\rm g~ cm}^{-3}$, %these densities are above WD densities ($10^{4}-10^{7}g cm^{-3}$, **Add citation**) and below NS densities ($10^{17}g cm^{-3}$, **Add citation**). 
intermediate between white dwarf and neutron star interiors. 

\begin{figure}
\includegraphics[width=0.48\textwidth]{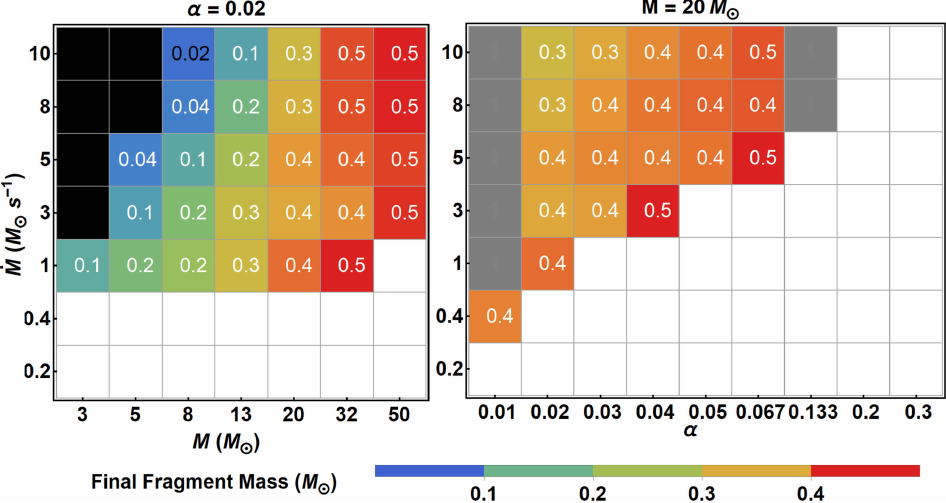}
\caption{The final fragment mass (in ${M}_{\odot}$), shown in two array plots. {\it Left} and {\it right} panels represent the same $\{M, \dot{M}, \alpha\}$ parameter choices as in Fig. \ref{fig: Lowest_Q_Radius_2_panel}. The rainbow color system represents the fragment final mass, as color-coded in the legend.  Black, white, and grey squares represent breakdowns in model assumptions or solutions as in Fig. \ref{fig: Lowest_beta_Radius_2_panel}. Numbers color coded in black represent fragment migration time bigger than $M/\dot{M}$. 
Fragment final masses tend to increase with decreasing accretion rate or increasing viscosity (though these trends are much weaker than in Fig. \ref{fig: InitialFragmentMass_2_panels}), and span a range of $0.04 \lesssim m_{\rm f}/M_\odot \lesssim 0.5$.  These final masses are usually in the range where stable NSs can exist, although in some instances (small $M$) they may be too low.}
\label{fig:FragmentMassGrowth_2PanelArray}
\end{figure}

These fragments will start to contract and increase their density which makes formation of an exotic NS plausible; however, the initial fragment masses lie well below the minimum stable masses for both cold and hot neutron stars \citep{HPY2007,Gondek+97,Gondek+1998,Dehman+2024}, so formation of an object in hydrostatic equilibrium is only possible with rapid mass accretion. Rapid accretion may help fragments gain mass faster than they can lose it (due to dynamically/thermally unstable surface layers), and more speculatively, may even confine a dynamically/thermally unstable surface by ram pressure.  More specifically, the fragments start with an initial mass $\sim10^{-3}M_\odot$ (as seen more generally in Fig. \ref{fig: Initial_Fragment_Mass_3_Panel}), but grow by a few orders of magnitude via Bondi-Hoyle accretion to $\sim1 M_\odot$ (as seen in more detail in Fig. \ref{fig:FragmentMassGrowth_2PanelArray}). At this point, they form a gap in the disk and continue their further migration inward with to the BH with much reduced mass growth.  The time elapsed between fragment formation and gap opening is typically $\sim 0.2$ s, though with wide variation. Portions of our migration model will sometimes become more questionable prior to gap opening: for example, it is common for the radial migration velocity $V_{\rm r}$ to become comparable to the azimuthal speed $V_\phi$, and on the other hand a small portion of the fragments migrate slowly enough for the assumption of a steady central mass to break down due to the central accretion rate.
%\clearpage

%NCS: add mass growth figure: 3 panels stacked vertically, same as previous figure
Fig. \ref{fig:FragmentMassGrowth_2PanelArray} shows the final fragment mass $({M}_{\odot})$ after mass growth via Bondi-Hoyle accretion, for two cross sections of our parameter survey (fixed $\alpha=0.02$ in one panel, and fixed $M=20 M_\odot$ in the other). %The left panel is an array plot with a constant viscosity parameter $\alpha=0.02$ varying the central mass from $3$ to $50{M}_{\odot}$ and the accretion rate from $0.2{M}_{\odot} {\rm s^{-1}}$ to $10{M}_{\odot} {\rm s^{-1}}$; the right panel is an array plot with a constant central mass $M=20{M}_{\odot}$ varying the viscosity parameter $\alpha$ from $0.01$ to $0.45$ and the accretion rate from $0.2{M}_{\odot} {\rm s^{-1}}$ to $10{M}_{\odot} {\rm s^{-1}}$. 
The figure shows that the final fragment mass range is $0.01 \lesssim m_{\rm f}/M_\odot \lesssim 0.5$.  We note that in the full parameter survey, presented in appendix Fig. \ref{app: BiggestFinalValidMass_9Plots}, the highest central BH masses can produce final fragment mass up to $\sim 1 {M}_{\odot}$. The final masses tend to weakly increase with decreasing accretion rate and/or increasing viscosity; qualitatively similar trends existed for the initial fragment masses shown in Fig. \ref{fig: InitialFragmentMass_2_panels}, but were much stronger there. The final outcome of fragment evolution depends highly on the EOS.  Possible outcomes include: (1) for fragment masses below the minimum stable mass, an explosion; (2) above the minimum stable mass, the formation of a NS (often an exotic, low-mass one); (3) mergers of fragments can lead to a merged mass above the maximum stable mass, forming a black hole.  Outcomes will also depend on timescales, as some fragments will migrate inwards quickly enough to tidally disrupt before any of the above possibilities are realized. 

We have made a numerical fitting function to the final fragment masses as a function of the central mass $M$, accretion rate $\dot{M}$ and the viscosity parameter $\alpha$:
\begin{equation}
    m_f\approx0.49\cdot\left(\frac{\alpha}{0.02}\right)^{0.64}\cdot\left(\frac{M}{20 M_\odot}\right)^{1.03}\cdot\left(\frac{\dot{M}}{M_\odot s^{-1}}\right)^{-0.46}
\end{equation}
The fitting was made to the full parameter survey of the final fragment mass (see appendix plot, Fig. \ref{app: BiggestFinalValidMass_9Plots}). The final fragment mass follows these trends: (1) almost linearly proportional to the central mass; (2) inversely related to the the central accretion rate; (3) positively correlated to the disk viscosity $\alpha$.
%NCS: add fragment final mass array plot, two panels side-by-side
% \clearpage

\section{Observable Predictions}
\label{sec:observables}
In order to estimate the observational consequences of collapsar disk fragmentation, and the subsequent interaction of fragments with the BH, the final outcomes of fragmentation must be understood. While the full nonlinear evolution of fragments embedded in collapsar disks is beyond the scope of this work, in the following sections we make simplifying assumptions to explore the broad contours of possible observational implications. 

A major uncertainty in understanding the evolution of fragmentation arises from the initial masses and densities of self-bound fragments. Fragment initial masses almost always fall below the minimum stable cold neutron star mass (see Figs. \ref{fig: Initial_Fragment_Mass_3_Panel}, \ref{fig: InitialFragmentMass_2_panels}), and the initial densities generally fall into a range intermediate between white dwarf and neutron star densities (namely $\sim 10^{8-11}~{\rm g/cm}^3$).  These objects cannot achieve a stable hydrostatic equilibrium in vacuum, even accounting for finite temperature effects \citep{Gondek+97}. In the absence of interaction with external media, the final fate of such objects is a delayed explosion \citep{Blinnikov1990,Sumiyoshi1998}\dd{, determined by the timescale of the $\beta$-decays in those environments \citep{Colpi1989,Sumiyoshi1998}}. 
%and their evolution in a vacuum environment would therefore be either collapse or expansion on a dynamical timescale.  More specifically, NS-like objects below the minimum stable mass will 
%can have interiors near a hydrostatic equililbrium state, but will steadily lose mass from their dynamically unstable ($\gamma < 4/3$) surface layers.  
However, such an explosion is known to be delayed for dozens of seconds \citep{Sumiyoshi1998}, and in the complex environment of a collapsar disk, it may be additionally stabilized by the external ram pressure.\footnote{When the fragment is still in its linear growth stage, ambient pressure from the collapsar disk will be important as well.}  The fragment may accrete enough mass to pass the minimum mass threshold before it explodes, or alternatively, migration to the ISCO may occur faster than the delayed explosion \dd{(\citealt{ChenAndMetzger25} likewise noted that $\beta$-decay timescales could exceed the system’s evolution timescales, leading to extended survivability.)}.  %ram pressure may keep a low-mass fragment stable for enough time to migrate closer to the ISCO.
%growth (growing the fragments mass above the minimum stable mass). %
%in the complex environment of a collapsar disk, such low mass objects may live longer or even become quasi-stable, due to the combined effects of mass replenishment and ram pressure

The enormous accretion rates $\dot{m}_{\rm f}$ we find may confine a nominally unstable low-mass compact object with ram pressure if the accretion is quasi-spherical.  Alternatively, if accretion onto the fragment is mediated through a highly aspherical mini-disk, ram pressure will not be able to confine the unstable fragment in its polar regions, but mass replenishment through the mini-disk may let the fragment gain mass faster than it loses it\footnote{Though investigating this possibility is beyond the scope of the present work, an analogous combination of aspherical accretion and outflow has been shown to solve (Eddington-related) stability problems for high-mass main sequence star formation \citep{Cunningham+11}.}.

The resolution to these questions is complicated and we defer their full investigation to future work.  For the purposes of this section, we optimistically assume that fragments will be able to grow according to the Bondi-Hoyle $\dot{m}_{\rm f}$ tracks we calculated in e.g. Fig. \ref{fig:  FragmentMassGrowth_3_Panel}, and consider observational consequences under this working hypothesis.

In particular, we investigate here the contribution of bound fragments to GRB variability (\S \ref{sec:variability}) and GW emission (\S \ref{sec:GWs}).  In order to make predictions for both of these effects, we need some estimate for fragment radii at different points in their evolution.  The reason for this is that the fragment will migrate quickly inwards (as specified in section \ref{sec: Fragment migration and mass accumulation}) until it is either tidally disrupted by the BH or (if the fragment becomes sufficiently massive and dense) it is directly captured at the ISCO.  %In the life time of the fragment it will emit GW radiation and vary the energy release by the central engine with the said tidal disruption. In the following sections, we will provide estimates for the gravitational wave frequency and strain, as well as an estimate on the variance in EM luminosity from the TDE.

If the fragment is below the minimum neutron star mass, we do not have a reliable way to estimate its radius for the reasons enumerated above. For fragment masses higher than this threshold, we use the simplest approximation for a fragment mass-radius relation: spherically symmetric hydrostatic equilibrium equations in GR \citep{Tolman1939,OppVol1939}, neglecting effects of the fragment spin and the ram pressure of the disk matter, using a zero-temperature limit for the EOS, namely the BSk24 model~\citep{bsk24}. The minimum mass for this EOS is about $0.09\,M_\odot$, so we only begin checking for fragment tidal disruption once the fragment exceeds this limit. However, the fragment temperatures can in principle be high enough to significantly affect the $R_{\rm f}(m_{\rm f})$ relationship; when relevant, finite-temperature effects {\it increase} the minimum NS mass~\citep[e.g.][]{Gondek+97,StrobelWeigel01,Dehman+2024}. Its value strongly depends on the temperature magnitude and profile and is generally greater than $0.5\,M_\odot$ for a high-temperature EOS. This high-temperature minimum stable mass can be higher than the {\it maximum} fragment mass we find in Fig.~\ref{fig:FragmentMassGrowth_2PanelArray}, but by varying the initial radius of fragmentation\footnote{When we consider more conservative, non-fiducial choices for the maximum $\beta$ that permits fragmentation, this often has the effect of moving the minimum radius of fragmentation outwards, and plots exploring results with $\beta_{\rm c}=10$ (e.g. Appendix \ref{app: beta 10}) can also be understood as sampling fragment evolution for those fragments which form at larger radii (see Fig. \ref{fig: Initial_Fragment_Mass_3_Panel}) in our {\it fiducial} models. } or considering different values of $\alpha$ (see appendix Fig. \ref{app: BiggestFinalValidMass_alpha002M20_beta10}) we can reach $m_{\rm f}$ values as large as $1\,M_\odot$. Since accurate calculation of the fragment temperature (as well as modelling of fragment formation and evolution) is beyond the scope of the present work, we avoid considering any finite-temperature effects on the fragment EOS here. Whether they are relevant in practice likely depends on the neutrino cooling times for these low-mass fragments.

Using our assumption of a mass to radius relation we can give an estimate on the fragment's tidal radius $r_{\rm t}$. For this we solve the tidal radius for equatorial geodesics equation \citep{Kesden2012}, via root-finding:
\begin{equation}\label{eq: tidal radius implicit eq}
r_{{\rm t}}=R_{{\rm f}}\left(\frac{2M}{m_{{\rm f}}}\left(1+\frac{3( \mathcal{L}/c-ar_{{\rm g}} \mathcal{E})^2}{2r_{{\rm t}}^2}\right)\right)^{1/3}.
\end{equation}

Here $\mathcal{E}$ is the dimensionless total specific energy (see appendix \ref{app:GR}). We farther explain our calculations of the tidal disruption in appendix section \ref{sec: Tidal disruption check}.

\subsection{GRB Variability}
\label{sec:variability}
Previous works \citep{Perna2006, DallOsso+17, ShahamatAbbassi20, Shahamat2021} have considered the tidal disruption of migrating fragments as a potential source of GRB variability.  Following disruption, a fragment returns its mass to the disk at a relatively small radius, creating a sizable overdensity in $\Sigma$ that then evolves viscously, temporarily enhancing the accretion rate onto the BH and modulating its jet power. \dd{The typical prompt-emission variability timescales for long GRBs span from $\sim 0.1 {\rm s}$, for the fastest spikes, up to tens of seconds for the slowest trends \citep{Magnus25}.}

Here we quantify this type of ``fragment disruption variability'' by calculating the ratio between the post-disruption accretion luminosity from the fragment ($L_{{\rm TD}}$, where TD stands for tidal disruption) to the steady state luminosity from the collapsar disk ($L_{{\rm \bullet}}$). We estimate this ratio as:
\begin{equation}
    \frac{L_{{\rm TD}}}{L_{{\rm \bullet}}}=\frac{\eta_{{\rm \bullet}}\dot{m}_{{\rm TD}}c^{2}}{\eta_{{\rm \bullet}}\dot{M}c^{2}}%=\frac{\dot{m}_{{\rm TD}}}{\dot{M}}\Big|_{{\rm r=r_{{\rm t}}}}
    \approx\dot{M}^{-1}\frac{m_{{\rm f}}(r_{\rm t})}{t_{\rm visc}(r_{\rm t})}. \label{eq: variability from TD}
\end{equation}
Here we estimate the accretion rate from the disrupted fragment ($\dot{m}_{{\rm TD}}$) as the fragment mass over the local viscous time, where the viscous time is approximated as $t_{{\rm visc}}^{-1}\approx\alpha(H/r_{{\rm t}})^{2}\Omega$. If the tidal radius is smaller than $r_{{\rm ISCO}}$ then we assume that the fragment will be swallowed whole rather than disrupted, leading to no electromagnetic variability.  %a 

After calculating the luminosity ratio in Eq. \ref{eq: variability from TD} across the $\{M, \dot{M}, \alpha\}$ parameter space we have surveyed, we have reached the conclusion that this type of accretion rate variability is generally modest.  In many cases, we find that fragments would already have disrupted before reaching $m_{\rm f} = 0.09 M_\odot$; when this occurs, it is at large enough radii that contributions to GRB variability would be completely negligible.  In other cases, particularly for high-mass BHs, fragments can be swallowed whole.  In between these two limits, when we can apply Eq. \ref{eq: variability from TD}, the resulting $L_{\rm TD}$ usually enhances $L_{\rm \bullet}$ at the $\sim 10\%$ level.  We present these results more quantitatively in Appendix \ref{app: GRB from TD}.

\subsubsection{Secondary jet formation}\label{subsec: Secondary jet}
\begin{figure*} % Use figure* to span both columns
    \centering
    \begin{subfigure}{\textwidth}
    \centering
        \includegraphics[width=0.99\textwidth]{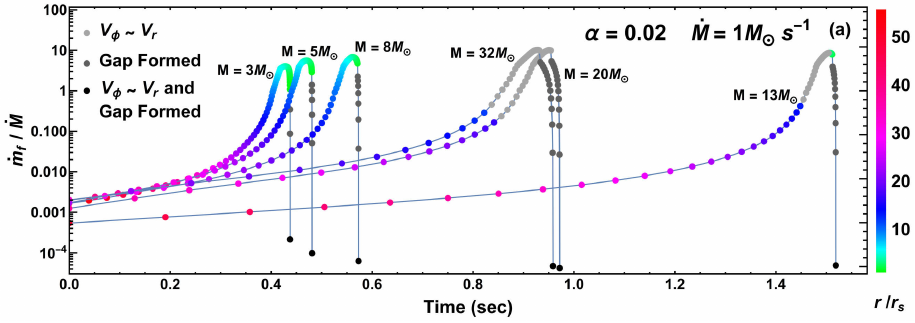}
         %\caption{}
        \label{fig: AccretionRateRatios_alpha002Mdot1}
    \end{subfigure}
    
    \begin{subfigure}{\textwidth}
    \centering
        \includegraphics[width=0.99\textwidth]{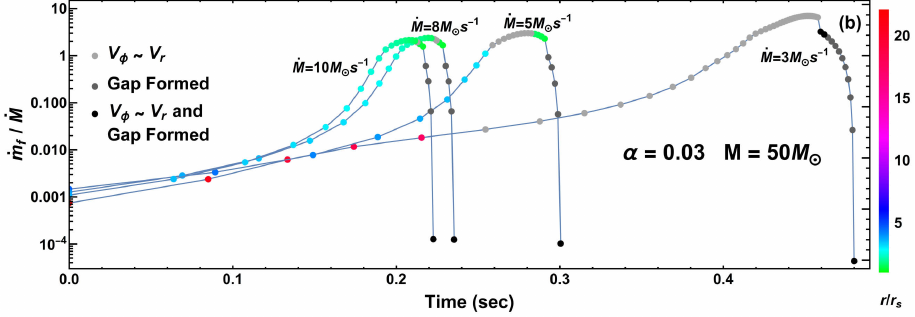}
        %\caption{}
        \label{fig: AccretionRateRatios_alpha003M50}
    \end{subfigure}

    \begin{subfigure}{\textwidth}
    \centering
        \includegraphics[width=0.99\textwidth]{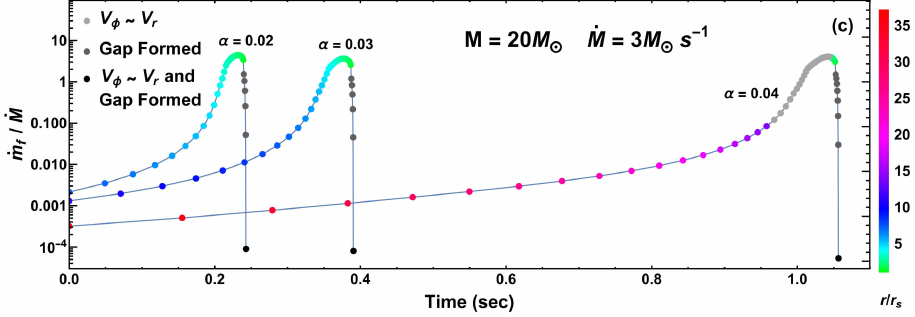}
        %\caption{}
        \label{fig: AccretionRateRatios_M20Mdot3}
    \end{subfigure}
    \caption{Bondi-Hoyle accretion rates onto migrating fragments, $\dot{m}_{\rm f}$, normalized by the collapsar accretion rate $\dot{M}$ and plotted as functions of time since fragmentation. The color scheme shows instantaneous orbital radius $r/r_{\rm S}$).  {\it Top panel}: results for fixed $\alpha=0.02$ and $\dot{M}=1 M_\odot~{\rm s}^{-1}$, with multiple values of central mass $M$ shown as different evolutionary tracks.  {\it Middle panel}: results for fixed $\alpha = 0.02$ and $M=20M_\odot$, with multiple values of $\dot{M}$ shown as different evolutionary tracks.  {\it Bottom panel}: results for fixed $M=20 M_\odot$ and $\dot{M}=3 M_\odot~{\rm s}^{-1}$, with multiple values of $\alpha$ shown as different evolutionary tracks.  Grey-scale points indicates failed assumptions, with the same color-coding as in Fig. \ref{fig:  FragmentMassGrowth_3_Panel}. The ratio $\dot{m}_{\rm f}/\dot{M}$ changes by multiple orders of magnitude during the fragment inspiral, rising from initial values $\sim 10^{-3}$ to final values $\sim 10^1$ (which are achieved briefly near the ISCO before terminating as the fragments plunge). At peak, $\dot{m}_{\rm f}$ generally exceeds $\dot{M}$, suggesting that secondary jets launched by the fragment may achieve significant luminosities (see \ref{subsec: Secondary jet}).  This plot does not account for fragment tidal disruption or the possibility of accretion rate suppression due to the neutrino Eddington limit.}
    \label{fig:AccretionRateRatios_3_Panel}
\end{figure*}

\begin{figure}
\includegraphics[width=0.48\textwidth]{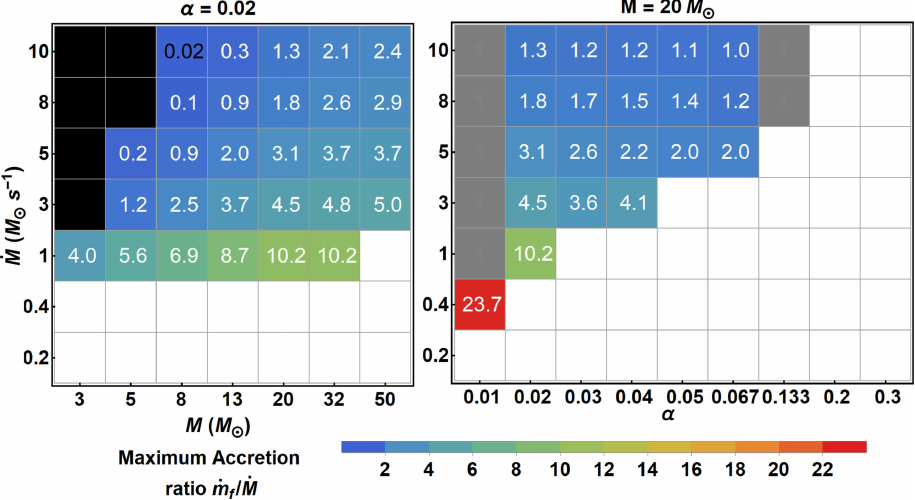}
\caption{The peak Bondi-Hoyle accretion rate, $\dot{m}_{\rm f}$, normalized by the collapsar accretion rate $\dot{M}$, shown in two array plots. {\it Left} and {\it right} panels represent the same $\{M, \dot{M}, \alpha\}$ parameter choices as in Fig. \ref{fig: Lowest_Q_Radius_2_panel}. The rainbow color system represents the peak of the accretion rate ratio, as color-coded in the legend.  Black, white, and grey squares represent breakdowns in model assumptions or solutions as in Fig. \ref{fig: Lowest_beta_Radius_2_panel}. Numbers color coded in black represent fragment migration time bigger than $M/\dot{M}$. 
The accretion rate ratio tends to increase with increasing central mass  or decreasing viscosity. The non-normalized $\dot{m}_{\rm f}$ tends to increase with the accretion rate, while the ratio $\dot{m}_{\rm f}/\dot{M}$ declines with increasing $\dot{M}$. It is common for the fragment accretion rate to exceed $\dot{M}$ by a factor of a few for a short duration ($\sim$ few ms, see Fig. \ref{app:DurationOfHighAccretion_alpha002M20_a0}). This plot does not include the possibility of accretion rate suppression due to the neutrino Eddington limit, which are examined in Figs. \ref{app: MaxAccretionRateRatio_alpha002M20_NoTrapsEta01} and \ref{app: MaxAccretionRateRatio_alpha002M20_NoTrapsEta0005}. }
\label{fig:MaxAccretionRateRatio_2PanelArray}
\end{figure}

Even though fragment disruption does not appear promising as a source of high-amplitude GRB variability, we notice that the instantaneous accretion rates $\dot{m}_{\rm f}$ onto embedded fragments seem to grow enormously during the final moments of fragment inspiral, when $r \sim r_{\rm ISCO}$.  %(see e.g. Fig. \ref{fig:AccretionRateRatios_3_Panel}). 
This leads us to consider a novel way in which embedded fragments may accomplish high-amplitude GRB variability: by launching a secondary jet. To quantify this further, we have explicitly measured $\dot{m}_{\rm f}(t)$ across our $\{M, \dot{M}, \alpha\}$ parameter space.  In Fig. \ref{fig:AccretionRateRatios_3_Panel}, we plot $\dot{m}_{\rm f}/\dot{M}(t)$ for the same representative parameter values as in Fig. \ref{fig:  FragmentMassGrowth_3_Panel}.  We see three robust trends in fragment accretion rate visible on this figure:
\begin{itemize}
    \item Fragment accretion rates grow by multiple (usually 3-4) orders of magnitude as fragments increase their mass (increasing $R_{\rm BH}$) and move inward to denser regions of the collapsar disk.
    \item Fragment accretion rates reach a maximum value just outside $r=r_{\rm ISCO}$, at a radius roughly matching the peak in disk surface density $\Sigma$.  
    \item The ratio between the peak fragment accretion rate $\dot{m}_{\rm f}$ and the collapsar accretion rate $\dot{M}$ is usually $\sim 1-10$ (see Fig. \ref{fig:MaxAccretionRateRatio_2PanelArray}), indicating that fragments can briefly (for a few milliseconds, see Fig. \ref{app:DurationOfHighAccretion_alpha002M20_a0}) accrete at even higher rates than the BH. This conclusion must be tempered by an important caveat, which is that the peak fragment accretion rates we calculate can sometimes exceed the (fragment) neutrino Eddington limit, although this depends on the neutrino radiative efficiency of the fragments. Figs. \ref{app: MaxAccretionRateRatio_alpha002M20_NoTrapsEta01} and \ref{app: MaxAccretionRateRatio_alpha002M20_NoTrapsEta0005} show how the normalized accretion rate would change if we consider the Eddington limit mediating the maximum accretion rate (we explain our choice of Eddington limits in section \ref{sec: Validity of Model}).
\end{itemize} 

Whether and how fragments launch their own relativistic jets is an important question that we cannot fully resolve in this paper, but our results suggest that the answer may sometimes be ``yes'' and also offer relevant constraints on the resulting secondary-jets.  A general conclusion of our work is that it will be extremely rare for embedded fragments to grow large enough to form BHs (though this might perhaps occur for extremely large progenitor stars that produce $M\gtrsim 100 M_\odot$, or if multiple fragments merge with each other during the migration process).  Therefore the classic Blandford-Znajek mechanism will not be relevant for fragment jet launching, and it is more likely that any jets powered by the fragments will be the result of either $\nu-\bar{\nu}$ annihilation \citep{Eichler1989, ZalameaBeloborodov11} or proto-magnetar spindown \citep{Usov92}.

\begin{comment}
For concreteness, we parametrize an approximate luminosity for the secondary jets in the following way:
\begin{subequations}
    \begin{align}  
    \eta_{\rm f}&\approx\epsilon_{\rm f}\left(\frac{\dot{m}_{f}}{\dot{M}}\right)^{\zeta}\\
    L_{\rm f}&=\eta_{\rm f}\dot{m}_{f}c^2\approx\epsilon_{\rm f}\left(\frac{\dot{m}_{f}}{\dot{M}}\right)^{\zeta+1} \dot{M} c^2\\
    L_{\rm steady}&=\eta_{\rm BZ}\dot{M}c^2\\
    L_{\rm tot}&=L_{\rm steady}+L_{\rm f}=L_{\rm steady}\left(1+\frac{\epsilon_{\rm f}}{\eta_{\rm BZ}}\left(\frac{\dot{m}_{f}}{\dot{M}}\right)^{\zeta+1}\right)
    \end{align}
\end{subequations}
If $\zeta>-1$ and $\frac{\epsilon_{\rm f}}{\eta_{\rm BZ}}$ is an order unity number, variability from our calculations should reach a factor of a few level.
\end{comment}

For concreteness, we parametrize an approximate luminosity for the secondary jets ($L_{\rm f}=\eta_{\rm f}\dot{m}_{f}c^2$) in the following way.  We assume that the radiative efficiency of the fragment is $\eta_{\rm f}=Gm_{\rm f}/(R_{\rm f}c^2)$ and that the Blandford-Znajek luminosity of the primary jet is $L_\bullet = \eta_\bullet \dot{M}c^2$.  Thus the total luminosity from both jets will be
\begin{equation}
    L_{\rm tot}=L_{\bullet}+L_{\rm f}=L_\bullet\left(1+\frac{\eta_{\rm f}}{\eta_\bullet}\left(\frac{\dot{m}_{f}}{\dot{M}}\right)\right).
\end{equation}
\begin{comment}
\begin{subequations}
    \begin{align}  
    \eta_{\rm f}&\approx \frac{G m_{\rm f} }{R_{\rm f} c^2}\\
    L_{\rm f}&\approx\frac{G m_{\rm f} }{R_{\rm f} c^2}\left(\frac{\dot{m}_{f}}{\dot{M}}\right) \dot{M} c^2\\
    L_{\bullet}&=\eta_{\rm \bullet}\dot{M}c^2\\
    L_{\rm tot}&=L_{\bullet}+L_{\rm f}=L_\bullet\left(1+\frac{\eta_{\rm f}}{\eta_\bullet}\left(\frac{\dot{m}_{f}}{\dot{M}}\right)\right)
    \end{align}
\end{subequations}
\end{comment}
Taking $\eta_\bullet \sim0.1-0.4$, $\eta_{\rm f}\sim 0.06$ (from our calculation for the final fragment), and $\dot{m}_{f}/\dot{M}\sim 1-10$, we estimate that the variability from the secondary jet could be between $15-600 \%$.

As the fragment migrates closer to the BH (and the accompanying central jet), its accretion rate, $\dot{m}_{\rm f}$, becomes comparable to the central accretion rate $\dot{M}$. The jet launched by fragment accretion could start to interact with the main jet in a number of non-trivial ways.  
When the instantaneous accretion rate onto the fragment (briefly) exceeds the central accretion rate $\dot{M}$, we hypothesize that the fragment's jet luminosity may become larger than the primary jet.  If the two are aligned, this could be visible to a distant observer as an inverted jet structure (i.e. more power off-axis). 
Both misalignment of the fragment jet (with respect to the central jet) and lateral spreading of either jet can cause an actual intersection between the two relativistic outflows (with likely different relativistic velocities). This intersection between the jets could lead to dissipation in shocks, or alternatively in a reconnection layer.  Either of these interactions could potentially accelerate a population of high energy particles  \citep{ResslerandCombi2025,GutierrezandCombi2024}.

For any variability from the secondary jet to be observable, it must reach the progenitor star's envelope {\it after} the primary jet has broken out.  The observed flatness of GRB durations at the shorter end of the {\it long} GRB duration distribution can be used to estimate typical breakout times \citep{Bromberg+12}.  Such observational estimates typically find $t_{\rm out} \sim 10$ s \citep{Bromberg+15}, modestly longer than the $\sim {\rm few}$ s timescales over which fragments migrate inwards and activate their secondary jets (not taking into account the secondary jet propagation time out of the broken stars envelope).  Theoretical simulations of jet breakout find qualitatively similar results \citep{Gottlieb+23}.  However, the breakout time varies as a function of the progenitor star mass and radius, the jet opening angle, and jet energetics, scaling as the jet isotropic equivalent luminosity $L_{\rm iso}^{-1/3}$ \citep{Bromberg+11}. Since fragmentation is only relevant in the highest $\dot{M}$ collapsar disks, and $\dot{M}$ is likely positively correlated to $L_{\rm iso}$, the collapsar disks capable of fragmenting and launching secondary jets may originate in a low-$t_{\rm out}$ tail of the breakout time distribution for which secondary jets activate after the primary has already broken out.  Secondary jet launching could also be favored in situations where migration is slow or fragment formation is delayed by a few seconds after the activation of the central engine.

Throughout this discussion, we have assumed that the fragment accretes mass at the geometrically reduced Bondi-Hoyle rate, $\dot{m}_{\rm RBH}$.  However, the extreme fragment accretion rates discussed in this subsection suggest that energy feedback from the fragment mini-disk into the broader collapsar environment may be significant.  A powerful secondary jet will heat a cocoon of surrounding gas as it exits the collapsar disk, thereby decreasing $\dot{m}_{\rm RBH}$.  At lower $\dot{m}_{\rm f}$ values, secondary jets may not be a major source of feedback, but sub-relativistic mini-disk winds could be.  Below an ``ignition'' threshold in $\dot{m}_{\rm f}$ required for efficient neutrino cooling \citep{Metzger+08,Aman+25}, most of the mini-disk will be geometrically thick and advective, potentially losing substantial mass to winds (\citealt{BlandfordBegelman99}; this both mechanically decreases fragment growth and also may do so indirectly, by heating collapsar disk gas and reducing $\dot{m}_{\rm RBH}$).  While these feedback effects are potentially important, we defer a full investigation of them to future work.

\begin{comment}
From Sean M. Ressler1 ,Luciano Combi 25:
The first mechanism occurs when the black holes power persistent jets and the spins are aligned. In this case, the two jets form an extended reconnection layer that dissipates magnetic energy and causes the jets to merge. This reconnection layer is characterized by a highly magnetized upstream flow (σ<<1 and β=1) with a moderately strong guide field of comparable strength to the reconnecting field. Reconnection in this regime is known to be a source of high-energy particles that would radiate at higher frequencies than the accretion disk.

From Eduardo M. Gutiérrez, Luciano Combi 24:
Magnetic reconnection accelerates particles in the collision region. These particles cool by synchrotron radiation and SSC. Because the jet interaction region is compact and highly magnetised, the emission is internally absorbed at low frequencies (≲ 1013 Hz) due to SSA and at high-energies (≳ 10 MeV) due to gamma gamma absorption.
\end{comment}

%\clearpage
\subsection{Gravitational Waves}
\label{sec:GWs}
As migrating fragments inspiral towards the BH, they emit high-frequency GWs. In this section, we estimate two characteristic quantities of the GWs needed to assess their detectability, their frequency ($f_{\rm GW}$) and strain ($h_{\rm GW}$) at the end of the inspiral, which will occur at either the ISCO ($r_{\rm ISCO}$) or the fragment tidal radius ($r_{\rm t}$), whichever is larger.  We therefore calculate these two quantities as follows (\citealt{peters1964, PiroPfahl07}):
\begin{subequations}
\begin{align}
f_{\rm GW}&=\Omega_{{\rm GW}}/\pi=\begin{cases} 
    \displaystyle \Omega(r_{{\rm t}})/\pi & r_{{\rm t}}>r_{{\rm Isco}}\\[1ex]
    \displaystyle \Omega(r_{{\rm Isco}})/\pi & r_{{\rm t}}\leq r_{{\rm Isco}}
\end{cases}\\[2ex]
h_{{\rm GW}}&=\Theta\frac{G^{5/3} m_{\rm GW} M^{2/3}\Omega_{{\rm GW}}^{2/3}}{c^{4}D_{{\rm obs}}}\\[2ex]
&\approx5\cdot 10^{-23}\Theta D_{{\rm 100}}^{-1} m_{{\rm 0.4}}{M_{\rm 20}}^{2/3}{f_{\rm 780}}^{2/3}. \label{eq: GW scaling}
\end{align}
\end{subequations}
Here $\Theta$ is a dimensionless factor describing orientation of the source with respect to the antenna (spanning $0-4$; hereafter we choose $\Theta=2.5$, taking the mean square expectation of  $\Theta$; \citealt{FinnChernoff93}), $m_{\rm GW}$ is the fragment mass at ${\rm max}(r_{{\rm t}}, r_{{\rm Isco}})$ where $m_{{\rm 0.4}}\equiv m_{\rm GW}/0.4{M}_{\odot}$, the distance from the observer is set to be $100~{\rm Mpc}$, defining $D_{{\rm 100}}\equiv D_{{\rm obs}}/100{\rm Mpc}$, $M_{{\rm 20}}\equiv M/20{M}_{\odot}$ and $f_{{\rm 780}}\equiv f_{\rm GW}/780 {\rm Hz}$ (as is appropriate for the ISCO frequency of the $a=0.95$ case). %We choose a distance of $100Mpc$, as it is in the range of GW detection by LIGO for NS-BH mergers. Also by using the volumetric rates for off axis GRBs (and also choked GRBs) we can estimate the rate of these events by: \dd{add volumetric rates (and citations) and rate calculations}

\begin{figure}
\includegraphics[width=0.48\textwidth]{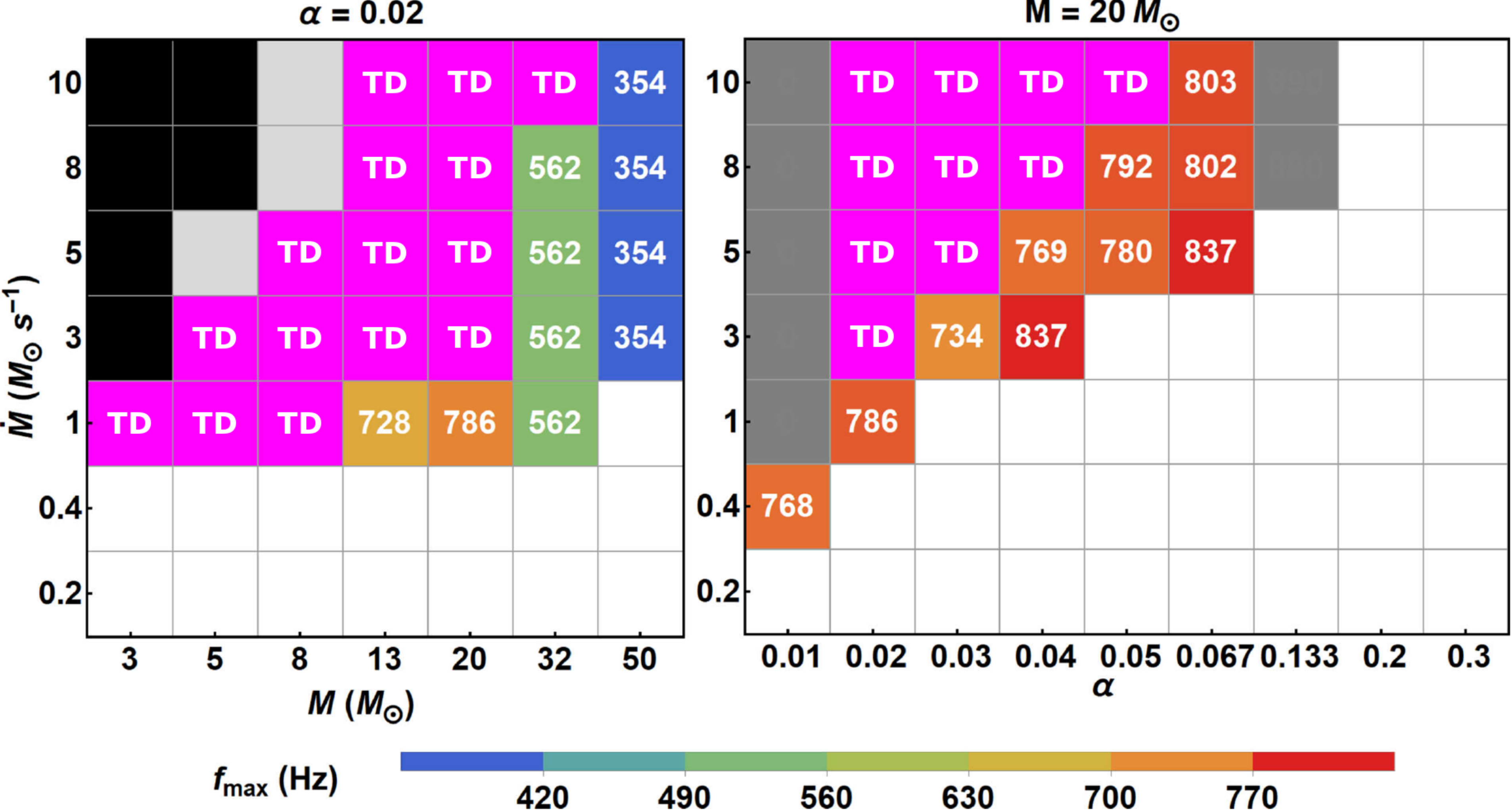}
\caption{The peak orbital frequency $f_{{\rm max}}$ of the GWs emitted by an inspiraling fragment. We compute this by evaluating the orbital frequency ($f_{\rm GW}$) at the maximum between the tidal disruption radius ($r_{\rm t}$) and the ISCO. {\it Left panel}: the viscosity parameter is held constant at $\alpha=0.02$ while the central mass $M$ and the accretion rate $\dot{M}$ vary. {\it Right panel}: the central mass is held constant at $M=20{M}_{\odot}$, while the viscosity parameter $\alpha$ and the accretion rate $\dot{M}$ vary. We do not estimate GW emission from cells color-coded as: (1) white, which do not support fragmentation; (2) black, which are points in parameter space where our numerical model failed to reach a solution; (3) light grey, where final fragment masses are below the minimal stable mass for a cold NS; (4) magenta, where fragments are tidally disrupted at larger radii (see \ref{sec: Tidal disruption check}), though this depends on the hot mass-radius relationship; and (5) dark grey, where the disk mass enclosed by the fragment's formation radius exceeds the central mass. The rainbow color system in other cells represents the frequency, as represented in the bar legend below the figure. The frequency generally increases as the central mass decreases or the viscosity increases, while it does not significantly depend on the accretion rate.}
\label{fig:AngularFrequancy_alpha002M20_PaperRas}
\end{figure}

\begin{figure}
\includegraphics[width=0.48\textwidth]{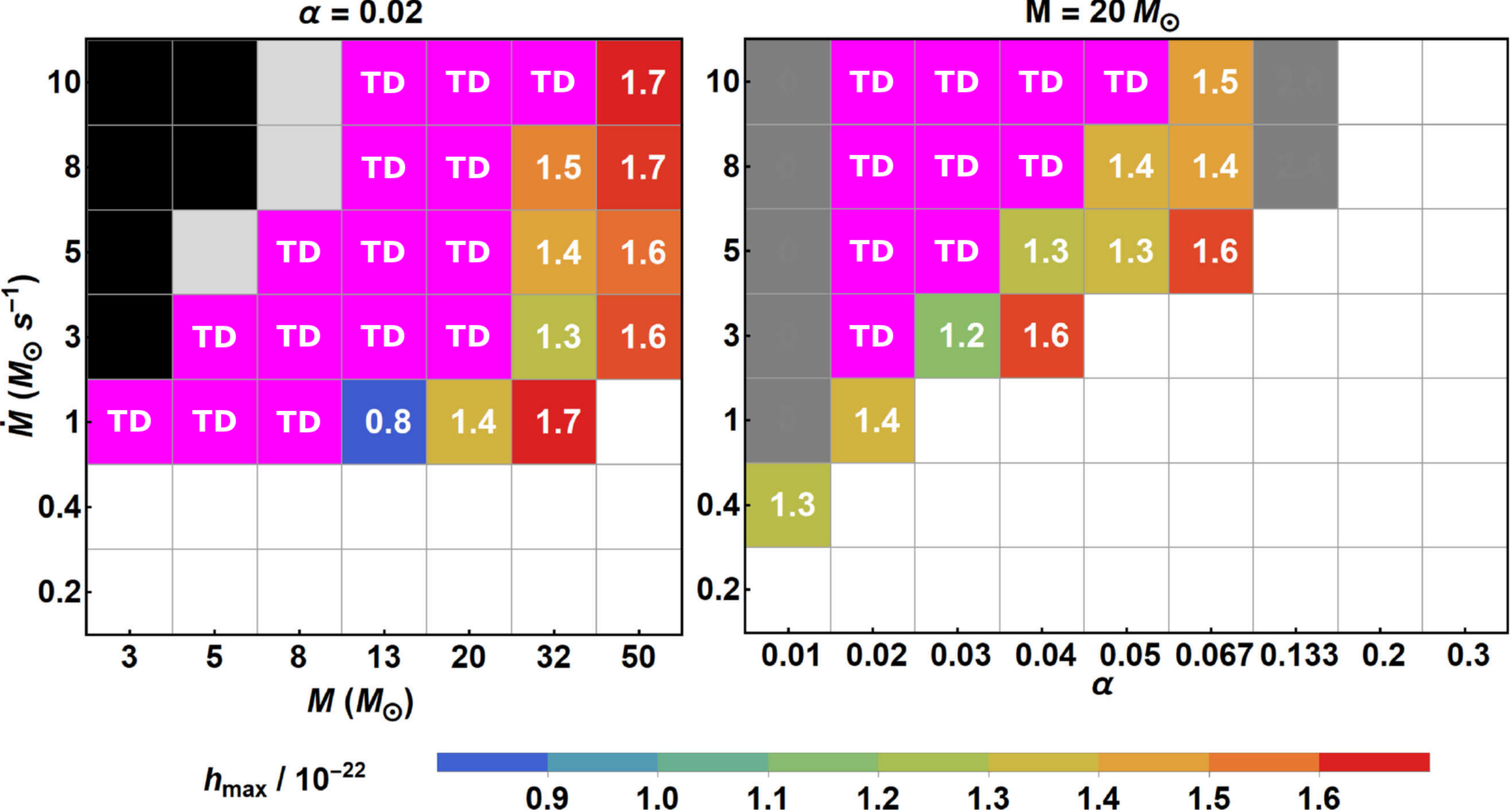}
\caption{The peak strain $h_{{\rm max}}$ of the GWs emitted by an inspiraling fragment is shown in the same parameter space as in Fig. \ref{fig:AngularFrequancy_alpha002M20_PaperRas}, normalized by $10^{-22}$.  As previously we compute this by evaluating the strain ($h_{\rm GW}$) at the maximum between the tidal disruption radius ($r_{\rm t}$) and the ISCO. {\it Left panel}: the viscosity parameter is held constant at $\alpha=0.02$ while the central mass $M$ and the accretion rate $\dot{M}$ vary. {\it Right panel}: the central mass is held constant at $M=20{M}_{\odot}$, while the viscosity parameter $\alpha$ and the accretion rate $\dot{M}$ vary. This figure has the same color scheme as in Fig. \ref{fig:AngularFrequancy_alpha002M20_PaperRas} for cells where we do not compute GW emission. 
The peak GW strain is generally insensitive to $M$, $\dot{M}$, and $\alpha$, at least in those parts of parameter space where we can estimate the GW emission.}
\label{fig:Strain_alpha002M20_PaperRas}
\end{figure}
Figs. \ref{fig:AngularFrequancy_alpha002M20_PaperRas} and \ref{fig:Strain_alpha002M20_PaperRas} show the GW strain and angular frequency, respectively, at the smallest radius we expect the fragment to survive at (i.e. the maximum of the tidal disruption radius, $r_{\rm t}$, and the ISCO).  % (as some fragments pass the ISCO undisrupted while some disrupt between the ISCO and the radius in which fragments exceed the minimum cold stable mass).
The maximum angular frequency is strongly correlated with the central mass because of the ISCO scaling. %distance, as the ISCO becomes smaller with decreasing the central mass the orbital frequency of the fragment increases.  
As the fragment migrates and grows with mass, it sweeps through an range of increasing angular frequencies.  Because fragments usually form at large distances, their initial inspiral will not depend strongly on the central BH spin, but the maximum achievable frequency can be a sensitive function of spin (particularly for the largest fragments, which do not tidally disrupt).  The range in maximum frequencies that we see in Fig. \ref{fig:AngularFrequancy_alpha002M20_PaperRas}, where $a=0.95$, are in between $354-837 \text{Hz}$. In contrast, the $a=0.001$ alternate case we have considered sees maximum frequencies between $88-350~{\rm Hz}$ (Fig. \ref{app: AngularFrequency_alpha002M20_a0}).  

The maximum strain is likewise a sensitive function of spin for high mass fragments that avoid tidal disruption.  As our scaling relation in Eq. \ref{eq: GW scaling} indicates, for a fiducial source at $D=100~{\rm Mpc}$, peak strains $h \sim 10^{-22}$ are achieved across most of the parameter space we consider.  However, it is important to note that only some parameters allow us to estimate GW properties during the fragment inspiral.  As discussed previously, fragmentation does not even occur for lower accretion rates ($\dot{M} \lesssim 0.4 M_\odot~{\rm s}^{-1}$) and higher effective viscosities ($\alpha \gtrsim 0.133$).  For many of the smallest BHs and largest accretion rates we consider, we are unable to estimate GW emission, usually because of uncertainties about tidal disruption at large radii where the mass-radius relationship for hot, low mass fragments is unclear.  For the smallest values of BH mass ($M\lesssim 10 M_\odot$), we sometimes encounter other problems, such as fragments that never exceed the minimum cold neutron star mass, or numerical non-convergence of our disk model.

Gravitational wave signals from fragments merging with the BH in a collapsar are not limited to events that produce a relativistic jet (i.e. LGRBs) but also can arise from choked collapsars: events where a merger generates a GW signal without an associated high-energy jet. While the LGRB rate is observationally constrained \citep{WandermanandPiran2010}, the total volumetric collapsar rate (the sum of jet-producing and choked collapsars) is highly uncertain, so we take a parametric approach to estimating the potential GW detection rate.  The conservative beaming corrected LGRB volumetric rate, $\dot{\rho}_{\rm LGRB} \approx 10 \, \text{Gpc}^{-3} \text{yr}^{-1}$ \citep{Gottlieb+24}, represents only an unknown fraction of the total collapsar volumetric rate $\dot{\rho}_{\rm Coll}$. We express this relationship as $\dot{\rho}_{\rm Coll} = \chi_{\rm choke} \dot{\rho}_{\rm LGRB}$ (where the enhancement factor $\chi_{\rm choke} \geq 1$). 

We have estimated a typical maximum strain of $\sim 1.5 \cdot 10^{-22}$. Taking a LIGO ${\rm O4}$ strain limit for detection\footnote{We estimate this limit crudely by taking a detection horizon of $160$ Mpc for the merger of two $1.4 M_\odot$ NSs, each with a radius of $10$ km and $\Theta=2.5$.} of $\sim 5 \cdot 10^{-23}$, the corresponding maximum detection distance for fragment-BH mergers is approximately $295 ~{\rm Mpc}$. 
From these estimates, we derive the LIGO detection rate ($\dot{\Re}_{\rm LIGO} $) of such GW signals as a function of $\chi_{\rm choke}$, the fraction of collapsars that generate fragments $f_{\rm f} \le 1$, and the mean number of fragments that merge with the BH $N_{\rm f} \geq 1$:
\begin{equation}
    \dot{\Re}_{\rm LIGO} = 1.08 \cdot \chi_{\rm choke} f_{\rm f}N_{\rm f} \, \text{yr}^{-1}
\end{equation}\label{eq: choked GW rate}
A plausible range of $\chi_{\rm choke}$ is between $10-10^{4}$ \citep{Gottlieb+24}, suggesting a significant yearly detection rate is possible (specifically, this requires $\chi_{\rm choke} f_{\rm f} N_{\rm f} \gtrsim 1$).

The above calculation is approximate in a number of ways and could be improved in the future with a more precise estimate of signal-to-noise ratios for different fragment inspiral waveforms.  However, even setting aside the level of approximation we have worked with here, there are notable uncertainties in the GWs produced by fragments that we defer for future work.
\begin{itemize}
    \item The GW signal will cut off sharply if fragments tidally disrupt outside the ISCO, but where and whether this occurs depends sensitively on the fragment mass-radius relationship, which we have modeled optimistically by taking a cold NS mass-radius relation (for fragments above the minimum cold NS mass).
    \item The maximum frequency and strain for non-disrupted fragments will depend sensitively on the ISCO location and thus on the spin of the BH.  In our fiducial estimates here we have optimistically assumed a rapidly spinning ($a=0.95$) central BH, although the pessimistic case ($a=0.001$) is considered in Appendix \ref{app: a0}.  In reality, the spin of the central BH will reflect a balance between spin-up through accretion and spin-down from other processes, such as Blandford-Znajek jet power; the time evolution and equilibria of this process are actively debated \citep{Lowell+25}.
    \item Fragments migrating through the dense environment of a collapsar disk will experience orbital evolution both to GW losses but also, potentially, from non-vacuum forces, such as the gravitational-hydrodynamic Type I torques or the mass accumulation torques that we have modeled earlier.  As long as these torques are negative (hastening the inspiral), the resulting GW signal will be weakened, but if they become positive (e.g. in a migration trap), then the GW signal could be prolonged and strengthened.  These hydrodynamic torques, regardless of sign, can also de-phase the waveform to the point where vacuum point-particle inspiral templates are no longer suitable for matched filtering \citep{Zwick+25}.
    \item Likewise, typical GW predictions (including those here) examine the inspiral of constant-mass objects. In our scenario, both masses grow during the inspiral. The central mass grows more modestly (typically by a range from a few 10s of percent), while the fragment itself grows rapidly, by 2-3 orders of magnitude. This growth would cause an accelerated chirp that would be substantially different from constant-mass templates, also causeing a breakdown of matched filtering.
    \item Finally, there may be other sources of GWs from the system, in particular from the inspiral of other fragments or from the generation of global instabilities due to disk self-gravity \citep{Gottlieb+24}.  More sources of GW radiation may produce ultimately higher strains but may also make the source harder to detect, by (again) limiting the applicability of matched filtering sources on point-particle templates.
\end{itemize}

\begin{figure} % Use figure* to span both columns
    \centering
    \begin{subfigure}{\textwidth}
    %\centering
        \includegraphics[width=0.42\textwidth]{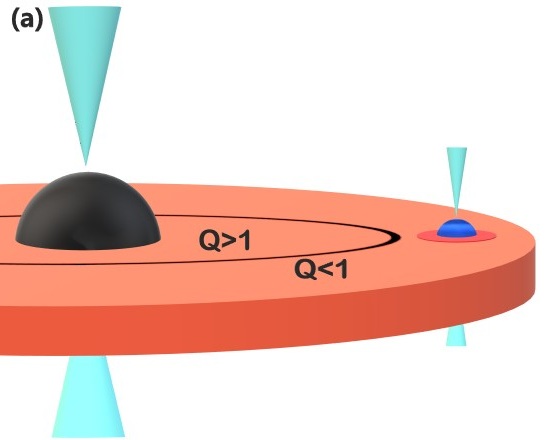}
         %\caption{}
        \label{fig: cartoon Toomre Q cropped}
    \end{subfigure}
    
    \begin{subfigure}{\textwidth}
    %\centering
        \includegraphics[width=0.42\textwidth]{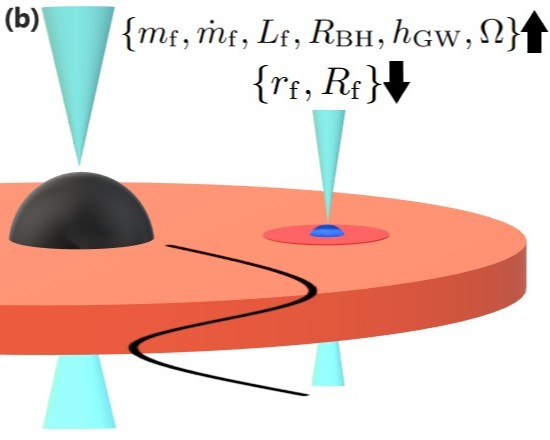}
        %\caption{}
        \label{fig: cartoon Increased cropped}
    \end{subfigure}

    \begin{subfigure}{\textwidth}
    %\centering
        \includegraphics[width=0.42\textwidth]{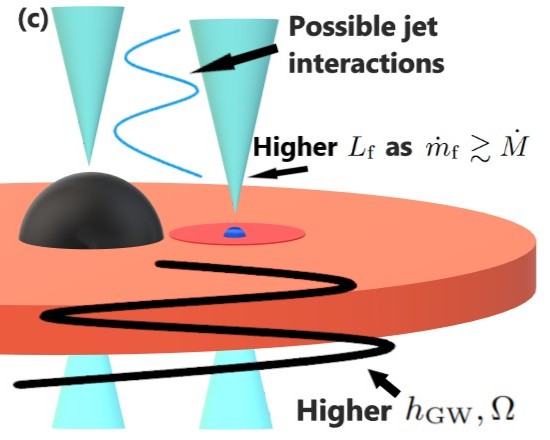}
        %\caption{}
        \label{fig: cartoon jet interactions}
    \end{subfigure}
    \caption{A schematic of the emission from primary and secondary jets (in cyan) in a fragmenting collapsar disk.  {\it Panel (a)}: a low mass fragment will form at large radii in the $Q<1$ zone; Bondi-Hoyle accretion will initially power a weak secondary jet.  {\it Panel (b)}: the fragment grows in mass $m_{\rm f}$ as it migrates inwards through the collapsar disk.  The Bondi-Hoyle accretion rate increases (increasing power of the secondary jet), due both to the increase in $m_{\rm f}$ and the increase in local gas density at smaller radii. A lower frequency and strain GW signal is shown (in black). {\it Panel (c)}: if the fragment can reach the inner disk without tidally disrupting, $\dot{m}_{\rm f}$ can briefly exceed $\dot{M}$, potentially creating GRB variability from a powerful secondary jet.  Jet-jet interactions may be important at this stage.  Gravitational wave strain and frequency become stronger throughout the fragment inspiral. The fragments size decreases trough the inspiral as speculated from the mass to radius relation. }
    \label{fig: Jet cartoon}
\end{figure}

We summarize the observable consequences of disk fragmentation with a schematic diagram in Fig. \ref{fig: Jet cartoon}, showing different phases of a fragment's inspiral.  Beginning with its initial formation as a low-mass object in the collapsar disk $Q<1$ zone, a fragment will grow through Bondi-Hoyle accretion as it migrates inwards.  As its inspiral takes it to smaller radii, the fragment accretion rate grows dramatically, increasing the power of any secondary jet it launches.  Both the power of the secondary jet and the GW emission will be maximized near the smallest radius the fragment can migrate to (i.e. the larger of the ISCO or the fragment tidal disruption radius). At small radii, jet-jet interactions may become important.

\section{Discussion}
\label{sec:discussion}

In our analysis of the fragment initial conditions, we have found that the fragments are born with initial masses $10^{-3} \lesssim M_{\rm f}/ M_\odot \lesssim 10^{-1}$ (see Fig. \ref{fig: InitialFragmentMass_2_panels}), typically orders of magnitude lower than the minimal NS stable mass (which is $\approx0.1{M}_{\odot}$ baryon mass for the cold EOS and $\approx0.7{M}_{\odot}$ baryon mass for the hot EOS).
Assuming that a fragment remains bound during, Bondi-Hoyle accretion from the surrounding collapsar disk, it will in many cases grow enough to reach the range of stable NS masses (see Fig. \ref{fig:FragmentMassGrowth_2PanelArray}). Past work studying fragmentation in collapsars usually estimated initial fragment masses in simpler, order-of-magnitude ways and reached more optimistic conclusions about immediately forming stable NSs \citep{PiroPfahl07, Metzger+24,Shahamat2021}.  Perhaps as a result, past work did not consider Bondi-Hoyle mass growth and its contribution to fragment masses or (via mass accumulation torques) migration.  In contrast to past examinations of this problem, we find that formation of stable fragments is unlikely without substantial mass accretion, which may need to fight against mass loss in (initially) unstable fragments.

%

%Which collapsars generate self-gravitating fragments?  How do these correspond to observed long-GRBs?  (Twiddle-level energetics/luminosity calculation as Tsvi suggested)  What durations are relevant?

In \S \ref{sec:instability_results}, we have seen that fragmentation is generally favored for high values of $\dot{M}$ and low values of $\alpha$; both these parameter choices increase the surface density of the collapsar disk and therefore the importance of its own self-gravity.  In contrast, the central mass $M$ has a weak impact on the profiles of both $Q$ and $\beta$ with respect to a dimensionless radius $r/r_{\rm S}$.  While low values of $\alpha$ emerge naturally in some MHD simulations of collapsar disks \citep{Siegel+22}, the peak $\dot{M}$ values are less clear, and depend sensitively on the central density and rotation profiles of the pre-collapse progenitor star \citep{Gottlieb+22a, Gottlieb+22b}.  As these profiles are not well constrained by existing stellar evolution calculations, a range of peak $\dot{M}$ emerge from core collapse supernova simulations, in many cases reflecting the initial conditions.

We can put some constraints on both the peak $\dot{M}$ and its steady accretion duration from observations of successful long GRBs.  Assuming a $\gamma$-ray radiative efficiency $\eta_{\rm eff}$ for the emitted jets, then the peak accretion rate $\dot{M} = L_{\gamma} / (\eta_{\rm eff} c^2)$ (here $L_\gamma$ is the peak $\gamma$-ray luminosity).  While luminosities in most long GRBs are measured only in an isotropic-equivalent way, the true beaming-corrected luminosity can be inferred for the minority of long GRBs with measured jet breaks \citep{Sari1998}.  The $\gamma$-ray radiative efficiency is not so easily determined observationally, but we will attempt to bound this by relating the (beaming-corrected) energy release $E_\gamma$ to the mass accreted through the collapsar disk, $\Delta M$: $E_\gamma = \eta_{\rm eff} \Delta M c^2$.  The two observables $E_\gamma$ and $L_\gamma$ can thus be combined to estimate the peak $\dot{M} = \Delta M L_\gamma / E_\gamma$ if $\Delta M$ is known, and also to estimate the typical time of peak accretion $t_{\rm peak} = E_\gamma / L_\gamma$.  

To estimate plausible ranges of $E_\gamma / L_\gamma$ we use the long GRB catalogs of \citet{Tsvetkova+17, Tsvetkova+21}, which compile 62 long GRBs with measured jet breaks and thus estimates for beaming-corrected $E_\gamma$ and $L_\gamma$.  In this sample, typical values for the beaming-corrected energy release are $E_\gamma \sim 10^{49-52}~{\rm erg}$ and typical values for the beaming-corrected peak $\gamma$-ray luminosity are $L_\gamma \sim 10^{49-51.5}~{\rm erg~s}^{-1}$.  The mean and median values of $t_{\rm peak} = E_\gamma / L_\gamma$ are 4.5 and 3.2 s, respectively; the 10th percentile and 90th percentile values for $t_{\rm peak}$ are 0.6 and 9.0 s, respectively.  This implies that typical values of $\dot{M} \sim 1 M_\odot~{\rm s}^{-1}$ will be achieved if $\Delta M \sim 1 M_\odot$; larger values will be reached if $\Delta M \sim 10 M_\odot$.  It is only if $\Delta M \ll 1 M_\odot$ that the accretion rates required for fragmentation will not be reached in collapsar disks.  Since we are primarily interested in central BHs with masses $\sim 10 M_\odot$ that have grown by rapidly accreting a large fraction of their mass from the collapsing star's core, high ($\gtrsim 1 M_\odot~{\rm s}^{-1}$) accretion rates are generally plausible unless the pre-collapse core was rotating so slowly that $\gtrsim 90 \%$ of the BH mass was acquired through radial free-fall rather than through collapsar disk accretion. 

\begin{comment}
can be sustained for a time $t_{\rm peak} = E_{\rm rad} / (\eta \dot{M} c^2)$, where $L_{\rm peak}$ is the peak luminosity and $E_{\rm rad}$ the radiated energy.  Equivalently, 
\begin{align}
    \dot{M} \approx& 0.6 M_\odot~{\rm s}^{-1} \left(\frac{\eta}{0.1}\right)^{-1} \left(\frac{L_{\rm peak} }{10^{53}~{\rm erg~s}^{-1}} \right) \\
    t_{\rm peak} \approx& 10~{\rm s}~\left(\frac{E_{\rm rad} }{10^{54}~{\rm erg} } \right) \left(\frac{L_{\rm peak} }{10^{53}~{\rm erg~s}^{-1}} \right)^{-1}.
\end{align}
Note that here both $L_{\rm peak}$ and $E_{\rm rad}$ have been used in a beaming-corrected sense; isotropic equivalent luminosities and energy releases would set upper limits on $\dot{M}$ and $t_{\rm peak}$, respectively.
Within this crude framework, we can use beaming-corrected GRB energetics to see that typical long GRBs are too dim to have accretion rates susceptible to fragmentation: observations infer an average $E_{\rm rad} \sim 10^{51}~{\rm erg~s}^{-1}$ \citep{Frail+01} and an average $L_{\rm peak}$
\end{comment}

% we are the only model that makes a time dependent migration calculation that also involves different types of torques combined. this lets us to be the only model that has m_f(t), maybe that too big of a clue with regards to a us having a waveform. also we use an approch to GRB variability of calculting the tidal radius which uses a mass to radius relation that past work dosen't. We are also the only ones solving a more complex diskmodel
\subsection{Possible implications}\label{sec: discussion Possible implications}

While fragment disruption provides only modest GRB variability, a more significant effect may arise from the rapid accretion onto embedded fragments as they migrate inward. As fragments approach the ISCO, their accretion rates increase by several orders of magnitude, sometimes exceeding the central accretion rate. This could power secondary jets via neutrino annihilation or proto-magnetar spin-down, introducing variability between $15\%$ and $600\%$. \dd{The secondary jets appear to exhibit two distinct timescales: the migration time of the fragment until it reaches gap formation—halting accretion and secondary jet power—which is roughly in the range of $\sim 0.1$–$1 {\rm s}$ (see Fig.~\ref{fig:AccretionRateRatios_3_Panel}); and the duration during which the accretion rate of the fragment exceeds the central accretion rate (see Fig.~\ref{app:DurationOfHighAccretion_alpha002M20_a0}), which is $\sim 1$–$10 {\rm ms}$. The migration timescale could plausibly be related to the variability timescales observed in long GRBs, where the fastest spikes in the prompt emission typically occur on timescales that are greater than $ \sim 0.1 {\rm s}$ \citep{Magnus25}.}

Both, misalignment between the two jets and their widening could lead to interacting ejecta with different relativistic velocities, potentially causing jet-jet shocks. As the jets interact further, they may form a reconnection layer, merging into a high-energy particle accelerator capable of emitting radiation at extremely high frequencies. If a fragment’s accretion rate temporarily surpasses that of the BH, its jet could outshine the primary jet, creating an inverted jet structure.

If the fragmentation timescale turns out to be small compared to both the duration of the entire collapse event (the average time for LGRBs is in tens of seconds, \citealt{Kouveliotou1993,Kumar2015,Bissaldi2017}) and the fragment migration times, we hypothesize that fragments might collide and scatter \citep{DOrazio2018} because of constant stochastic creation and rapid migration at different radii. For example, if a fragment formed at an outer radius migrates inwards, it could overtake fragments that formed at lower radii. Close encounters \dd{can lead to binary formation via gas dissipation \citep{Liu2023}, in turn producing} mergers of high-density and high mass (following accretion fragments in a gas-rich environment.  This exciting possibility could produce non-vacuum GWs with electromagnetic counterparts \citep{PiroPfahl07,Fedrow2017,Metzger+24}, while also adding to GRB variability. Also, the presence of multiple bound fragments will lead to few-body scatterings \citep{Stone2019} that can eject single fragments or even binary fragments from the system. Since the end product of fragmentation with these extreme initial densities and fast mass growth is unclear, these ejected fragments could become compact objects such as BHs and NSs. An ejection of a compact binary might let them merge in a less gaseous environment, which would help produce the observed vacuum chirp signature seen by GW detectors.

If unstable low mass fragments %(with masses $\lesssim 0.67 M_\odot$) 
are ejected to infinity or even to the outer regions of the collapsing massive star, their subsequent explosion \citep{Colpi1989} could create a novel astrophysical site for r-process nucleosynthesis \citep{Kajino2019}.  Here the precise meaning of ``low mass'' will depend on the cooling rate of the fragments; if they cool rapidly, the cold NS EoS minimum mass ($\approx 0.1 M_\odot$) will be the upper limit for fragment explosions, but if they cannot cool quickly enough, the hot NS EoS minimum mass ($\approx 0.6-1 M_\odot$) will instead be that upper limit.  Rapid deconfinement of an unstable, low mass fragment will begin in its surface layers but will eventually run away and deconfine the neutron-rich core, providing a potentially novel site for the r-process to occur.  While the total r-process yield from these fragments is quite unclear (as it will depend on both typical fragment masses as well as the total number of ejections), the indirect signatures may resemble those of other collapsar \citep{Siegel+19} or ``hypernova'' \citep{Mosta+18} models for r-process production: high yields, low volumetric occurrence rate, short delay time distributions, and enrichment occurring even in dwarf galaxies with low escape velocities (that could not e.g. easily retain binary neutron star systems).

%How does fragment collapse/contraction proceed?  (We don't know , will investigate in the future, mention things like ram pressure).  What range of mass-radius relationships is possible?  What range of minimum/maximum stable masses is possible? 
The process by which fragments collapse or contract remains uncertain and requires further study. In vacuum, fragments below the minimum stable neutron star mass cannot maintain hydrostatic equilibrium and will either collapse or expand dynamically. However, in the complex environment of a collapsar disk, factors such as ram pressure and mass replenishment through Bondi-Hoyle accretion may allow these low-mass objects to persist longer or even become quasi-stable (see discussion in Sec. \ref{sec:observables}). If accretion is highly aspherical, ram pressure may not be sufficient to stabilize the fragment’s polar regions, but continuous mass replenishment through a mini-disk could enable the fragment to gain mass faster than it loses it. The exact stability and evolution of these fragments depend on how these competing effects interact, and further investigation is needed.  

For fragments above the minimum neutron star mass, their mass-radius relationship is estimated using the BSk24 equation of state (EOS). In the zero-temperature approximation, this model predicts a minimum stable mass of about $0.09\,M_\odot$. However, if the fragment temperature is high, finite-temperature effects can significantly increase the minimum stable mass, potentially exceeding $0.5\,M_\odot$. Since accurate modeling of fragment temperatures is beyond the scope of the present study, the exact impact of these effects remains uncertain. Whether a fragment ultimately collapses, becomes tidally disrupted, or reaches a stable neutron star configuration depends on temperature, neutrino cooling timescales, and the detailed interplay of accretion and mass loss.

%What volumetric rate is plausible for fragmentation collapsar disks?  What LIGO horizon results from that?
%Our discussion of a GW signal from a merger between, a fragment and the BH, in a collapsar system isn't limited to collapsar events that produce a jet but also to chocked events that produce the GW signal from the merger but not a highly energetic detectable jet. As the volumetric rate of the total collapsar event rate (LGRB rate combined with choked collapsar rate) is highly debatable we take a different approach in inferring anything about the possible detection rate of these events. The maximum distance in which our average maximum strain ($\sim 1.5\cdot10^{-22}$) could be detected by LIGO's A+ minimum sensitivity ($\sim 2\cdot10^{-23}$) is roughly $750 Mpc$. As stated above the LGRB volumetric rate ($\dot{\rho}_{\rm LGRB}\approx1.3 Gpc^{-1}year^{-1}$, \citealt{WandermanandPiran2010}) is an unknown fraction of the total collapsar volumetric rate ($\dot{\rho}_{\rm Coll}$) meaning that we can write the following equation $\dot{\rho}_{\rm Coll}=\chi_{\rm choked} \dot{\rho}_{\rm LGRB}$ (Where $\chi_{\rm choked}\geq1$). From all the data specified above we reach an estimate on the LIGO detection rate of these GW signals as a function of $\chi_{\rm choked}$ to be $\dot{\Re}_{\rm LIGO}=1.36 \cdot {\chi}_{\rm choked} \cdot year^{-1}$

We have also estimated a significant GW detection rate from fragmentation in collapsars (As specified in Eq. \ref{eq: choked GW rate} and under certain ranges of the highly unconstrained value of $\chi_{\rm choke} f_{\rm f} N_{\rm f}$). The derived detection rate $\dot{\Re}_{\rm LIGO} = 1.08 \cdot \chi_{\rm choke} f_{\rm f} N_{\rm f} , \text{yr}^{-1}$ suggests that fragment mergers within choked collapsars could substantially contribute to a new GW signal population. This highlights the potential of LIGO and future detectors to probe the dynamics of core collapse beyond traditional gamma-ray bursts.

As the fragments migrate inward due to various torques they emit GWs. The resulting GW signal will be quite exotic in comparison to previously observed compact binary inspirals: the mass ratio will be extreme, and the secondary (i.e. the fragment) will potentially be a very low mass ($\sim 0.1 M_\odot$) neutron star, which tidally disrupts at a large radius, cutting off the waveform at a low frequency.  Because of the influence of torques from gas interactions (see section \ref{sec: Fragment migration}), the gravitational waveform of these fragments would differ from a vacuum waveform template (for which the migration velocity is given only by Eq. \ref{eq: type GW rate of migration}). This difference could cause these mergers to evade matched filtering searches for GW signals, though if very strong they could still be detected in a burst search. As most fragments form a gap in the disk at the end of the migration period, which turns off torques from the gaseous environment, we expect that the end of the merger might look more like a vacuum signal.

\subsection{Validity of Model}\label{sec: Validity of Model}\label{subsec: Validity of Model}
The results presented in this section rely on a number of assumptions that we have made either explicitly or implicitly.  Here we perform {\it post hoc} consistency checks to see when these assumptions are valid and when they may break down.  Note that we will not recapitulate here the earlier discussions of assumptions behind ``fundamental'' results in Sec. \ref{sec:results} (e.g. choice of critical $\beta$ for fragmentation, parameter values, etc.).
\begin{itemize}
%\item Bondi-Hoyle accretion geometry: 

\item Eddington limits to fragment growth: in this work, we have so far assumed that fragments in the collapsar disk will accrete at the Bondi-Hoyle rate, or by a version of this that is reduced by geometrical factors if $R_{\rm BH}$ is larger than $H$ or $R_{\rm H}$ (Eq. \ref{eq: Bondi-Hoyle accretion rate}).  As with the collapsar disk itself, the implied fragment accretion rates $\dot{m}_{\rm f}$ are many orders of magnitude above the photon Eddington limit; however, we disregard this Eddington limit because the photons are effectively trapped \citep{Begelman78}.  However, there are situations where the reduced Bondi-Hoyle $\dot{m}_{\rm f}$ may approach or even exceed a {\it neutrino} Eddington limit \citep{DiMatteo2002}, and it is not obvious that this can be ignored.  The final outcomes of super-Eddington accretion remain poorly understood even for the photon Eddington limit, with some radiation-hydrodynamics simulations finding super-Eddington accretion rates and luminosities \citep{Jiang2014}; others finding that very high mass accretion rates can be accommodated by a reduction in the disk radiative efficiency \citep{Sadowski+14, SadowskiNarayan16, Jiang2019}; and still others find that outflows regulate super-Eddington accretion rates to the Eddington limit set by naive radiative efficiencies.  

Setting aside this fundamental question, we can also ask more practically: in what parts of parameter space will our previously computed mass growth rates be reduced by a neutrino Eddington limit, should that apply? We have examined this in a separate calculation where we set our fragment accretion rate to be $\min(\Dot{m}_{{\rm f}},\Dot{m}_{{\rm Edd}})$, where %. We have checked two limits of approximated Eddington accretion rates ($\Dot{m}_{{\rm Edd}}$). Where the Eddington accretion rate is approximated as $\Dot{m}_{{\rm Edd}}\approx L_{\rm Edd}(\eta_{\rm Edd}c^2)^{-1}$ and 
the neutrino Eddington luminosity ($L_{\rm Edd}$) is given by the following equation:
\begin{equation}
    L_{\rm Edd}=\frac{4\pi G c H\rho m_{{\rm f}}}{\tau_{\rm \nu}+\tau_{\rm \thickbar{\nu}}}
\end{equation}
Where $\tau_{{\rm \nu}}$ and $\tau_{{\rm \widetilde{\nu}}}$ are the neutrino and anti neutrino optical depths where we use the definitions from CB07. The approximate neutrino Eddington accretion rate is $\Dot{m}_{{\rm Edd}}\approx L_{\rm Edd}\eta_{\rm f }^{-1}c^{-2}$.  As we expect $\eta_{\rm f} \approx G m_{\rm f} / (R_{\rm f} c^2)$, we consider the limiting cases of $\eta_{\rm f}=0.005$ and $\eta_{\rm f}=0.1$, as would be appropriate for accretion onto a very low-mass fragment ($m_{\rm f} \sim 0.1 M_\odot$, $R_{\rm f} \sim 35$ km) and a more neutron star-like fragment ($m_{\rm f} \sim 1 M_\odot$, $R_{\rm f} \sim 10$ km), respectively.  We find that if $\eta_{\rm f} = 0.005$, all of our results are unchanged, while if $\eta_{\rm f}=0.1$, then the neutrino Eddington limit has a significant effect on the maximum fragment accretion rate and the final fragment masses. This effect is especially strong in regions of parameter space where the central mass is lower than $20 M_\odot$ and the accretion rate is relatively high ($\gtrsim 3 M_\odot {\rm s}^{-1}$; see appendix section \ref{app: Eddington limit}).  Final fragment masses are not reduced dramatically for the highest $M$ and lowest $\dot{M}$ combinations.

Although we do not show the results here, we have also performed a side calculation with $\eta_{\rm f}=0.01$ (appropriate for $m_{\rm f} \approx 0.15 M_\odot$) and as with the $\eta_{\rm f}=0.005$ case, the results are largely unchanged from our fiducial scenario.  In summary, the neutrino Eddington limit will not affect early stages of fragment growth, but could potentially begin to alter fragment accretion rates for $m_{\rm f} \gtrsim 0.2M_\odot$, as $\eta_{\rm f}$ increases.

%\item Lack of heavy nuclei: 

\item Timescales: throughout our work, we employ variations on the steady state collapsar disk models of \citet{CB2007}.  Core collapse simulations generally show that collapsar disk accretion rates $\dot{M}$ evolve (downwards) swiftly, so some of our results can only be valid over a limited radial range and duration.  Likewise, we have assumed a fixed central mass $M$, but accretion onto the central black hole will increase $M$ over time.  As we do not explicitly model time evolution in $M$ or $\dot{M}$,  we have only accounted for these caveats with a post-hoc comparison of relevant timescales, namely the viscous time at $r_{\rm Q}$, the mass doubling time for the BH $M/\dot{M}$, the migration time $\Delta t$, and the fragmentation time $t_{\rm frag}  \approx Q/\Omega $.  In general, $t_{\rm frag}$ is shorter than all other timescales in the problem and thus can be ignored (i.e. a linearly unstable overdensity will always successfully fragment given favorable $\beta$).  We consider our disk model to be able to self-consistently handle\footnote{The prefactor is chosen based on the relatively fast propagation of disturbances through 1D viscous disk models, e.g. \citealt{Pringle81}.} the inner part of a Toomre-unstable zone if $t_{\rm visc}(r_{\rm Q}) \le 5 M/\dot{M}$.  Finally, we have mark results based on fragment migration as questionable if $\Delta t > M/\dot{M}$.  Across the $\{M, \dot{M}, \alpha\}$ parameter space we have explored, there are substantial regions of interest for fragment formation and evolution; it is possible that other zones (especially at lower $\dot{M}$) could emerge if larger radii could be treated accurately, but for the simple reasons listed here, this would likely require time-dependent disk models.

\item Migration in a collapsar disk: in modeling the migration of the fragments, we used pieces of physics (such as gap formation, type I and type II torques, and fragment cooling) which were originally derived and calibrated with regards to protoplanetary disks. In some cases these processes have been studied in other contexts, such as AGN disks, but they have received little detailed study in the context of collapsar disks.  The very different microphysics of collapsar disks (compared to protoplanetary or AGN disks) may imprint itself into the macrophysics of migration and cooling via different equations of state or emission mechanisms; while a full exploration of this is beyond the scope of our work, it is important to note that future investigations of fragment evolution in collapsars may alter the migration formulae and cooling criterions we have used here.%  As these are functions of the object interacting with the surrounding gas it isn't obvious that the correlation to the microphysics of a collapsar disk should apply.

\item Quasi-circular inspirals: our treatment of fragment migration further assumes that fragment semimajor axes shrink adiabatically through a slow, quasi-circular inspiral.  While it is generally true that Type I and GW torques act to damp out initial eccentricities, we do observe situations where the migration rate is so high that the radial velocity $V_{\rm a} \sim V_{\phi}$, the azimuthal velocity.  We flag these results as problematic, as here the validity of our migration and Bondi-Hoyle formulae becomes more doubtful; this problem generally emerges in the final stages of migration, on radial scales $r \lesssim 6 r_{\rm g}$.

\item Validity of Kerr metric: throughout this paper, we have used a Kerr metric framework for calculating the disk dynamics (e.g. angular frequency $\Omega$) and the migratory evolution of the fragment (i.e the GW torques). As we go farther from the BH, the mass in the disk becomes significant compared to $M$, making our use of the Kerr metric less reliable.  In theory, the use of the Kerr metric could also be problematic if $m_{{\rm f}}/M \sim 1$, but in our results this ratio never exceeds $5\%$.
\end{itemize}

\section{Conclusions}
\label{sec:conclusions}

We have explored the linear stability of collapsar disks to self-gravitational perturbations, and (more approximately) the non-linear consequences of fragmentation.  By employing a modified \citet{CB2007} model to describe the structure of steady state, 1D, general relativistic hyperaccretion disks, we have performed a broad parameter survey of gravitational instability in the parameter space of central mass $M$, accretion rate $\dot{M}$, and effective viscosity $\alpha$.  In parts of disk parameter space where self-gravity generates Toomre instability ($Q<1$), and cooling is fast enough to permit fragmentation ($\beta < \beta_{\rm c})$, we estimated initial fragment masses and studied their subsequent time evolution as (i) they grow in mass through accretion, (ii) migrate inwards through the collapsar disk, and (iii) in some cases, terminate their lives via tidal disruption.

The physics of collapsar disks are complex, and our 1D models are subject to inherent limitations.  Most notably, the transport of angular momentum by MHD processes has been reduced to the standard $\alpha$-prescription, non-axisymmetric instabilities may further drive angular momentum transport, and our treatment of neutrino losses is approximate.  We have also neglected magnetic pressure support in the accretion disk, which if present can stabilize massive disks against fragmentation.  To the extent possible, we have aimed to parametrize our ignorance by e.g. exploring a range of $\alpha$ values, but ultimately 1D models such as these must be tested and calibrated against more realistic magnetohydrodynamic simulations, with realistic neutrino microphysics.  Our models also break down at large radii, where either our assumption of inflow equilibrium (i.e. constant $\dot{M}$, requiring  $t_{\rm visc}(R) \lesssim M / \dot{M}$), our assumption of the Kerr metric (requiring that the enclosed disk mass is subdominant), or our assumption of no compound nuclei beyond helium (requiring high temperatures) will eventually fail.  These three assumptions can at least be checked {\it post hoc} in the context of our model, and we did so to estimate its range of validity.

With these caveats in mind, our primary conclusions are as follows:
\begin{enumerate}
    \item Toomre instability ($Q<1$) is a generic feature of collapsar disks with high accretion rates ($\dot{M} \gtrsim 1 M_\odot~{\rm s}^{-1}$).  For some combinations of $\{M, \alpha\}$, Toomre instability can emerge down to accretion rates of $\dot{M} \approx 0.2 M_\odot~{\rm s}^{-1}$.  While these accretion rates are extreme, averaged over the lifetime of a long GRB, we emphasize that they only need to be sustained for short periods of time (fragmentation of a gravitationally unstable zone occurs in $\sim$ few ms, and inward migration of fragments in $\sim 100$ ms).  Toomre instability generally arises in the outer, advective regions of collapsar disks, beyond a radius $R \sim 100 r_{\rm S}$.  However, for large central masses $M\gtrsim 30 M_\odot$, a second zone of Toomre instability can emerge at very small radii as well.
    \item Toomre-unstable zones in collapsar disks feature a wide range of cooling times.  It is uncommon for the dimensionless cooling time $\beta$ to be less than 3, the classical criterion for fragmentation of self-gravitating regions \citep{Gammie2001edit2}.  However, it is relatively common to find $\beta <10$, a threshold that has been found to produce a low but finite probability of fragmentation for stiff equations of state \citep{Hennebelle2021}, and very common to find $\beta < 35$, a threshold for fragmentation with soft equations of state \citep{Chen+23}.
    \item The range of initial fragment masses formed in $Q<1$, $\beta< \beta_{\rm c}$ zones ranges roughly from $10^{-3} \lesssim M_{\rm f,0} / M_\odot \lesssim 10^{-1}$.  Fragments can grow quickly through Bondi-Hoyle accretion, sometimes reaching final masses $m_{\rm f} \gtrsim 0.5 M_\odot$.
    \item Fragments migrate inwards rapidly, initially due to Type I torques.  At small radii, Type I migration will eventually become unimportant, usually because the mass accumulation torque comes to dominate Type I migration%\footnote{In our investigations, we found that a major uncertainty in the accumulation torque is whether or not fragments can accrete at rates above the neutrino Eddington limit.}
    .  At the smallest radii ($r\lesssim ~{\rm few}~ \times R_{\rm g}$), GW losses will dominate the orbit evolution; at this stage, gaps sometimes open in the disk as well.  The final outcome of fragment migration can be either tidal disruption or merger into the BH, depending on the fragment radius.
    \item A major uncertainty in fragment evolution, beyond the scope of this paper to address, concerns the way in which gravitationally unstable fragments will undergo a contraction-collapse sequence to achieve hydrostatic equilibrium.  As a limiting case, we consider a situation in which a ($M_{\rm f} \gtrsim 0.1 M_\odot$) fragment is able to contract to the cold NS mass-radius relationship.  This scenario sets a lower limit on the fragment tidal disruption radius and therefore an upper limit on (a) $\gamma$-ray variability from accretion of a disrupted fragment and (b) GW luminosity from fragment inspiral.  Under this (optimistic) assumption, we find typical contributions to GRB variability, from the fragment tidal disruption, at the $\sim 10\%$ level, and peak GW strains of $h_{\rm GW} \sim 1.5 \cdot 10^{-22} (d / 100~{\rm Mpc})^{-1}$.  
    \item In addition to fragment tidal disruption, a stronger source of electromagnetic GRB variability will be launching of secondary jets from accreting fragments embedded within the collapsar disk.  The peak accretion rates onto these fragments can exceed the total accretion rate onto the BH by a factor $\sim 10$ for a duration of $\sim 1-10$ ms (see fig \ref{app:DurationOfHighAccretion_alpha002M20_a0}).  If the fragments can convert accretion power into jet power via (e.g.) neutrino-antineutrino annihilation, their secondary jets will be able to briefly overpower the primary jet.  Speculatively, jet-jet interactions may be an important source of particle acceleration. \dd{}
    %\item Because fragments can either be born or grow to a mass $0.1 \lesssim M_{\rm f} / M_\odot \lesssim 1$, this mechanism represents a potential channel to produce ultra-low mass NSs, beyond the mass range accessible to both inert iron core collapse in a massive star, and accretion-induced collapse of a white dwarf.  If ultra-low mass NSs can form in this way, they may exit the collapsar and escape to the outside Universe via 3-body scatterings.
\end{enumerate}

The exact outcomes of conclusions (iv-vi) depend critically on fragment evolution after formation.  While the dynamical collapse of inert iron cores is a well-studied problem in high energy astrophysics, less attention has been paid to the ultra-low mass objects here, with initial densities between that of a white dwarf and that of a NS.  The conditions in the collapsar disk add additional uncertainty to the problem, as high Bondi-Hoyle accretion rates may increase the importance of ram pressure confinement compared to standard iron core collapse.  If low-mass fragments can quickly cool and contract/collapse, they will achieve small physical radii that maximize the resulting GW and $\gamma$-ray signals.  Conversely, if cooling is inefficient, then the fragments will obey a hot equation of state, and contraction/collapse will slow.  In this case, the fragments will either tidally disrupt at large radii (producing weaker GW signals and GRB variability) or become internally unstable and explode (particularly at low masses; the hot NS EoS has a substantially higher minimum mass than the cold EoS).  We defer a detailed investigation of fragment evolution for future work.

Finally, we will speculate briefly on the survivability of fragments after the end of the collapsar phase.  As mentioned above, some fragments may in principle be ejected from the collapsar in 3-body scatterings (similar fragment-fragment interactions may also produce collisions or mergers within the disk).  In other situations, the fragment migration time can exceed the duration of a long GRB (and perhaps the associated collapsar disk lifetime).  If the fragment formed is larger than the minimum stable NS mass, it can undergo a slower, GW-driven inspiral, and create a delayed merger with the central BH, producing a delayed GW signal and perhaps a second (this time, short) GRB.  Alternatively, if the fragment is too low mass, it can become dynamically unstable and undergo catastrophic expansion.  The resulting deconfinement of low-$Y_{\rm e}$ fragment matter creates a natural environment for r-process nucleosynthesis to occur.

Although our work represents the first broad parameter survey of fragmentation in collapsar disks, the limitations of 1D models raise a number of questions, many of which can be best answered by future multi-dimensional MHD simulations.  Even if conditions for fragment formation are not achievable in the majority of collapsars, and require e.g. very high-mass, low-metallicity progenitor stars, the observational implications of collapsar fragmentation are potentially quite dramatic, and may be observable even if they only occur infrequently.

\section*{Acknowledgements}

YL and NCS gratefully acknowledge support from the Binational Science Foundation (grant Nos. 2019772 and 2020397) and the Israel Science Foundation (Individual Research Grant Nos. 2565/19 and 2414/23).  DDO was supported by Advanced ERC grant MultiJets and by the Simons Collaboration on Extreme Electrodynamics of Compact Sources (SCEECS).  We thank Shahram Abbassi, Shmuel Gilbaum, Sivan Ginzburg, Ore Gottlieb, Brian Metzger, Tsvi Piran, Narges Shahamat, and Lorenz Zwick for helpful discussions, and comments on earlier versions of the manuscript.

\section*{Data Availability}
The data and code used to make the figures in this paper will be shared upon reasonable request.  Inquiries should be directed to Yonatan Lerner (yonatan.lerner@mail.huji.ac.il).

\bibliographystyle{mnras}
\bibliography{IMS_refs} 

\appendix

\section{Clarification to the CB07 model} 
\label{app: Clarification to the CB07 model}
\begin{comment}
\begin{figure*}
\caption{Table of constants.} \label{tab: Table of constants}
\includegraphics[width=0.65\textwidth]{constants_table (1)-cropped.pdf}
\label{fig: constants_table}
\end{figure*}
\end{comment}
In this appendix, we describe typo corrections, changes and clarifications to the microphysics introduced in CB07. In order to introduce these thermodynamic quantities we will use the Fermi-Dirac (FD) distribution:
\begin{equation}\label{eq: Fermi-Dirac distribution}
    \begin{aligned}
   f(\epsilon,\mu)=\frac{1}{\exp((\epsilon-\mu)/k_{\rm b} T)+1}.
    \end{aligned}
\end{equation}
Where $\epsilon$ is the particle energy and $\mu$ is the chemical potential. For simplicity, we perform a change in variables to $\theta=k_{\rm b} T/m_{\rm e} c^2$, $\eta=\mu/k_{\rm b} T$ and $E$=$\epsilon/m_{\rm e} c^2$ that changes the FD distribution to:
\begin{equation}\label{eq: changed Fermi-Dirac distribution}
    \begin{aligned}
   f(E,\eta)=\frac{1}{\exp(E/\theta-\eta)+1}.
    \end{aligned}
\end{equation}
$\theta$ is a dimensionless temperature parameter that relates the average kinetic energy of the material $k_{\rm b} T$ to the electron rest mass energy $m_{\rm e} c^2$. Likewise, $\eta$ is a dimensionless degeneracy parameter.

From the FD statistics we derive the $e^{\pm}$ energy densities, which are denoted as $U_{e^{\pm}}$:
\begin{equation}
    \begin{split}
      U_{e^{\pm}}&=\frac{(m_{\rm e}c)^{3}}{\pi^{2}\hslash^{3}}m_{\rm e}c^{2}\\
      &\times\int_{0}^{\infty}f\left(\sqrt{p^{2}+1},\mp\eta_{\rm e}\right)\left(\sqrt{p^{2}+1}-1\right)p^{2}{\rm d}p.\label{eq: electron positron energy density}
    \end{split}
\end{equation}
In the CB07 paper it seems that there was a typo writing this equation. Specifically, the second parenthetical term only has $\sqrt{p^{2}+1}$ instead of $\sqrt{p^{2}+1}-1$. It is worth noting that the correct equation reproduces the CB07 plots.
CB07 have introduced also charge neutrality: $n_{\rm e^{-}}-n_{\rm e^{+}}=n_{\rm p}$ (where $n_{\rm e^{-}}, n_{\rm e^{+}}$ and $n_{\rm p}$ are electron, positron and proton number densities respectively). With the definition of $Y_{\rm e}\equiv(n_{\rm n}/n_{\rm p}+1)^{-1}$, where $n_{\rm n}$ is neutron number density, and the assumption that the mass is dominated by only the baryons ($m_{\rm e}\ll m_{\rm p}\simeq m_{\rm n}$). we can write charge neutrality as:
\begin{equation}\label{eq: charge neutrality}
    \begin{aligned}
    n_{\rm e^{-}}-n_{\rm e^{+}}=Y_{\rm e} \frac{\rho}{m_{\rm p}}.
    \end{aligned}
\end{equation}
If we want to keep the equation for charge neutrality as given in CB07, some of the other definitions written in their paper need to change in order to not have a contradiction. The right side of Eq. \ref{eq: charge neutrality} suggests that $n_{\rm p}$ and $n_{\rm n}$ are the number densities of protons and neutrons including the ones who are occupied by $\alpha$-particles. This lets us define:
\begin{subequations}
    \begin{align}
       n_{\rm p}&= Y_{\rm e}\frac{\rho}{m_{\rm p}}\\
      n_{\rm n}&= (1-Y_{\rm e})\frac{\rho}{m_{\rm p}}\\
      n_{\rm \alpha}&= \rho\frac{1-X_{\rm f}}{4m_{\rm p}}
    \end{align}
\end{subequations}
Here $1-X_{\rm f}$ is the mass fraction of $\alpha$-particles. In order for these definitions to be correct, while being consistent with the the equations involving electron positron capture (meaning that electron positron capture is suppressed for protons and neutron that are occupied by $\alpha$-particles). The equations governing neutrino emission from the neutrino transparent region in CB07 (Eqs. 22-25 in CB07) need to undergo the following change in order to have no contradiction with Eq. \ref{eq: charge neutrality}:
\begin{subequations}
    \begin{align}
       Y_{\rm e}\frac{\rho}{m_{\rm p}}\text{ }\rightarrow\text{ }n_{\rm p}-2n_{\rm \alpha}\\
      (1-Y_{\rm e})\frac{\rho}{m_{\rm p}}\text{ }\rightarrow\text{ }n_{\rm n}-2n_{\rm \alpha}.
    \end{align}
\end{subequations}
Furthermore, CB07 didn't specify their approach for calculating the neutrino (and antineutrino) number density in the transparent region. They only gave the equations for the number densities in the neutrino opaque region assuming thermal equilibrium ($n_{\nu/\thickbar{\nu}}^{\rm opaque}$, in Eq. 31-32 in CB07). In order to be more self consistent in the transparent region we defined the following transition:
\begin{subequations}
\begin{align}
    n_{\nu/\thickbar{\nu}}&=    \begin{cases}\dot{S}_{\nu/\thickbar{\nu}}^{\rm{trans}}/c & x_{\nu/\thickbar{\nu}}<1 / 2  \vspace{1mm} \\ n_{\nu/\thickbar{\nu}}^{\rm opaque}  & x_{\nu/\thickbar{\nu}} \geq 1 / 2\end{cases}\label{neutrino number density transtion}.
   \end{align}
\end{subequations}
Where $\dot{S}_{\nu/\thickbar{\nu}}$ are the number fluxes of neutrinos or anti-neutrinos (shown at Eq. \ref{eq: lepton number advection equation} and in Eqs. 24-25 in CB07, though we rewrite their $\dot{N}$ as $\dot{S}$). $x_{\nu/\thickbar{\nu}}$ is a transition criteria defined in Eq. 33 in CB07.

It is worth noting that implementing all of these changes (excluding the changed lepton equation, specified in Eq. \ref{eq: lepton number advection equation}) reproduces the plots in the CB07 paper, meaning that they were probably already implemented in the CB07 model, just not explicitly described.

\section{Relativistic Disk Theory} 
\label{app:GR}
In this appendix, we describe the relevant aspects of the Kerr metric and how these are incorporated into our disk model.  This includes both standard formulae as well as those with more contested definitions.
\subsection*{Innermost stable circular orbit (ISCO)
}
We calculate the innermost stable circular orbit in the standard way:
\begin{subequations}\label{eq: Marginally stable radius}
    \begin{align}
        Z_1&=1+(1-a^2)^\frac{1}{3} \left((1-a)^\frac{1}{3}+(1-a)^\frac{1}{3}\right)\\
      Z_2&=\sqrt{3a^2+\text{$Z_1$}^2}\\
       r_{\rm ISCO}&=\frac{G M}{c^2} \left(3+\text{$Z_2$}\mp \sqrt{(3-\text{$Z_1$}) (3+\text{$Z_1$}+2 \text{$Z_2$})}\right).
    \end{align}
\end{subequations}
The $\mp$ in the equation is negative for $a>0$ (prograde) and positive for $a<0$ (retrograde).
\subsection*{GR functions
}
The following supplemental functions define the general kinematics of geodesic motion in the Kerr metric, including the azimuthal frequency $\Omega$, the gravitational radius $r_{\rm g}$, the Schwarzschild radius $r_{\rm s}$.
\begin{subequations}\label{eq: terms for s(r)}
    \begin{align}
      \Omega&=\left(\frac{r^3}{GM}^{1/2}+a\frac{GM}{c^3}\right)^{-1}\\
      r_{\rm g}&=G M c^{-2}\\r_{\rm s}&=2 r_{\rm g}\\r_{\rm *}&=\frac{r}{r_{\rm g}}\\x&=\sqrt{r_{\rm *}}\\
      x_0&=\sqrt{\frac{c^2}{G M}r_{\rm ISCO}}\\
      x_1&=2 \cos \left(\frac{\cos ^{-1}(a)}{3} -\frac{\pi }{3}\right)\\
      x_2&=2 \cos \left(\frac{\cos ^{-1}(a)}{3} +\frac{\pi }{3}\right)\\
      x_3&= - 2 \cos \left(\frac{\cos ^{-1}(a)}{3} \right)\\
      \mathcal{E}&=\frac{a r_{\rm g}^{3/2}-2 \sqrt{r} r_{\rm g}+r^{3/2}}{r^{3/4} \sqrt{2 a r_{\rm g}^{3/2}-3 \sqrt{r} r_{\rm g}+r^{3/2}}}\\
      \mathcal{L}&=\frac{\sqrt{r_{\rm g}} \left(a^2 r_{\rm g}^2-2 a \sqrt{r} r_{\rm g}^{3/2}+r^2\right)c}{r^{3/4} \sqrt{2 a r_{\rm g}^{3/2}-3 \sqrt{r} r_{\rm g}+r^{3/2}}}\label{specific angular momentum GR}.
          \end{align}
\end{subequations}
$\mathcal{L}$ is the specific angular momentum parallel to the symmetry axis and $\mathcal{E}$ is the dimensionless total specific energy.
\begin{equation}\label{eq: curly GR corrections}
    \begin{aligned}
      \mathcal{S}&=
      \\&\frac{1+a x^{-3}}{x \sqrt{1-3x^{-2}+2a x^{-3}}} \left(x-x_0-\frac{3}{2} a \ln \left(\frac{x}{x_0}\right)\right.\\
      &-\frac{3 (x_1-a)^2 }{x_1(x_1-x_2)(x_1-x_3)}\ln \left(\frac{x-x_1}{x_0-x_1}\right)\\
      &-\frac{3 (x_2-a)^2 }{x_2 (x_2-x_1) (x_2-x_3)}\ln \left(\frac{x-x_2}{x_0-x_2}\right)\\
      &-\frac{3 (x_3-a)^2 }{x_3(x_3-x_1) (x_3-x_2)}\ln \left(\frac{x-x_3}{x_0-x_3}\right) \biggr)
    \end{aligned}
\end{equation}
$\mathcal{S}$ is the dimensionless multiplicative factor arising from a zero-torque inner boundary condition. The calculation was taken from \cite{PageThorne74}.
\begin{subequations}
    \begin{align}
      \mathcal{C}&=1-\frac{3}{r_{\rm *}}+\frac{2a}{{r_{\rm *}}^{3/2}},\\
      J&=\frac{1-4a {r_{\rm *}}^{-3/2}+3a^2{r_{\rm *}}^{-2}}
      {\mathcal{C}}.
      \end{align}
\end{subequations}

\section{Tidal disruption check}\label{sec: Tidal disruption check}
%\begin{comment}
The equation for the tidal radius (Eq. \ref{eq: tidal radius implicit eq}) is an implicit function: $r_{\rm t}$ is a function of the fragment mass, $m_{\rm f}(r)$, and the fragment radius $R_{\rm f}(r)$, which all evolve over the fragment's migration track. The tidal radius equation needs to be solved numerically via root-finding for each fragment mass and radius, which in turn gives a corresponding tidal radius. The mass and fragment radius evolve as the fragment migrates inwards, so we solve the equation for each distance $r$ and check whether the tidal radius $r_{\rm t}$ is bigger or smaller than $r$. The ratio $\Phi=r_{\rm t}/r$ between the distance and the equivalent tidal radius lets us check whether a fragment migrates safely through a radius $r$, or whether it is tidally disrupted. 

\begin{comment}
At each radius, $\Phi$ is evaluated and its value means: (1) equal to 1, a TD is will occur at $r$ unless the duration in which the fragment passes through the tidal disruption region (where $\Phi\geq 1$ and $r<r_{\rm min}$), $\Delta t_{\rm TD}$, is smaller than its dynamical time at the start of the TD region, $t_{\rm dyn}=\sqrt{R_{\rm f}^3/G m_{\rm f}}$; (2) smaller than 1, for this specific mass and radius the possible TD is farther inwards in the migration track; (3) bigger than 1, a TD has already occurred farther out unless $\Delta t_{\rm TD}<t_{\rm dyn}$. A TD can occur either before reaching the $r_{\rm min}$ the radius where the fragment reaches the minimum stable cold NS mass ($m_{\rm f}(r_{\rm min})=0.09 M_{\odot}$, we can't calculate a tidal radius before that point as we don't have an estimate on the mass to radius relation) or later in the migration closer to the ISCO with a significant mass causing a more significant observable implications.
\end{comment}

At each radius, the parameter $\Phi$ is evaluated. Its value indicates the likelihood of a tidal disruption (TD): 
\begin{comment}
\begin{enumerate}
    \item If $\Phi = 1$, a TD will occur at radius $r$, unless the time the fragment spends in the tidal disruption region ($\Phi \geq 1$) % with $r < r_{\rm min}$, the radius at which the fragment reaches the minimum stable cold neutron star mass. where $m_{\rm f}(r_{\rm min}) = 0.09\,M_{\odot}$), denoted $\Delta t_{\rm TD}$, 
    is shorter than its own dynamical time, $t_{\rm dyn}=\sqrt{R_{\rm f}^3/G m_{\rm f}}$.
    
    \item If $\Phi < 1$, for the given mass and radius, a potential TD would occur further inward along the migration track.
    
    \item If $\Phi > 1$, a TD has already occurred at a larger radius, unless $\Delta t_{\rm TD} < t_{\rm dyn}$.
\end{enumerate}

A TD can occur at different regions along the migration track:

\begin{itemize}
    \item Before reaching $r_{\rm min}$. We cannot compute a tidal radius prior to this point, as we lack an estimate of the mass-radius relation. We exclude these from our plots.
    
    \item Just after passing $r_{\rm min}$, where the fragment may have a small mass, leading to less significant observable implications. We also exclude these from our plots.
    
    \item Closer to the innermost stable circular orbit (ISCO), where the fragment may retain significant mass, resulting in more substantial observational signatures.
\end{itemize}
\end{comment}
$\Phi = 1$ indicates that the fragment is at the classic tidal disruption radius, and it will disrupt there, unless one of a few possible exceptions arise.  The first possibility concerns rapid migration of the fragment: if the time spent in a given region $\Delta t_{\rm TD}<t_{\rm dyn}=\sqrt{R_{\rm f}^3/G m_{\rm f}}$, then the fragment will not disrupt until it moves inwards (to a larger value of $\Phi$).  

The other possible exceptions relate to $r_{\rm min}$, the outermost radius at which the fragment has reached the minimum cold neutron star mass.  The fragment's radius beyond this point is beyond the scope of this work, so we do not attempt to calculate it.  Even for $r<r_{\rm min}$, it is far from clear that the fragment will follow a cold neutron star mass-radius relationship (this is likely only possible with efficient neutrino cooling), but for concreteness we use this as a rough guide to fragment survivability.  If $r_{\rm min} > r_{\rm t}$, then the TD check is straightforward.  Conversely, if $r_{\rm min} < r_{\rm t}$, we only permit disruption after $\Delta t_{\rm TD} < t_{\rm dyn}$.  Because of the extreme sensitivity of the cold neutron star radius to $m_{\rm f}$ when $m_{\rm f}$ is near the minimum stable mass, this condition often excludes immediate disruption at $r= r_{\rm min}$.

\begin{figure}
\includegraphics[width=0.48\textwidth]{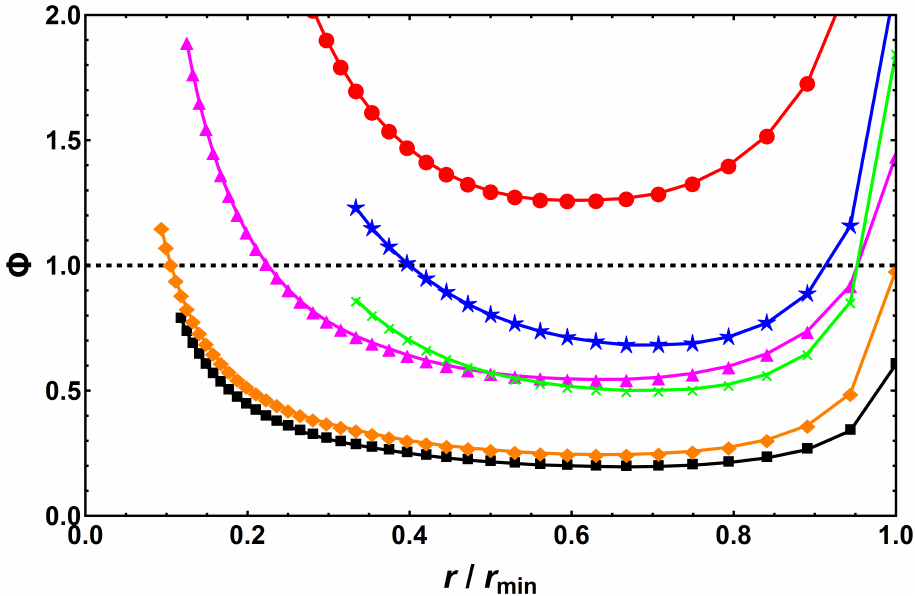}
\caption{The ratio $\Phi$ between the tidal radius $r_{\rm t}(r)$ and the distance $r$, shown as a function of $r$ (normalized by $r_{\rm min}$), for six 
migration tracks of fragments in disks with different mass $M$, accretion rate $\dot{M}$, and viscosity parameter $\alpha$. The red line ($M=8{M}_{\odot}$, $\dot{M}=1{M}_{\odot} {\rm s^{-1}}$ and $\alpha=0.02$) corresponds to a fragment that has a TD before $r_{\rm min}$. The black solid line ($M=32{M}_{\odot}$, $\dot{M}=0.4{M}_{\odot} {\rm s^{-1}}$ and $\alpha=0.01$) shows no TD. The orange line ($M=20{M}_{\odot}$, $\dot{M}=0.4{M}_{\odot} {\rm s^{-1}}$ and $\alpha=0.01$) shows a fragment that has a later TD close to the ISCO with high $m_f$. The blue line ($M=20{M}_{\odot}$, $\dot{M}=3{M}_{\odot} {\rm s^{-1}}$ and $\alpha=0.02$) shows a fragment that has a TD before $r_{\rm min}$ as $t_{TD} > t_{dyn}$. The magenta line ($M=13{M}_{\odot}$, $\dot{M}=0.4{M}_{\odot} {\rm s^{-1}}$ and $\alpha=0.01$) shows a fragment that has a later TD close to the ISCO with high $m_f$ after not being tidally disrupted near $r_{\rm min}$ as $t_{TD} < t_{dyn}$. The green line ($M=32{M}_{\odot}$, $\dot{M}=3{M}_{\odot} {\rm s^{-1}}$ and $\alpha=0.02$) shows no TD of the fragment after not being tidally disrupted near $r_{\rm min}$ as $t_{TD} < t_{dyn}$.}
\label{app: Tidal_disruption_examples}
\end{figure}

In our calculations, we see the following six options for the final outcome (examples of each are shown in Fig. \ref{app: Tidal_disruption_examples}):
\begin{enumerate}
    \item $\Phi > 1$ everywhere between $\{r_{\rm ISCO}, r_{\rm min}\}$: a TD occurs before the fragment ever reaches $r_{\rm min}$ (red line in Fig.~\ref{app: Tidal_disruption_examples}).
    \item $\Phi < 1$ everywhere between $\{r_{\rm ISCO}, r_{\rm min}\}$: no TD occurs, and the fragment is eventually swallowed whole (solid black line in Fig.~\ref{app: Tidal_disruption_examples}).
    \item $\Phi = 1$ somewhere between $\{r_{\rm ISCO}, r_{\rm min}\}$ and $\Phi(r_{\rm min}) < 1$: a TD occurs close to the ISCO with a high fragment mass $m_{\rm f}$ (orange line in Fig.~\ref{app: Tidal_disruption_examples}).
    \item $\Phi = 1$ somewhere between $\{r_{\rm ISCO}, r_{\rm min}\}$ and $\Phi(r_{\rm min}) > 1$ with $\Delta t_{\rm TD} > t_{\rm dyn}$: a TD occurs before $r$ reaches $r_{\rm min}$ (blue line in Fig.~\ref{app: Tidal_disruption_examples}).
    \item $\Phi = 1$ somewhere between $\{r_{\rm ISCO}, r_{\rm min}\}$ and $\Phi(r_{\rm min}) > 1$ with $\Delta t_{\rm TD} < t_{\rm dyn}$, and $\Phi$ crosses 1 again: a TD occurs close to the ISCO with high $m_{\rm f}$ (magenta line in Fig.~\ref{app: Tidal_disruption_examples}).
    \item $\Phi = 1$ somewhere between $\{r_{\rm ISCO}, r_{\rm min}\}$ and $\Phi(r_{\rm min}) > 1$ with $\Delta t_{\rm TD} < t_{\rm dyn}$, and $\Phi$ does not cross 1 again: no TD occurs (green line in Fig.~\ref{app: Tidal_disruption_examples}).
\end{enumerate}
\begin{comment}
$\Phi >1$ everywhere between $\{r_{ISCO}, r_{\rm min}\}$, Means TD before $r_{\rm min}$ (red line in fig \ref{app: Tidal_disruption_examples}). $\Phi <1$ everywhere between $\{r_{ISCO}, r_{\rm min}\}$, Means no TD (solid black line in fig \ref{app: Tidal_disruption_examples}). $\Phi=1$ somewhere between $\{r_{ISCO}, r_{\rm min}\}$, leading to a few possibilities:
\begin{itemize}
    \item $\Phi(r_{\rm min})<1$, Means a TD close to the ISCO with high $m_f$ (orange line in fig \ref{app: Tidal_disruption_examples}).
    \item $\Phi(r_{\rm min})>1$, leading to a few possibilities:
    \begin{itemize}
        \item $t_{TD} > t_{dyn}$. Means TD before $r_{\rm min}$ (blue line in fig \ref{app: Tidal_disruption_examples}).
        \item $t_{TD} < t_{dyn}$. Means TD before $r_{\rm min}$, leading to further possibilities:
        \begin{itemize}
            \item $\Phi$ passes through 1 again. Means a TD close to the ISCO with high $m_f$ (magenta line in fig \ref{app: Tidal_disruption_examples}).
            \item $\Phi$ doesn’t pass through 1 again. Means no TD (green line in fig \ref{app: Tidal_disruption_examples}).
        \end{itemize}
    \end{itemize}
\end{itemize}
\end{comment}

%\clearpage

\section{Numerical method}
\label{app:numerics}

As the energy and lepton number conservation equations (Eqs. \ref{eq: collapsar energy relation} and \ref{eq: lepton number advection equation}) are differential equations they need to be discretized. We start by building a logarithmic grid of $r_{\rm i}$ $(\rm i=0,...,N)$, ranging inwards ($r_{\rm i}<r_{\rm i-1}$), from $r_{\rm 0}=r_{\rm out}$ to $r_{\rm N}=r_{\rm ISCO}$. Then the discretized equations, Where we replaced all the derivatives in Eqs. \ref{eq: collapsar energy relation} and \ref{eq: lepton number advection equation} with 1st order finite differences, become:
\begin{equation}\label{eq: discrete equations}
\begin{split}
F^{+}-F^{-}=
&u^{r}\left[\frac{U_{\rm i} H_{\rm i}-U_{\rm i-1} H_{\rm i-1}}{r_{\rm i}-r_{\rm i-1}}-\right.\vspace{2mm}\\
&\qquad -\frac{(U+P)}{\rho} \frac{\left(\rho_{\rm i} H_{\rm i}-\rho_{\rm i-1} H_{\rm i-1}\right)}{r_{\rm i}-r_{\rm i-1}}\biggr] 
\end{split}
\end{equation}
\vspace{1mm}
\begin{equation}\label{eq: discrete lepton number advection equation}
\begin{split}
\dot{S}_{\rm \nu}-\dot{S}_{\rm \thickbar{\nu}}&=-\left(\frac{u^{\rm r}}{ r}\right)\frac{rH}{r_{\rm i}-r_{\rm i-1}}\biggl[\\
&\qquad\frac{r_{\rm i}-r_{\rm i-1}}{r}\left(Y_{e}\frac{\rho}{m_{p}}+(n_{\nu}-n_{\thickbar{\nu}})\right)\\
&\qquad+\frac{H_{\rm i}-H_{\rm i-1}}{H}\left(Y_{e}\frac{\rho}{m_{p}}+(n_{\nu}-n_{\thickbar{\nu}})\right)\\
&\qquad+(Y_{e}(r_{\rm i})-Y_{e}(r_{\rm i-1}))\frac{\rho}{m_{p}}\\
&\qquad+(\rho_{\rm i}-\rho_{\rm i-1})\frac{Y_{e}}{m_{p}}\\
&\qquad+(n_{\nu}-n_{\thickbar{\nu}})_{\rm i}-(n_{\nu}-n_{\thickbar{\nu}})_{\rm i-1} \biggr]
\end{split}
\end{equation}
Each quantity in the above equations that does not have an index explicitly written is defined as the average of the quantity between $r_{\rm i}$ and $r_{\rm i-1}$. 
% Old sentence: Where each quantity in the above equations that doesn't have an index related to it is defined as the average of the quantity between $r_{\rm i}$ and $r_{\rm i-1}$.

In CB07, Eq. \ref{eq: discrete lepton number advection equation} was not discretized; instead it was suggested that knowing $\eta_{\rm e}(r_{\rm i-1}),$ $ \theta(r_{\rm i-1})$ and $Y_{\rm e}(r_{\rm i-1})$ at the previous step $r_{\rm i-1}$, gives $Y_{\rm e}(r_{\rm i})$, at the next step, directly using their version of the equation (see Eq. \ref{eq: lepton number advection equation CB}). Even for the CB07 version of lepton number conservation, this approach is only correct if one neglects the difference between the neutrino and antineutrino number densities $n_{\nu}-n_{\thickbar{\nu}}$. This assumption may be appropriate in some regions but does not hold in the neutrino opaque zone where the non-zero chemical potentials %$\mu_{\rm \nu}=-\mu_{\rm \thickbar{\nu}}$ 
gives rise to bigger differences between the values of the neutrino and antineutrino number densities. This assumption will fail more broadly for the revised lepton number conservation equation, which incorporates a larger number of radial derivatives that must be discretized (see Eq. \ref{eq: lepton number advection equation} and the discussion below Eq. \ref{eq: lepton number advection equation CB}).

The change to the lepton number equation therefore requires a 3D root finding approach, wherein we need to find the correct $\eta_{\rm e}(r_{\rm i}),$ $\theta(r_{\rm i})$ and $Y_{\rm e}(r_{\rm i})$ that satisfy our three master equations. This is in contrast to the more approximate method by CB07 that searched in the 2D $\eta_{\rm e}$ and $\theta$ plane. The modified method we employ in this paper is specified in the flow chart in Fig. \ref{app: Flow chart algo}.

\tikzstyle{startstop} = [rectangle, rounded corners, minimum width=3cm, minimum height=1cm,text centered, text width=7cm, draw=black, fill=blue1]
\tikzstyle{log} = [rectangle, rounded corners, minimum width=3cm, minimum height=1cm, text centered, text width=7cm, draw=black, fill=orange!30]
\tikzstyle{outer} = [rectangle, rounded corners, minimum width=3cm, minimum height=1cm, text centered, text width=7cm, draw=black, fill=orange!30]
\tikzstyle{root} = [rectangle, rounded corners, minimum width=3cm, minimum height=1cm, text centered, text width=7cm, draw=black, fill=orange!30]
\tikzstyle{test} = [trapezium, trapezium left angle=80, trapezium right angle=100, minimum width=2.5cm, minimum height=1cm, text centered, text width=4cm, draw=black, fill=blue!30]
\tikzstyle{true} = [rectangle, rounded corners, minimum width=1.8cm, minimum height=1cm, text centered, text width=1.6cm, draw=black, fill=green!30]
\tikzstyle{2D} = [rectangle, rounded corners, minimum width=3cm, minimum height=1cm, text centered, text width=5cm, draw=black, fill=red!30]
\tikzstyle{end} = [rectangle, rounded corners, minimum width=3cm, minimum height=1cm,text centered, text width=6cm, draw=black, fill=blue2!70]
\tikzstyle{arrow} = [very thick,->,>=stealth]
\begin{figure}
    \centering
\begin{tikzpicture}[node distance=2.1cm]
<TikZ code>
\node (start) [startstop] {Start by choosing:
$\alpha$, $a$, $M$, $\dot{M}$, $r_{\rm out}$, an acceptable error ($N_{\rm er}$) and a numerical push ($N_{\rm p}$).};
\node (log) [log, below of=start, yshift=0.7cm] {Build a logarithmic grid of $N+1$ radial cells, where $r_{\rm i}$ ranges from $r_{\rm ISCO}=r_{\rm N}$ to $r_{\rm out}=r_{\rm 0}$ .};
\node (outer) [outer, below of=log, yshift=0.8cm] {Solve for $\eta_{\rm e}$ and $\theta$ at $r_{\rm out}$ using the pressure equation, $Y_{\rm e}=0.5$ and $U/\rho=G M/r_{\rm out}$.};
\node (root) [root, below of=outer, yshift=0.3cm] {Try solving for $\eta_{\rm e}(r_{\rm i}),$ $\theta(r_{\rm i})$, and $Y_{e}(r_{{\rm i}})$ at $r_{{\rm i}}$ with the $\eta_{\rm e}(r_{\rm i-1}),$ $\theta(r_{\rm i-1})$, and $Y_{\rm e}(r_{{\rm i-1}})$ from the previous $r_{{\rm i-1}}>r_{{\rm i}}$ using 3D root-finding, when solving for the pressure, energy and lepton conservation equations. };
\node (test) [test, below of=root, yshift=-0.2cm,,xshift=-1.2cm] {Check whether root-finding worked by testing that for the lepton equation
$\Big|\frac{RHS({\rm i})-LHS ({\rm i})}{LHS ({\rm i})}\Big|<N_{{\rm er}}$};
\node (true) [true, right of=test,xshift= 1.9cm] {Continue to the next step where ${\rm i=i+1}$.};
\node (2D) [2D, below of=test, yshift=-0.6cm, xshift=-0.2cm] {Set $Y_{\rm e}(r_{\rm i})=Y_{\rm e}(r_{\rm i-1})+N_{\rm p}$ and solve for $\eta_{\rm e}(r_{\rm i}),$ $\theta(r_{\rm i})$ using a 2D root-finding solving for the pressure and energy equations. Then use the $Y_{\rm e}(r_{\rm i})$ and the $\eta_{\rm e}(r_{\rm i}),$ $\theta(r_{\rm i})$ found in the 2D root-finding to solve for the next step. Where $\rm i=i+1$.};
\node (end) [end, below of=2D, yshift=-0.2cm, xshift=1.5cm] {Collect all the calculated $\eta_{\rm e}(r_{\rm i}),$ $\theta(r_{\rm i})$, and $Y_{e}(r_{{\rm i}})$ for all $r_{\rm i}$'s. Now we can calculate all of the disks properties, like temperature, density, scale height, pressure and more.};
\draw [arrow] (start) -- (log);
\draw [arrow] (log) -- (outer);
\draw [arrow] (outer) -- (root);
\draw [arrow] (root) -- (test);
\draw [arrow] (root) -- node[anchor=east] {While  ${\rm i}\leq {\rm N}$ \hspace{7mm}} (test);
\draw [arrow] (test) -- node[anchor=north] {True} (true);
\draw [arrow] (true) -- (root);
\draw [arrow] (test) -- (2D);
%\draw[thick,-latex] (true.north) -- ++(0,0)  |- (root) node[near start,anchor=below]{Non};
\draw[very thick,-latex] (root.east) -- ++(0.3,0)  |- (end)  node[near start,anchor=east] {When $\rm i>N$};
\draw[very thick,-latex] (2D.west) -- ++(-0.15,0)  |- (root)  node[near start,anchor=east] {};
\draw [arrow] (test) -- node[anchor=east] {False} (2D);
\end{tikzpicture}
\caption{A flow chart of the numerical algorithm used to solve our model. RHS(i) and LHS(i) are the right and left-hand sides (respectively) of the lepton equation with values from the 3D root finding at step $r_{\rm i}$. $N_{\rm p}$ is a small numerical push set to $-10^{-15}$, which was needed because root finding had trouble at the outer regions where changes in $Y_{\rm e}$ are small. $N_{\rm er}=10^{-6}$.}\label{app: Flow chart algo}
\end{figure}
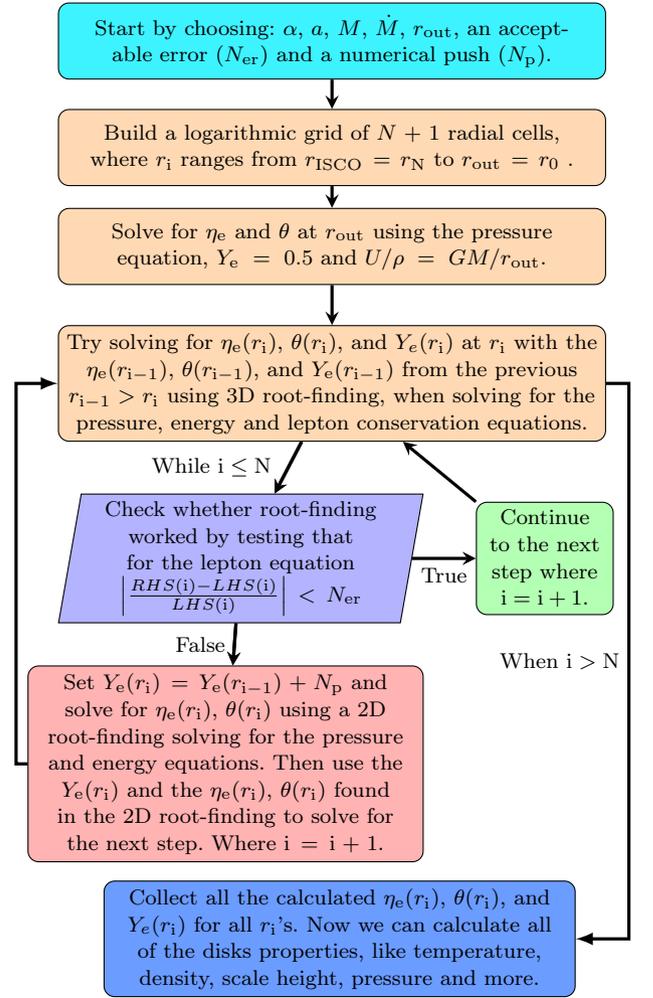
%\end{comment}

\section{Migration calculations}
\label{app: migration calculations}
\subsection*{Type I torque}
We use the recent 3D calibration of the type I torque from \citet{JimenezMasset17}, where the dimensionless prefactor
\begin{subequations}
    \begin{flalign}
      C_{\rm I} &= C_{\rm L} + -0.46 + 0.96\nabla_{\rm \Sigma} - 1.8\nabla_{\rm T}/{\gamma} && \\
      C_{\rm L} &= \left(2.34 - 0.1\nabla_{\rm \Sigma} + 1.5\nabla_{\rm T}\right)f_{\rm \gamma} && \\
      f_{\rm \gamma} &= 
      \begin{cases}
        1 & \tau_{{\rm \nu}} + \tau_{{\rm \widetilde{\nu}}} < 1 \\
        \frac{\sqrt{x/2} + 1/\gamma}{\sqrt{x/2} + 1} & \tau_{{\rm \nu}} + \tau_{{\rm \widetilde{\nu}}} \geq 1
      \end{cases} && \\
      x &= \frac{\chi_{{\rm \nu}}}{H^{2}\Omega} .
 \end{flalign}
\end{subequations}
Here $x$ is a dimensionless version of the thermal diffusivity and $\nabla_{\zeta}=-{\rm d}\ln(\zeta)/{{\rm d}\ln(r)}$.
\begin{equation}
          \chi_{{\rm \nu}} = \frac{7\sigma_{{\rm s}}T^{3}}{6\kappa_{{\rm \nu}} c_{{\rm \nu}}\rho^2}  ,
\end{equation}
which depends on the ratio of specific heats 
\begin{equation}
          c_{{\rm \nu}} =\frac{U}{T\rho}
\end{equation}
as well as the neutrino opacity
\begin{equation}
         \kappa_{{\rm \nu}} \approx\frac{2(\tau_{{\rm \nu}}+\tau_{{\rm \widetilde{\nu}}})}{\Sigma}\label{eq: neutrino opacity} .
\end{equation}

%      D_{{\rm \nu}}&=\frac{7\sigma_{{\rm s}}T^{3}}{6\kappa_{{\rm \nu}}\rho} && \\
%&&
%    \end{flalign}
%\end{subequations}
Here we changed the radiation diffusion calculations, which were originally presented in \citet{JimenezMasset17} for the case of photon diffusion.  These have been altered to focus on neutrino diffusion using an approximation for the neutrino opacity $\kappa_{{\rm \nu}}$ (as seen in Eq. \ref{eq: neutrino opacity}), and accounting for the slight differences in radiation fields for fermionic versus bosonic radiation. We take the definitions of $\tau_{{\rm \nu}}$ and $\tau_{{\rm \widetilde{\nu}}}$ from CB07. A negative value of $C_{\rm I}$ means that the type I torque drives an outspiral. As $C_{\rm I}$ was calibrated from 3D hydrodynamical simulations with a different equation of state than ours, we treat Type I migration approximately and set $C_{\rm I}=\max(C_{\rm I},0.5)$ to prevent a theoretically possible, but more speculative positive torque. If a gap opens (see below) and migration transitions to Type II , we follow \citealt {Kanagawa+18} and set $C_{\rm I}=1$ (see Eqs. \ref{eq: Sigma gap ratio} and \ref{eq: Torque gap ratio} and the discussion before them, for all the relevant changes for when a gap forms).
\begin{comment}
   \subsection*{Accumulated gas torque derivation}
\begin{subequations}
    \begin{align}
        \dot{l}_{{\rm f}}&=\frac{d(L_{{\rm f}}/m_{{\rm f}})}{dt}=\frac{1}{m_{{\rm f}}}\frac{dL_{{\rm f}}}{dt}-\frac{L_{{\rm f}}}{m_{{\rm f}}^{2}}\dot{m}_{{\rm f}}\\
        \dot{l}_{{\rm f}}&=\frac{\dot{m}_{{\rm f}}}{m_{{\rm f}}}(l_{{\rm gas}}-l_{{\rm f}})=\Lambda(a)m_{{\rm f}}(l_{{\rm gas}}-l_{{\rm f}})\\
        l_{{\rm f}}&=av_{{\rm k}}\\
        l_{{\rm gas}}&=av_{{\rm gas}}=av_{{\rm k}}\left(1-\nabla_{P}\left(\frac{H}{a}\right)^{2}\right)\\
        \begin{split}
        \left(\frac{da}{dt}\right)_{{\rm Acc}}&=2a\frac{\dot{l}_{{\rm f}}}{l_{{\rm f}}}=2a\Lambda(a)m_{{\rm f}}(\frac{l_{{\rm gas}}}{l_{{\rm f}}}-1)\\&=-2a\Lambda(a)m_{{\rm f}}\nabla_{P}\left(\frac{H}{a}\right)^{2}
        \end{split}
    \end{align}
\end{subequations}
\end{comment}

\subsection*{Gap formation
}
\label{app: gap formation}
The way in which a bound fragment migrates depends on whether or not it will form a gap in the accretion disk. Using the criteria developed in \cite{Kanagawa2017}, we say that a gap forms if:
\begin{equation}
\kappa_{{\rm gap}}=\left(\frac{m_{{\rm f}}}{M}\right)^{2}\left(\frac{2H}{r}\right)^{-5}\alpha^{-1}\geq20
\end{equation}
In order to know whether to use type I or type II torque in our calculation we check at each radius whether or not a gap forms.

\clearpage

\section{Additional Results}
\label{app: Results}

\subsection{GRB variability from tidal disruption of fragment}
\label{app: GRB from TD}
\begin{figure}
\includegraphics[width=0.48\textwidth]{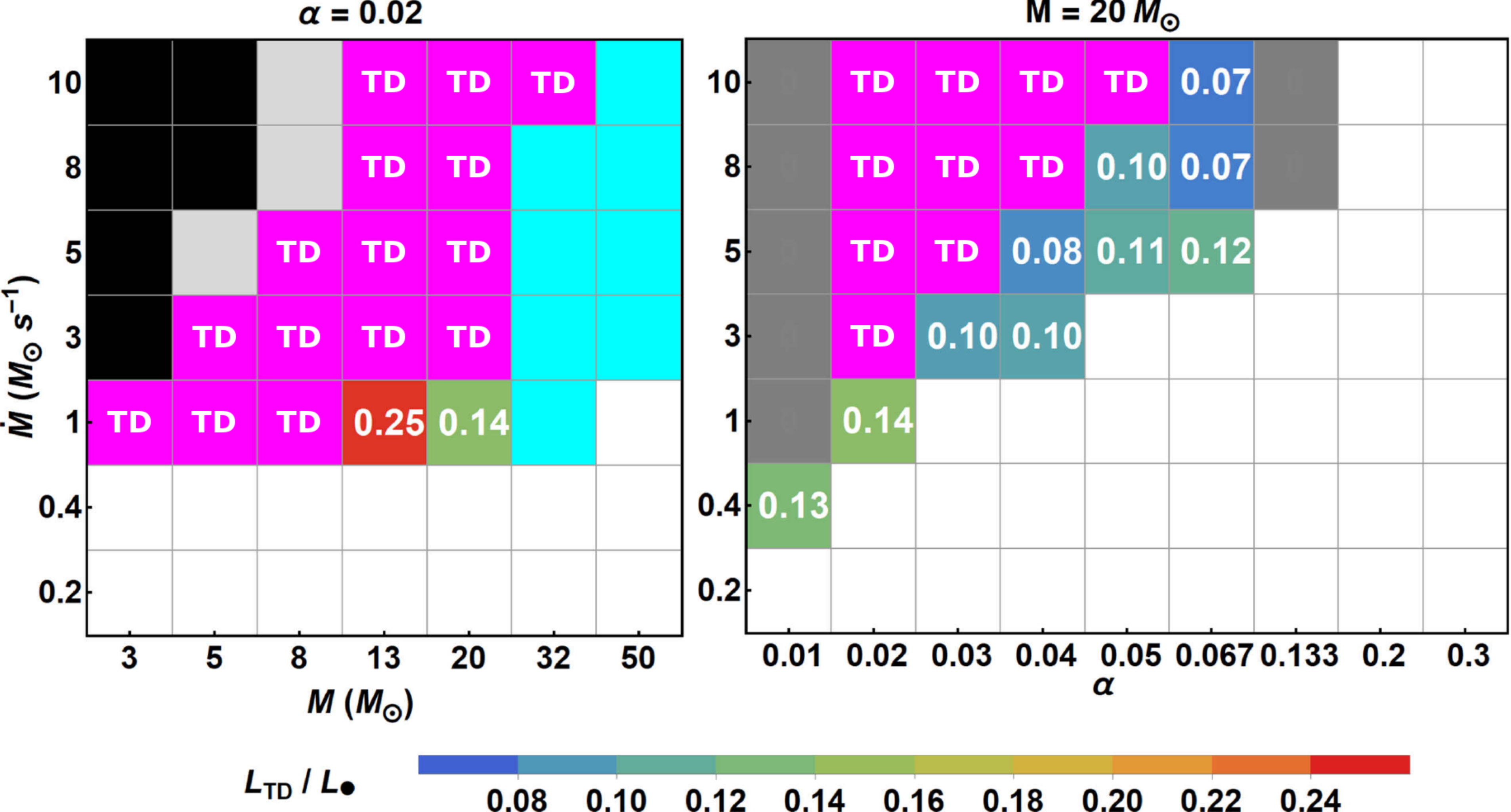}
\caption{The ratio between the luminosity from the tidal disruption of the fragment to the steady state luminosity from BH accretion ($L_{\rm TD}/L_{\bullet}$). {\it Left} and {\it right} panels represent the same $\{M, \dot{M}, \alpha\}$ parameter choices as in Fig. \ref{fig: Lowest_Q_Radius_2_panel}. White, black, gray and magenta squares are colored with the same color scheme as \ref{fig:AngularFrequancy_alpha002M20_PaperRas}.
Squares that: (1) have a number, represent the ratio between the luminosities as colored by the rainbow color system represented by the bar legend below the figure; (2) are filled in cyan represent that the tidal radius is smaller than the ISCO, meaning no TDE; (3) are filled in light grey, represent final fragment masses below the minimal stable mass for the cold EOS.  The luminosity from the tidal disruption is roughly $\sim10\%$ of the steady luminosity.}
\label{app: GRBVariabilityTD_alpha002M20_TD}
\end{figure}

In this appendix, we show results regarding GRB variability from fragment tidal disruption, as we briefly discussed in the beginning of section \ref{sec:variability}. We plot in figure \ref{app: GRBVariabilityTD_alpha002M20_TD} our estimate for the ratio between the TD luminosity of the accreted fragment and the steady luminosity of the collapsar ($L_{\rm TD}/L_{\rm \bullet}$, see Eq. \ref{eq: variability from TD}). 

Figure \ref{app: GRBVariabilityTD_alpha002M20_TD} supports the conclusion from the beginning of section \ref{sec:variability} that variability from fragment tidal disruption is generally modest. In most cases, fragments either disrupt at large radii—resulting in negligible GRB variability—or are swallowed whole after crossing the ISCO, causing no variability at all. Only in a narrow intermediate regime do disruptions lead to a mild enhancement in accretion luminosity, typically around $\sim10\%$, reinforcing our conclusion that this mechanism plays a limited role in shaping GRB variability.

\subsection{Eddington limit}
\label{app: Eddington limit}
In our discussion of the fragment accretion rate (see \ref{subsec: Validity of Model}) we suggested that the neutrino Eddington limit could cap the maximum accretion rate, though we do not quantify this possibility in the main text.  %. The possibility of the the accretion rate being capped by the neutrino Eddington limit is a function of the radiative efficiencies of the accreted matter onto the fragment. To quantify how the neutrino Eddington limit might change fragment mass growth, 
Here we recalculate our main findings regarding migration and mass accumulation (as presented in sections \ref{sec:results} and \ref{sec:observables}) with the fragment accretion rate now capped by the estimated neutrino Eddington rate, $\Dot{m}_{{\rm Edd}}$ (see \ref{subsec: Validity of Model}). 

\begin{figure}
\includegraphics[width=0.48\textwidth]{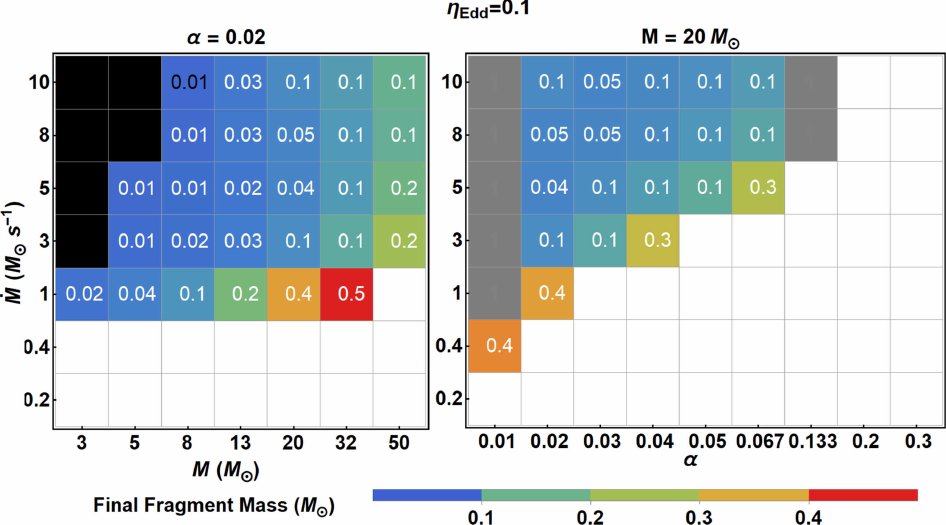}
\caption{An alternate version of Fig. \ref{fig:FragmentMassGrowth_2PanelArray}, plotting the final fragment mass (in ${M}_{\odot}$) if we cap the fragment accretion rate by the neutrino Eddington limit, assuming a high radiative efficiency of $\eta_{\rm Edd}=0.1$ (see \ref{subsec: Validity of Model}). The results are shown in two array plots. {\it Left} and {\it right} panels represent the same $\{M, \dot{M}, \alpha\}$ parameter choices as in Fig. \ref{fig: Lowest_Q_Radius_2_panel}, with the same color scheme as Fig. \ref{fig:FragmentMassGrowth_2PanelArray}.
The corresponding final fragment masses are mostly smaller (compared to the uncapped result in Fig. \ref{fig:FragmentMassGrowth_2PanelArray}) by factors of a few $\sim2-10$, while some regions in parameter space (with lower accretion rate or high BH masses) are unchanged.}
\label{app: BiggestFinalValidMass_alpha002M20_NoTrapseta01}
\end{figure}

\begin{figure}
\includegraphics[width=0.48\textwidth]{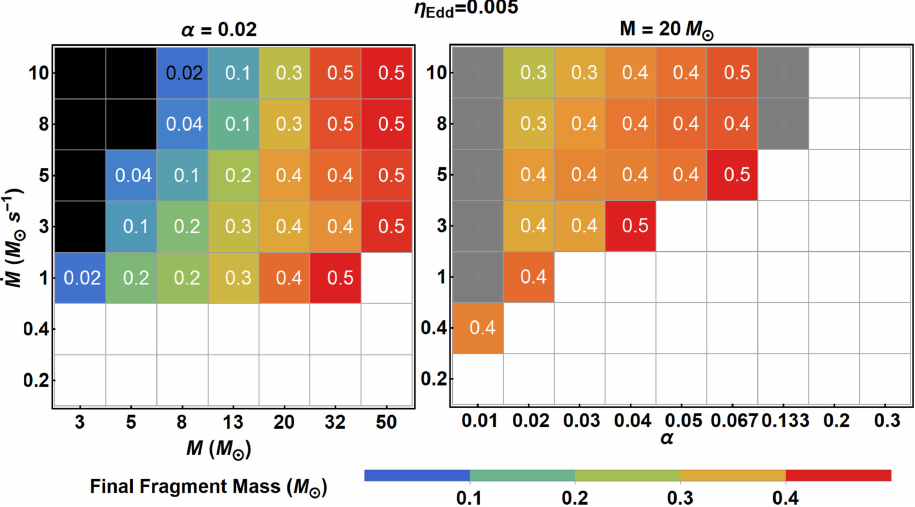}
\caption{An alternate version of Fig. \ref{app: BiggestFinalValidMass_alpha002M20_NoTrapseta01}, plotting the final fragment mass (in ${M}_{\odot}$) for Eddington-capped growth, but now assuming a radiative efficiency of $\eta_{\rm Edd}=0.005$ (see \ref{subsec: Validity of Model}). The results are shown in two array plots. {\it Left} and {\it right} panels represent the same $\{M, \dot{M}, \alpha\}$ parameter choices as in Fig. \ref{fig: Lowest_Q_Radius_2_panel}, with the same color scheme as Fig. \ref{fig:FragmentMassGrowth_2PanelArray}.
The corresponding final fragment masses are mostly unchanged %(other than $\{M=3 M_\odot, \dot{M}=1 M_\odot s^{-1}, \alpha=0.02\}$ and $\{M=13 M_\odot, \dot{M}=8 M_\odot s^{-1}, \alpha=0.02\}$) 
compared to the uncapped result in Fig. \ref{fig:FragmentMassGrowth_2PanelArray}, showing that high $\eta_{\rm Edd}$ values are necessary for the neutrino Eddington limit to strongly modify fragment growth.}
\label{app: BiggestFinalValidMass_alpha002M20_NoTrapsEta0005}
\end{figure}

\begin{figure}
\includegraphics[width=0.48\textwidth]{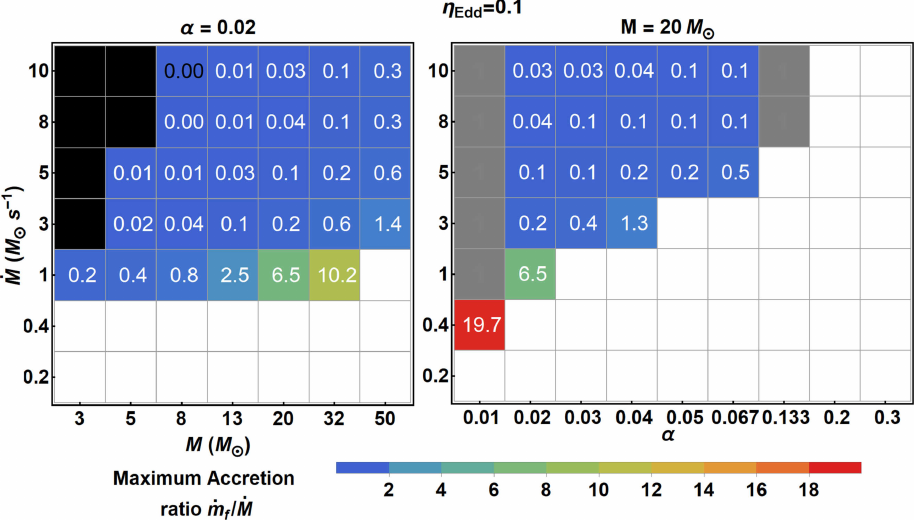}
\caption{An alternate version of Fig. \ref{fig:MaxAccretionRateRatio_2PanelArray}, plotting the peak fragment accretion rate, $\dot{m}_{\rm f}$, normalized by the collapsar accretion rate $\dot{M}$.  Unlike in the main text, we now cap the Bondi-Hoyle accretion rate by the neutrino Eddington limit, assuming a high radiative efficiency of $\eta_{\rm Edd}=0.1$ (see \ref{subsec: Validity of Model}). The results are shown in two array plots. {\it Left} and {\it right} panels represent the same $\{M, \dot{M}, \alpha\}$ parameter choices as in Fig. \ref{fig: Lowest_Q_Radius_2_panel}, with the same color scheme as Fig. \ref{fig:MaxAccretionRateRatio_2PanelArray}. The rainbow color system represents the peak of the accretion rate ratio, as color-coded in the bar legend below. The corresponding peak accretion rate ratios are greatly reduced (compared to the uncapped result in Fig. \ref{fig:MaxAccretionRateRatio_2PanelArray}), by typical factors $\sim10-50$, though some regions in parameter space (with lower accretion rate or high BH masses) exhibit a smaller reduction, by $\sim2-4$.}
\label{app: MaxAccretionRateRatio_alpha002M20_NoTrapsEta01}
\end{figure}

\begin{figure}
\includegraphics[width=0.48\textwidth]{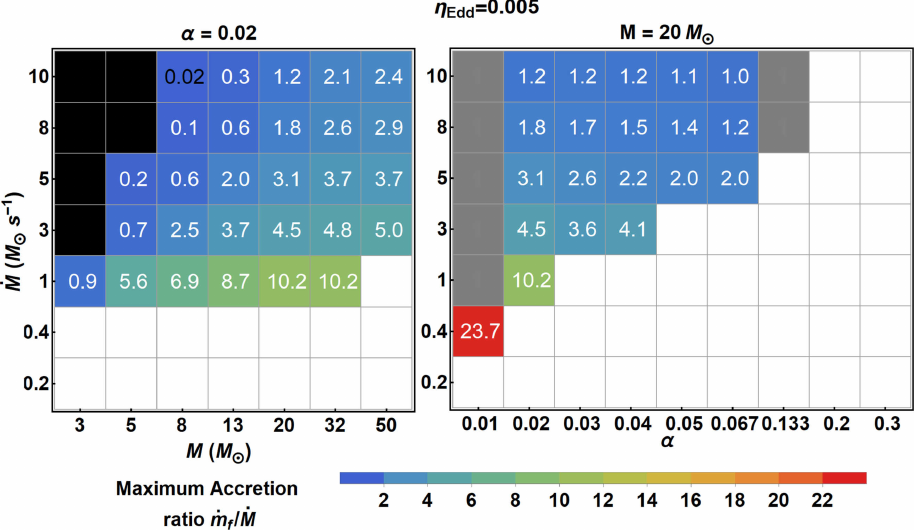}
\caption{An alternate version of Fig. \ref{app: MaxAccretionRateRatio_alpha002M20_NoTrapsEta01}, plotting the peak accretion rate, $\dot{m}_{\rm f}$, normalized by the collapsar accretion rate $\dot{M}$, now assuming a low neutrino radiative efficiency of $\eta_{\rm Edd}=0.005$ (see \ref{subsec: Validity of Model}). The results are shown in two array plots. {\it Left} and {\it right} panels represent the same $\{M, \dot{M}, \alpha\}$ parameter choices as in Fig. \ref{fig: Lowest_Q_Radius_2_panel}, with the same color scheme as Fig. \ref{fig:MaxAccretionRateRatio_2PanelArray}. The rainbow color system represents the peak of the accretion rate ratio, as color-coded in the bar legend below. The corresponding peak accretion rate ratios are mostly unchanged (other than a few examples with lower BH masses) compared to the uncapped result in Fig. \ref{fig:MaxAccretionRateRatio_2PanelArray}.}
\label{app: MaxAccretionRateRatio_alpha002M20_NoTrapsEta0005}
\end{figure}

Figs. \ref{app: BiggestFinalValidMass_alpha002M20_NoTrapseta01}-\ref{app: MaxAccretionRateRatio_alpha002M20_NoTrapsEta0005} demonstrate how a neutrino Eddington limit might effect our predictions for two extreme radiative efficiencies ($\eta_{\rm f}=0.1$ and $\eta_{\rm f}=0.005$). The peak accretion rate and the fragment final mass are largely unaffected for low radiative efficiencies ($\eta_{\rm f} \lesssim 0.005$). On the other hand, for a larger radiative efficiency, $\eta_{\rm f}=0.1$ (corresponding to a neutron star-like fragment with $m_{\rm f} \sim 1 M_\odot$ and $R_{\rm f} \sim 10 {\rm km}$), we see a significant suppression of the peak accretion rate and the fragment final mass. These results indicate that the neutrino Eddington limit does not hinder initial fragment formation, but may play an important role in capping growth as fragments become more compact and massive. 

\subsection{Critical cooling parameter \texorpdfstring{$\beta_{\rm c}=10$}{beta=10}}
\label{app: beta 10}
A key parameter in our collapsar disk models is the critical dimensionless cooling time, $\beta_{\rm c}$, that is assumed to  separate fragmenting from non-fragmenting $Q<1$ regions. The classical assumption that disk fragmentation requires $\beta< \beta_{\rm c}$ has been increasingly challenged by recent studies (see \S \ref{subsubsec: cooling time beta}), and in the main text we selected a large (optimistic) value of $\beta_{\rm c}=35$, following \citet{Chen+23}.  In this appendix, we follow \citet{Hennebelle2021} instead, and repeat our calculations with a lower (more conservative) choice of $\beta_{\rm c} = 10$.

\begin{figure}
\includegraphics[width=0.48\textwidth]{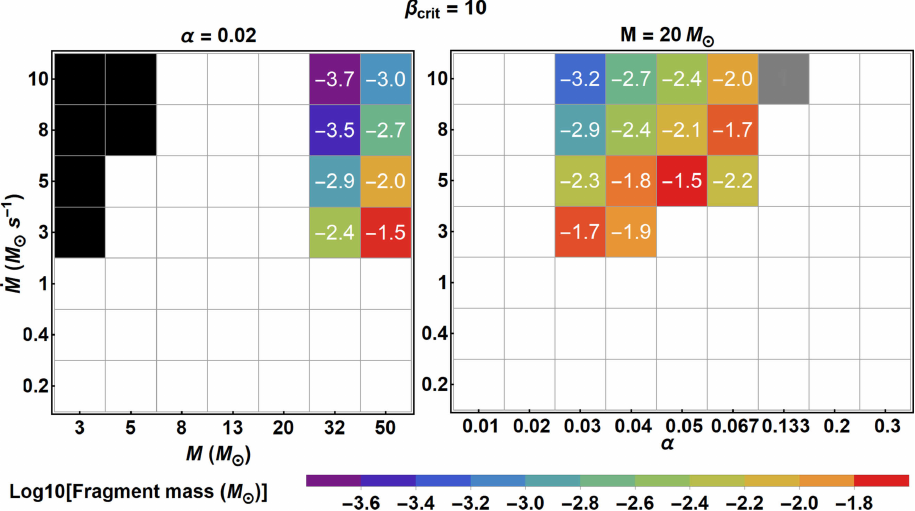}
\caption{An extended version of Fig. \ref{fig: InitialFragmentMass_2_panels} in the main text, plotting the color-coded values for the largest initial fragment mass (represented as $\log_{10}(m_{\rm f,i}/M_\odot)$) in the fragmenting $Q<1$ region, but now for $\beta_{\rm c}=10$. The results are shown in two array plots. {\it Left} and {\it right} panels represent the same $\{M, \dot{M}, \alpha\}$ parameter choices as in Fig. \ref{fig: Lowest_Q_Radius_2_panel}, with the same color scheme as Fig. \ref{fig: InitialFragmentMass_2_panels}. The rainbow color system represents the $\log_{10}$ of initial fragment mass, as color-coded in the legend. For the lower $\beta_{\rm c}$ value chosen here, fragmentation is prevented in many regions of our parameter space, especially for lower accretion rates, BH masses and viscosities. The initial fragment mass decreases by a factor of $\sim 2-15$ compared to original calculation in Fig. \ref{fig: InitialFragmentMass_2_panels}.
}
\label{app: BiggestInitialMass_alpha002M20_beta10}
\end{figure}

\begin{figure}
\includegraphics[width=0.48\textwidth]{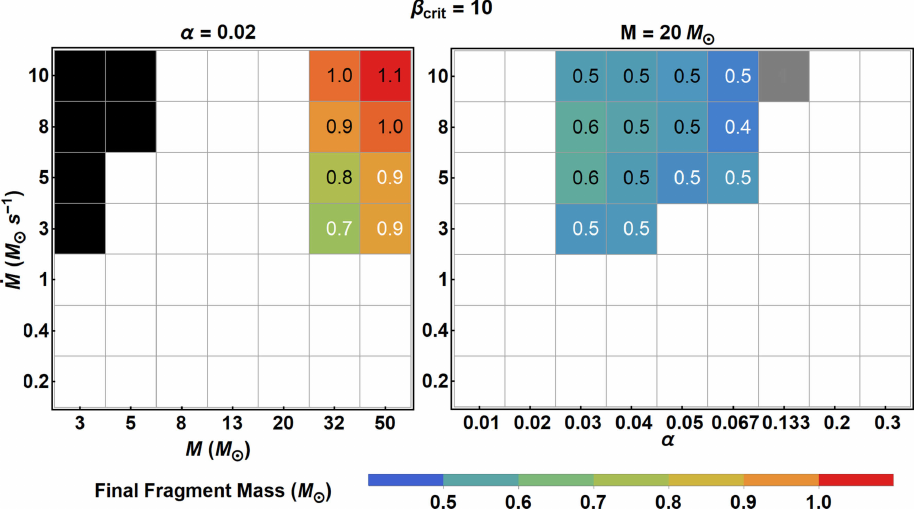}
\caption{An alternate version of Fig. \ref{fig:FragmentMassGrowth_2PanelArray}, plotting the final fragment mass (in ${M}_{\odot}$), but now for $\beta_{\rm c}=10$. The results are shown in two array plots. {\it Left} and {\it right} panels represent the same $\{M, \dot{M}, \alpha\}$ parameter choices as in Fig. \ref{fig: Lowest_Q_Radius_2_panel}, with the same color scheme as Fig. \ref{fig:FragmentMassGrowth_2PanelArray}.
The corresponding final fragment masses are bigger (compared to the higher $\beta_{\rm c}$ result in Fig. \ref{fig:FragmentMassGrowth_2PanelArray}) by factors of a few $\sim1.25-2.2$, with some regions in parameter space remain unchanged.}
\label{app: BiggestFinalValidMass_alpha002M20_beta10}
\end{figure}

\begin{figure}
\includegraphics[width=0.48\textwidth]{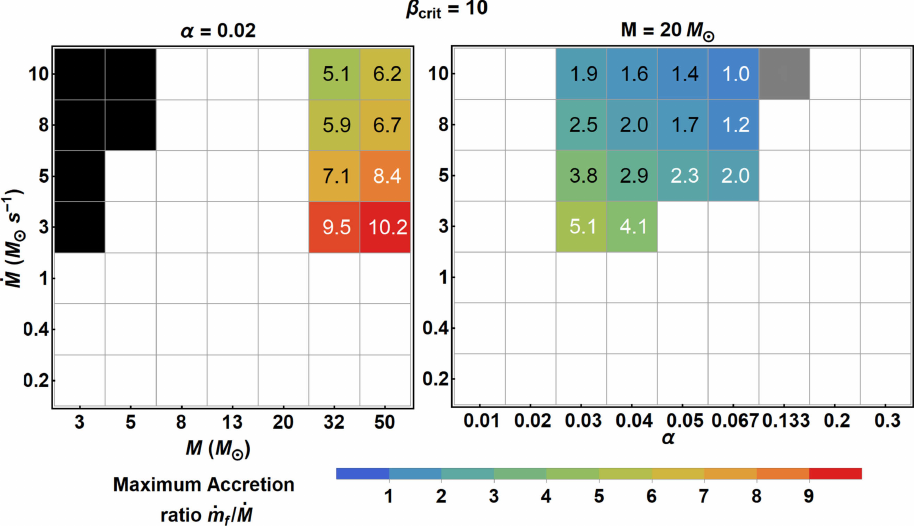}
\caption{An alternate version of Fig. \ref{app: MaxAccretionRateRatio_alpha002M20_NoTrapsEta01}, plotting the peak Bondi-Hoyle accretion rate, $\dot{m}_{\rm f}$, normalized by the collapsar accretion rate $\dot{M}$, but now for $\beta_{\rm c}=10$. The results are shown in two array plots. {\it Left} and {\it right} panels represent the same $\{M, \dot{M}, \alpha\}$ parameter choices as in Fig. \ref{fig: Lowest_Q_Radius_2_panel}, with the same color scheme as Fig. \ref{fig:MaxAccretionRateRatio_2PanelArray}. The rainbow color system represents the peak of the accretion rate ratio, as color-coded in the bar legend below. As the fragments become bigger in mass (Fig. \ref{app: BiggestFinalValidMass_alpha002M20_beta10}) the peak accretion rate ratio also increases, as is expected from Bondi-Hoyle accretion physics (Eq. \ref{eq: Bondi-Hoyle accretion rate}).}
\label{app: MaxAccretionRateRatio_alpha002M20_NoTrapsBeta10}
\end{figure}

In Figs. \ref{app: BiggestInitialMass_alpha002M20_beta10}-\ref{app: MaxAccretionRateRatio_alpha002M20_NoTrapsBeta10} we see that choosing a lower $\beta_{\rm c}$ affects our results in the following ways:
\begin{itemize}
    \item Fragmentation now fails to occur in many regions of our parameter space where it previously did, especially for lower accretion rates, BH masses and viscosities.
    \item Choosing a lower $\beta_{\rm c}$ will often move the minimum radius of fragmentation, $r_{\rm  min}$, outwards (see Fig. \ref{fig: Initial_Fragment_Mass_3_Panel}) to regions with lower disk densities. The lower density decreases the initial fragment mass, usually by a factor $\sim 2-15$ (Fig. \ref{app: BiggestInitialMass_alpha002M20_beta10}).
    \item Even though the initial fragment mass is usually lower, the longer migration distance enables more fragment growth, causing the final fragment mass to be significantly {\it larger} (by a factor of $\sim 1.25-2.20$, see Fig. \ref{app: BiggestFinalValidMass_alpha002M20_beta10}). The greater migration distance also causes many of the remaining points in parameter space to have a longer migration time than the mass doubling time of the BH ($M/\dot{M}$), reducing the self-consistency of our calculations.
    \item As the fragments become greater in mass, the peak accretion rate ratio increases also (see Fig. \ref{app: MaxAccretionRateRatio_alpha002M20_NoTrapsBeta10}), as is expected from Bondi-Hoyle accretion physics (Eq. \ref{eq: Bondi-Hoyle accretion rate}).
\end{itemize}

\subsection{Black hole Spin \texorpdfstring{$a=0.001$}{a=0.001}}
\label{app: a0}
The spin of the BH, characterized by the dimensionless parameter $a$, reflects a balance between spin-up through accretion and spin-down from other processes. It has a crucial impact on the Kerr metric and thus on the microphysics of the disk. This impact is strongest close to the ISCO and weak at $r\gg r_{\rm ISCO}$, so most of our ``initial'' results (gravitational instability, cooling time, and initial fragment mass calculations) are highly insensitive to BH spin. As fragments migrate inwards, the importance of spin increases. Because the spins of BHs in collapsar disks are highly uncertain and currently a subject of much debate \citep[e.g.][]{Lowell+25}, in this appendix we explore how our results would change if we take the other limiting case of a non-spinning BH (we choose $a=0.001$ for numerical reasons).

\begin{figure}
\includegraphics[width=0.48\textwidth]{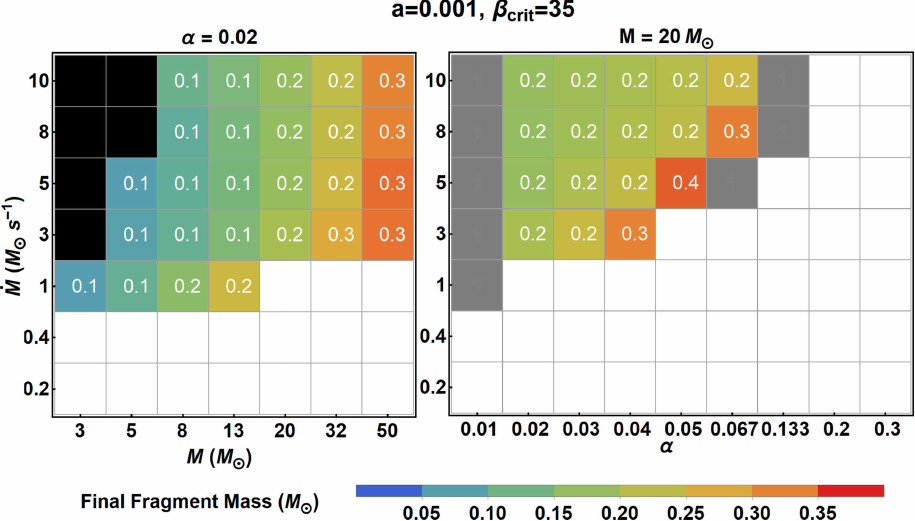}
\caption{An alternate version of Fig. \ref{fig:FragmentMassGrowth_2PanelArray}, plotting the final fragment mass (in ${M}_{\odot}$), but now for a spin parameter $a=0.001$. The results are shown in two array plots. {\it Left} and {\it right} panels represent the same $\{M, \dot{M}, \alpha\}$ parameter choices as in Fig. \ref{fig: Lowest_Q_Radius_2_panel}, with the same color scheme as Fig. \ref{fig:FragmentMassGrowth_2PanelArray}. The rainbow color system represents the final fragment mass, as color-coded in the legend. The corresponding final fragment masses are smaller (compared to the higher spin result in Fig. \ref{fig:FragmentMassGrowth_2PanelArray}) by factors of a few $\sim1.6-3$.}
\label{app: BiggestFinalValidMass_alpha002M20_a0}
\end{figure}

\begin{figure}
\includegraphics[width=0.48\textwidth]{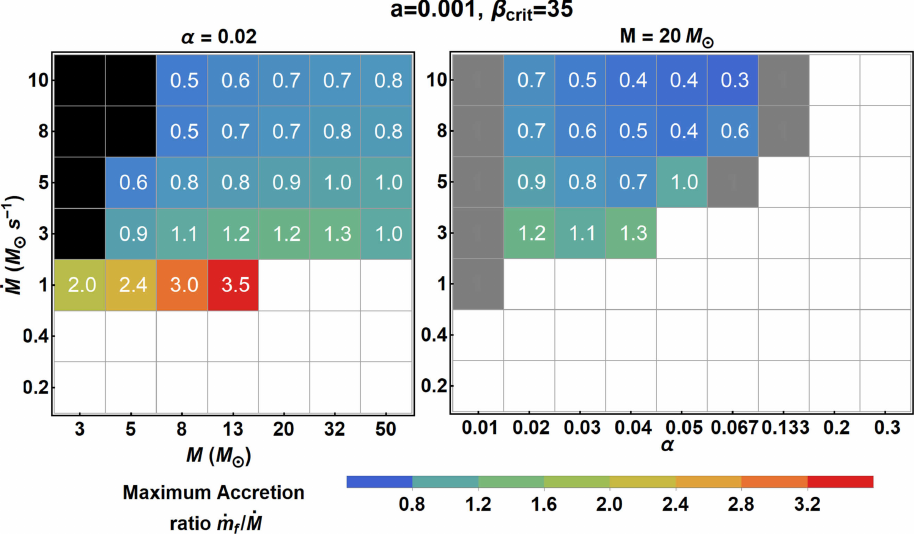}
\caption{An alternate version of Fig. \ref{app: MaxAccretionRateRatio_alpha002M20_NoTrapsEta01}, plotting the peak Bondi-Hoyle accretion rate, $\dot{m}_{\rm f}$, normalized by the collapsar accretion rate $\dot{M}$, but now for a spin parameter $a=0.001$. The results are shown in two array plots. {\it Left} and {\it right} panels represent the same $\{M, \dot{M}, \alpha\}$ parameter choices as in Fig. \ref{fig: Lowest_Q_Radius_2_panel}, with the same color scheme as Fig. \ref{fig:MaxAccretionRateRatio_2PanelArray}. The rainbow color system represents the peak of the accretion rate ratio, as color-coded in the bar legend below. The peak accretion rate ratios are lower (except for a few points with low BH mass and high accretion rate) by a factor of $\sim 2-3$ compared to the results with higher spin, in Fig. \ref{fig:MaxAccretionRateRatio_2PanelArray}.}
\label{app: MaxAccretionRateRatio_alpha002M20_NoTraps_a0}
\end{figure}

\begin{figure}
\includegraphics[width=0.48\textwidth]{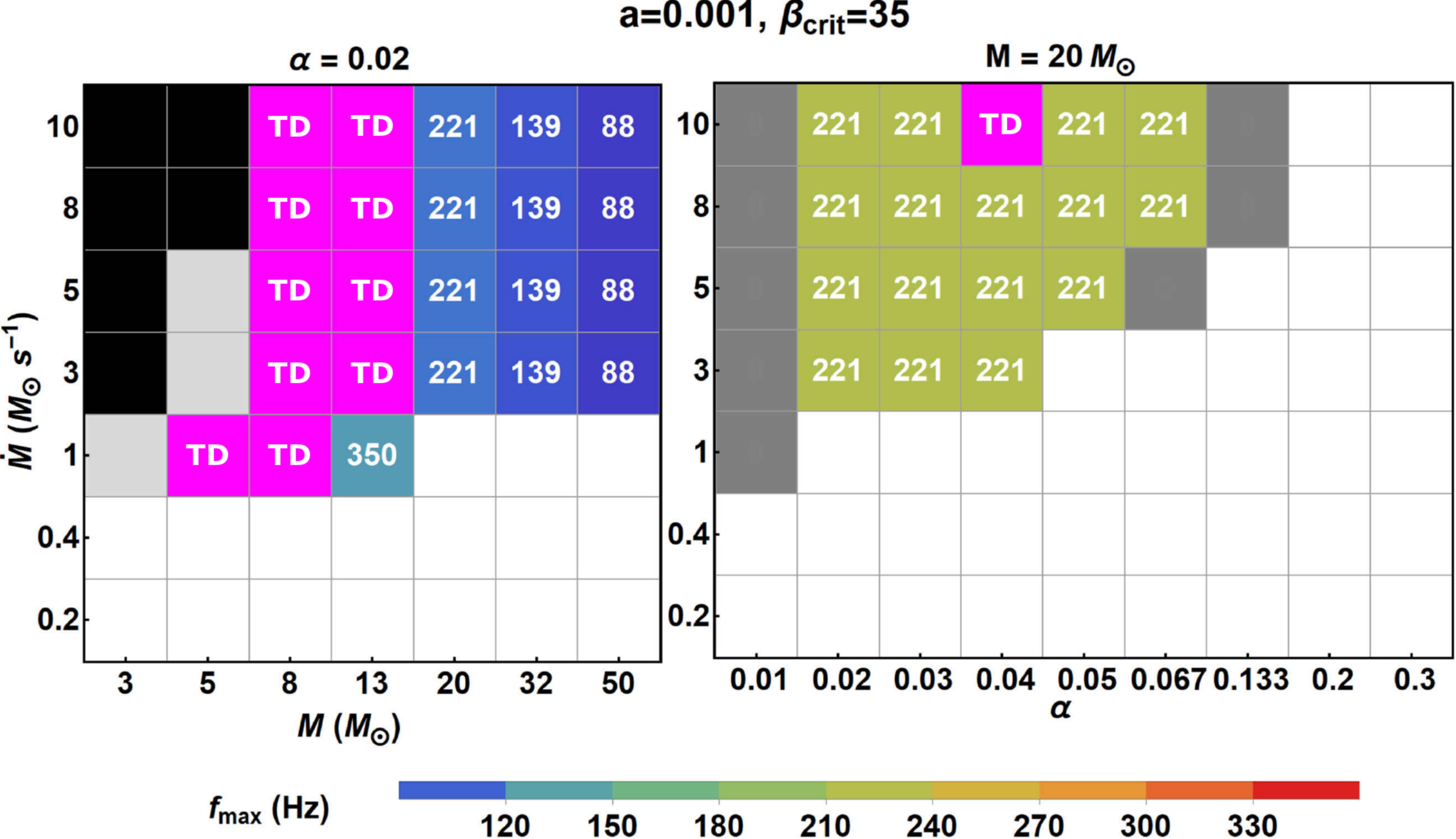}
\caption{An alternate version of Fig. \ref{fig:AngularFrequancy_alpha002M20_PaperRas}, plotting the peak orbital frequency $f_{{\rm max}}$ of the GWs emitted by an inspiraling fragment, but now for a spin parameter $a=0.001$. The results are shown in two array plots. {\it Left} and {\it right} panels represent the same $\{M, \dot{M}, \alpha\}$ parameter choices as in Fig. \ref{fig: Lowest_Q_Radius_2_panel}, with the same color scheme as Fig. \ref{fig:AngularFrequancy_alpha002M20_PaperRas}. The rainbow color system represents the frequency, as represented in the bar legend. Less regions in parameter space exhibit a TD compared to the higher spin, especially for lower viscosities. For the lower spin value the fragments are either tidally disrupted far from the ISCO (magenta squares) or pass the ISCO without a TD (causing the peak frequency to be the ISCO frequency, which is related only to the BH mass). As the ISCO is farther from the BH for the non spinning case, the peak frequency is lower by roughly a factor of $\sim 4$. 
}
\label{app: AngularFrequency_alpha002M20_a0}
\end{figure}

\begin{figure}
\includegraphics[width=0.48\textwidth]{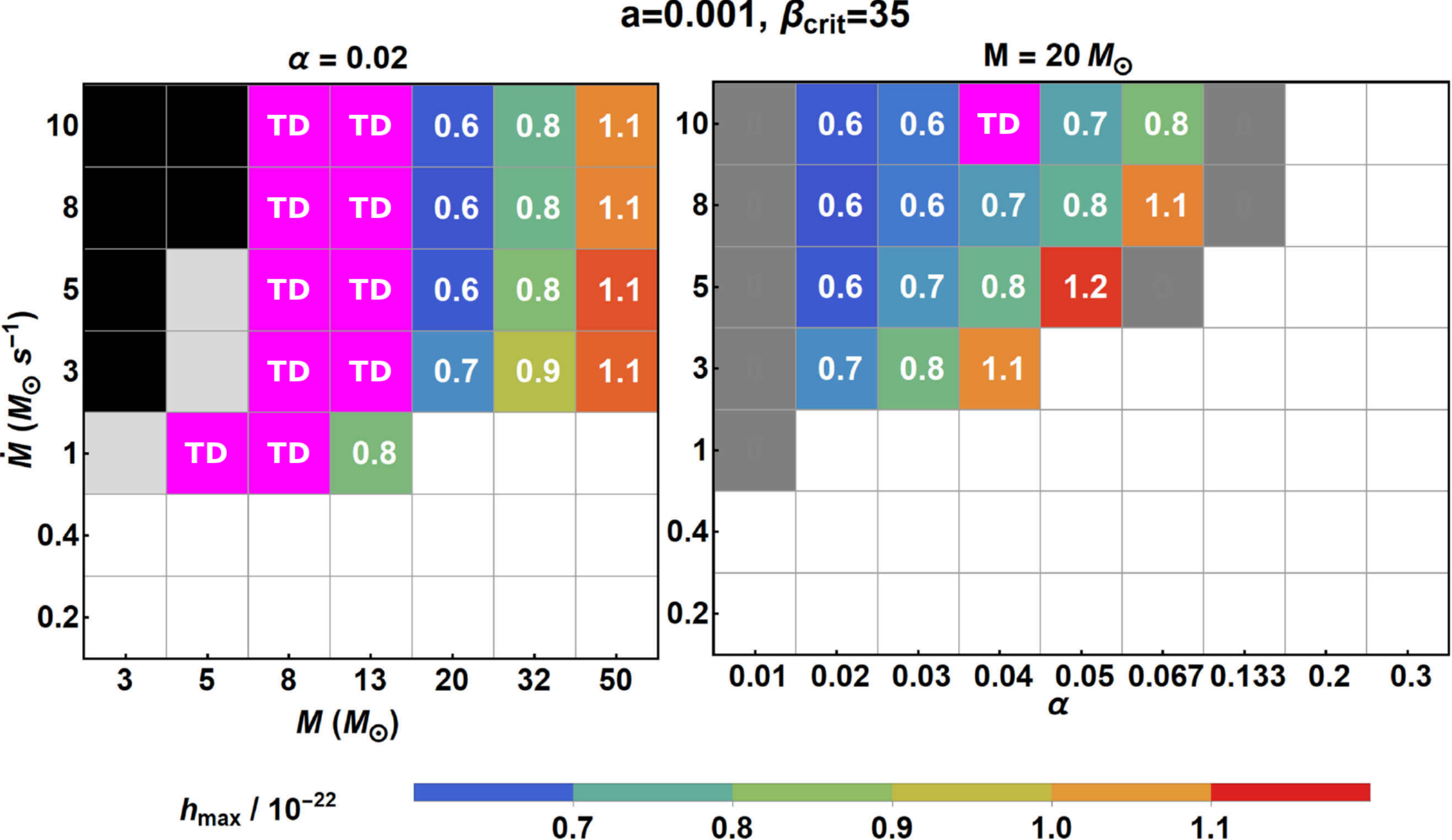}
\caption{An alternate version of Fig. \ref{fig:Strain_alpha002M20_PaperRas}, plotting the peak strain $h_{{\rm max}}$ of the GWs emitted by an inspiraling fragment normalized by $10^{-22}$, but now for a spin parameter $a=0.001$. The results are shown in two array plots. {\it Left} and {\it right} panels represent the same $\{M, \dot{M}, \alpha\}$ parameter choices as in Fig. \ref{fig: Lowest_Q_Radius_2_panel}, with the same color scheme as Fig. \ref{fig:Strain_alpha002M20_PaperRas}. The rainbow color system represents the strain, as represented in the bar legend below the figure. For the lower spin value the fragments are either tidally disrupted far from the ISCO (magenta squares) or pass the ISCO without a TD. The peak GW strain is lower by up to a factor of $\sim 2$.}
\label{app: Strain_alpha002M20_a0}
\end{figure}

In Figs. \ref{app: BiggestFinalValidMass_alpha002M20_a0}-\ref{app: Strain_alpha002M20_a0} we show how choosing a different value for BH spin ($a=0.001$) changes some of our results, which we summarize here:
\begin{itemize}
    \item The ISCO radius for a non rotating BH is much larger than in a highly spinning case ($r_{\rm ISCO}=6r_{\rm g}$ for our near-Schwarzschild alternate scenario, while $r_{\rm ISCO}\approx 1.9 r_{\rm g}$ in the main text), causing a decrease in both the peak density of the accretion disk and the orbital frequency at ISCO (seen in Fig \ref{app: AngularFrequency_alpha002M20_a0}).
    \item Fewer regions in parameter space exhibit a TD compared to the higher spin case, especially for lower viscosities.
    \item All fragments that are not tidally disrupted at $r_{\rm min}$ now pass through the ISCO, setting the peak frequency to be the ISCO frequency, which is related only to the BH mass .
    \item Both the lower density in the disk and the shorter migration distance cause the fragments to accrete less mass, making the final fragment masses smaller (compared to the higher spin result in Fig. \ref{fig:FragmentMassGrowth_2PanelArray}) by factors of a few $\sim1.6-3$ (see Fig. \ref{app: BiggestFinalValidMass_alpha002M20_a0}).
    \item The peak GW strain is lowered by a factor of $\sim 2$, and the peak GW frequency is also reduced by a factor of a few.
\end{itemize}

\subsection{Duration of \texorpdfstring{$\dot{m}_{\rm f}>\dot{M}$}{mdot>Mdot}}
\label{app: Duration Of High Accretion}
\begin{figure}
\includegraphics[width=0.48\textwidth]{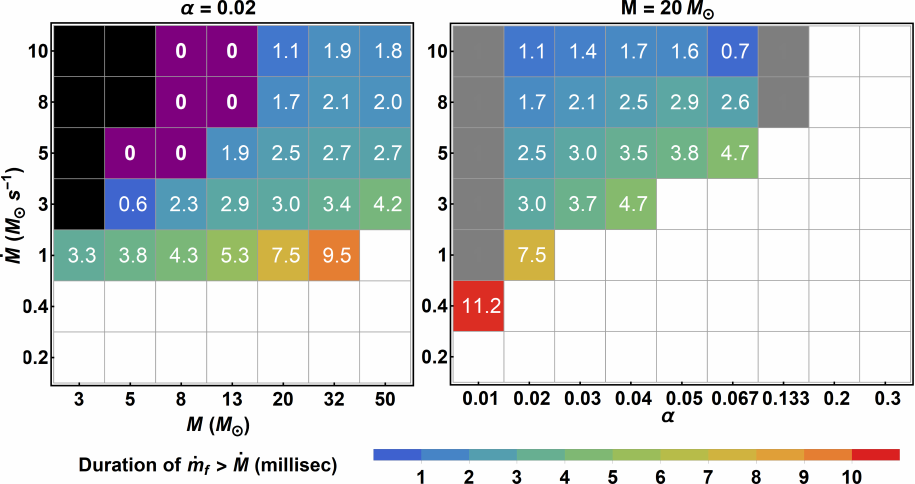}
\caption{Duration of the period when the fragment accretion rate $\dot{m}_{\rm f}$ exceeds the central accretion rate $\dot{M}$ (shown in ms). {\it Left} and {\it right} panels represent the same $\{M, \dot{M}, \alpha\}$ parameter choices as in Fig. \ref{fig: Lowest_Q_Radius_2_panel}. White, black, and gray squares are colored with the same color scheme as \ref{app: AngularFrequency_9Plots}. Purple squares indicate when $\dot{m}_{\rm f}$ never exceeds $\dot{M}$. The duration is roughly on the order of 5 ms, usually increasing with increasing BH mass, decreasing BH accretion rate, and increasing viscosity.}
\label{app:DurationOfHighAccretion_alpha002M20_a0}
\end{figure}

The fragment accretion rate $\dot{m}_{\rm f}$ commonly exceeds the BH accretion rate  $\dot{M}$ for a short duration of time (as is visible in the main text, in Figs. \ref{fig:AccretionRateRatios_3_Panel} and \ref{fig:MaxAccretionRateRatio_2PanelArray}). In this appendix we plot (in Fig. \ref{app:DurationOfHighAccretion_alpha002M20_a0}) two cross section of our parameter survey, in order to explore the duration of this period (which may be observable as extreme GRB variability, as a secondary jet from the accreting fragment may outshine the primary jet from the central BH).  Fig. \ref{app:DurationOfHighAccretion_alpha002M20_a0}shows that this duration generally ranges between $0.5-11.2 ms$.

%In Fig. \ref{app:DurationOfHighAccretion_alpha002M20_a0}, we see that for the highest accretion rates and low central masses, $\dot{m}_{\rm f}$ doesn't exceed $\dot{M}$ at all, while for the rest of parameter space, this duration ranges between $0.5-11.2 ms$, usually increasing with increasing the BH mass, decreasing the BH accretion rate and increasing viscosity.

\subsection{Extended parameter survey}
\label{app: Results Extended parameter survey}
In sections \ref{sec:results} and \ref{sec:observables} (results and observable predictions) we have focused only on two cross sections (a constant BH mass of $M=20 M_{\odot}$ and a constant viscosity of $\alpha=0.02$) out of our entire 3 dimensional parameter space, varying $\{M,\dot{M},\alpha\}$. In this section we give the full parameter space for some of the relevant findings shown in the main body of the text (Figs. \ref{fig: InitialFragmentMass_2_panels}, \ref{fig:FragmentMassGrowth_2PanelArray}, \ref{fig:MaxAccretionRateRatio_2PanelArray}, \ref{fig:AngularFrequancy_alpha002M20_PaperRas} and \ref{fig:Strain_alpha002M20_PaperRas}).

\begin{figure*}
\includegraphics[width=0.96\textwidth]{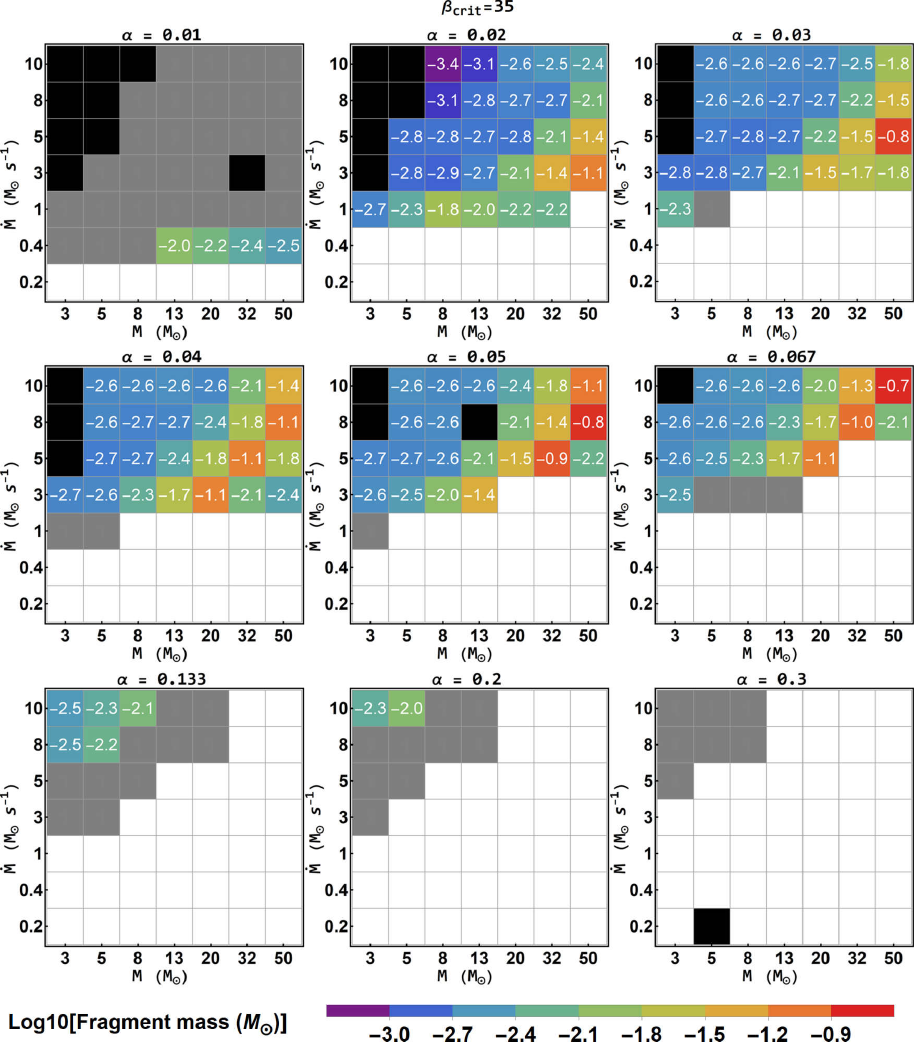}
\caption{An extended version of Fig. \ref{fig: InitialFragmentMass_2_panels} in the main text, plotting the color-coded values for the largest initial fragment mass (represented as $\log_{10}(m_{\rm f,i}/M_\odot)$) in the fragmenting $Q<1$ region, but now exploring the full 3 dimensional parameter space of $\{ M, \dot{M}, \alpha \}$.
Each of the panels represents a different $\alpha$ value, ranging from $0.01-0.3$, while varying $\{M, \dot{M}\}$.  Black, white, and grey squares represent breakdowns in model assumptions or solutions as in Fig. \ref{fig: Lowest_beta_Radius_2_panel}.
We see that fragment initial masses span a wide range, from $4\times10^{-4} \lesssim m_{\rm f}/M_\odot \lesssim 2\times10^{-1}$.  The initial masses tend to increase with declining accretion rate, or with increasing central mass or viscosity.}
\label{app: BiggestInitialMass_9Plots}
\end{figure*}

In the full survey of the largest initial fragment mass (Fig. \ref{app: BiggestInitialMass_9Plots}) we see a few examples of initial fragment masses above the minimum cold NS stable mass ($ m_{\rm f}\geq 0.09 M_\odot$. These examples occur only for the highest BH masses accompanied with the highest BH accretion rates. This extended parameter survey also shows that for both the lowest viscosity panel ($\alpha=0.01$) the highest viscosity ($\alpha=0.133-0.3$) panels, few self-consistent fragmentation zones exist because many of the solutions contain too much disk mass inside $r_Q$ (as indicated in gray squares).

\begin{figure*}
\includegraphics[width=0.96\textwidth]{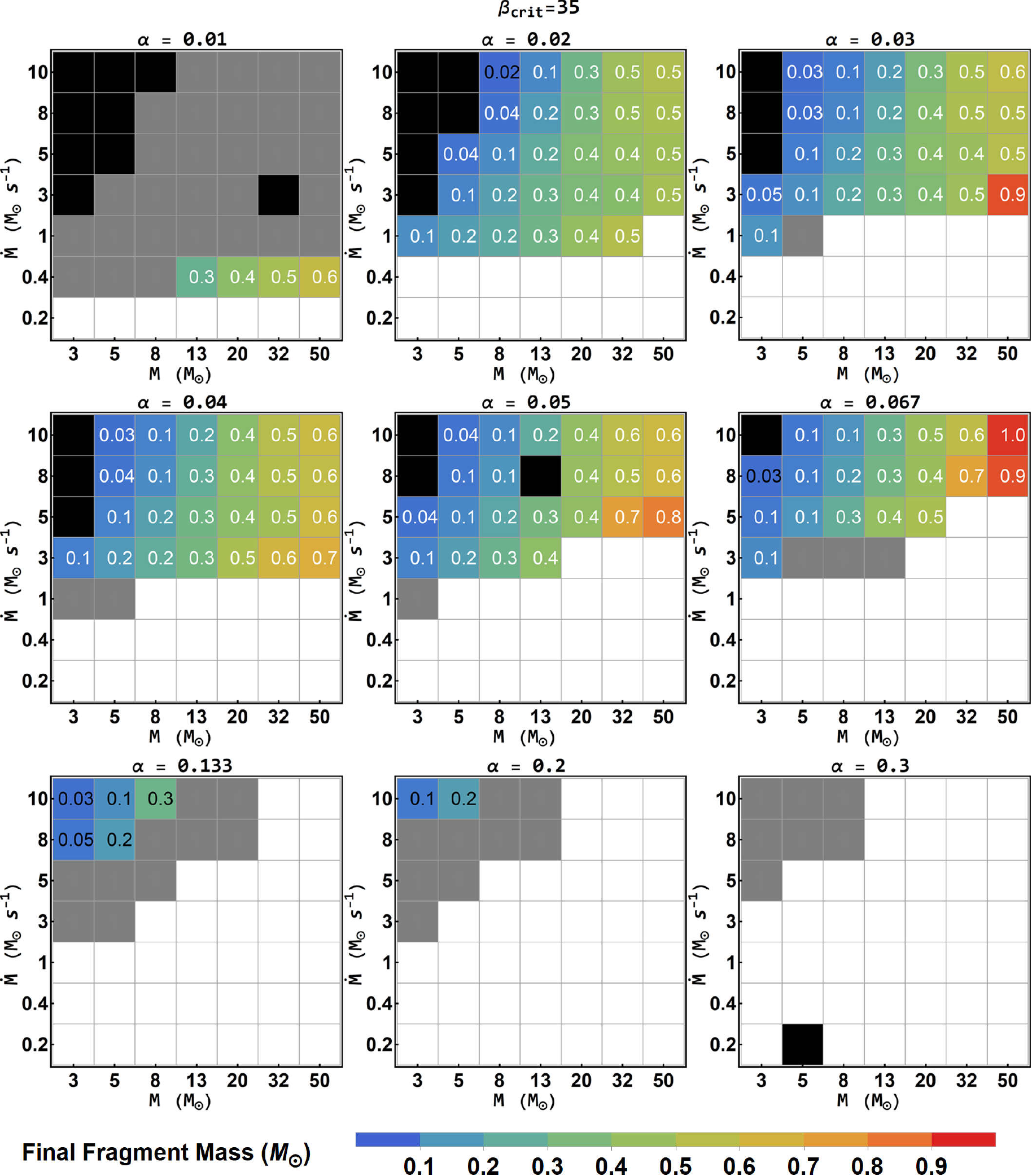}
\caption{An extended version of Fig. \ref{fig:FragmentMassGrowth_2PanelArray} in the main text, plotting the final fragment mass (in ${M}_{\odot}$), shown in 9 array plots. Panels represent the same $\{M, \dot{M}, \alpha\}$ parameter choices as in Fig. \ref{app: BiggestInitialMass_9Plots}. The rainbow color system represents the fragment final mass, as color-coded in the legend below. Black, white, and grey squares represent breakdowns in model assumptions or solutions as in Fig. \ref{fig: Lowest_beta_Radius_2_panel}. Numbers color coded in black represent fragment migration time bigger than $M/\dot{M}$. 
Fragment final masses tend to increase with decreasing accretion rate or increasing viscosity, and span a range of $0.04 \lesssim m_{\rm f}/M_\odot \lesssim 0.5$.  These final masses are usually in the range where cold stable NSs can exist, although in some instances (small $M$) they may be too low.}
\label{app: BiggestFinalValidMass_9Plots}
\end{figure*}

In the full survey of the final fragment mass (Fig. \ref{app: BiggestFinalValidMass_9Plots}) we see a few examples of final masses above the minimum hot NS stable mass ($ m_{\rm f}\geq 0.67 M_\odot$, \citealt{Gondek+97,Gondek+1998,LattimerSwesty1991}). These examples occur only for the highest BH masses accompanied by high BH accretion rates. This extended parameter survey also shows that for the highest $\alpha (=0.133-0.3)$ panels, the migration time exceeds the mass doubling time of the BH, reducing the reliability of our estimates for Bondi-Hoyle mass accumulation (denoted with black filled numbers).

\begin{figure*}
\includegraphics[width=0.96\textwidth]{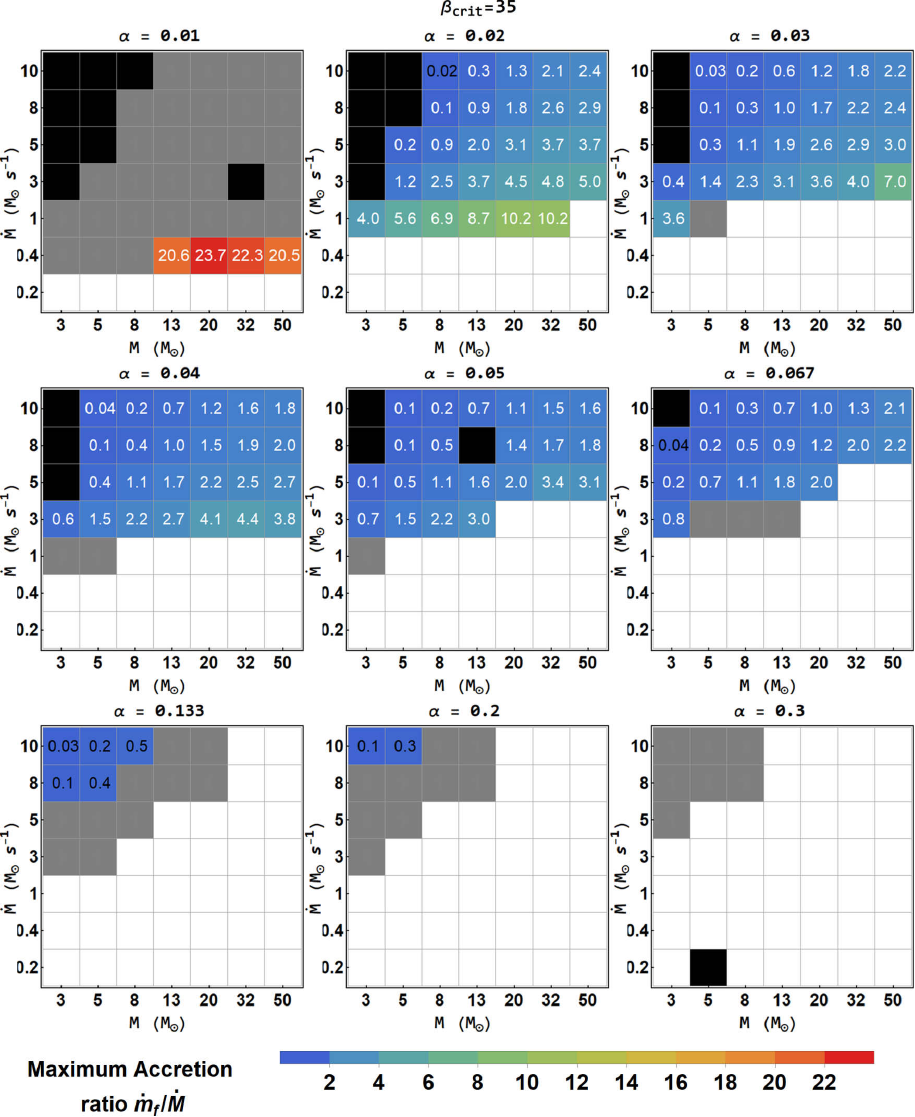}
\caption{
An extended version of Fig. \ref{fig:MaxAccretionRateRatio_2PanelArray} in the main text, plotting the peak Bondi-Hoyle accretion rate, $\dot{m}_{\rm f}$, normalized by the BH accretion rate $\dot{M}$, shown in 9 array plots. Panels represent the same $\{M, \dot{M}, \alpha\}$ parameter choices as in Fig. \ref{app: BiggestInitialMass_9Plots}. The rainbow color system represents the peak Bondi-Hoyle accretion rate, as color-coded in the legend below. The color scheme of the squares and the numbers is the same as in Fig. \ref{app: BiggestFinalValidMass_9Plots}.
The accretion rate ratio tends to increase with increasing central mass or decreasing viscosity. The non-normalized peak $\dot{m}_{\rm f}$ tends to increase with the accretion rate, while the ratio $\dot{m}_{\rm f}/\dot{M}$ declines with increasing $\dot{M}$. It is common for the fragment accretion rate to exceed $\dot{M}$ by a factor of a few.}
\label{app: MaxAccretionRatio_9Plots}
\end{figure*}

In the full parameter survey of the peak Bondi-Hoyle accretion rate ratio (Fig. \ref{app: MaxAccretionRatio_9Plots}) we see a few more examples of extreme (above a factor of 20) accretion rate ratios, but only for the the lowest viscosity disks ($\alpha=0.01$).  Our results are otherwise qualitatively similar to the smaller parameter slice in the main text.

\begin{figure*}
\includegraphics[width=0.96\textwidth]{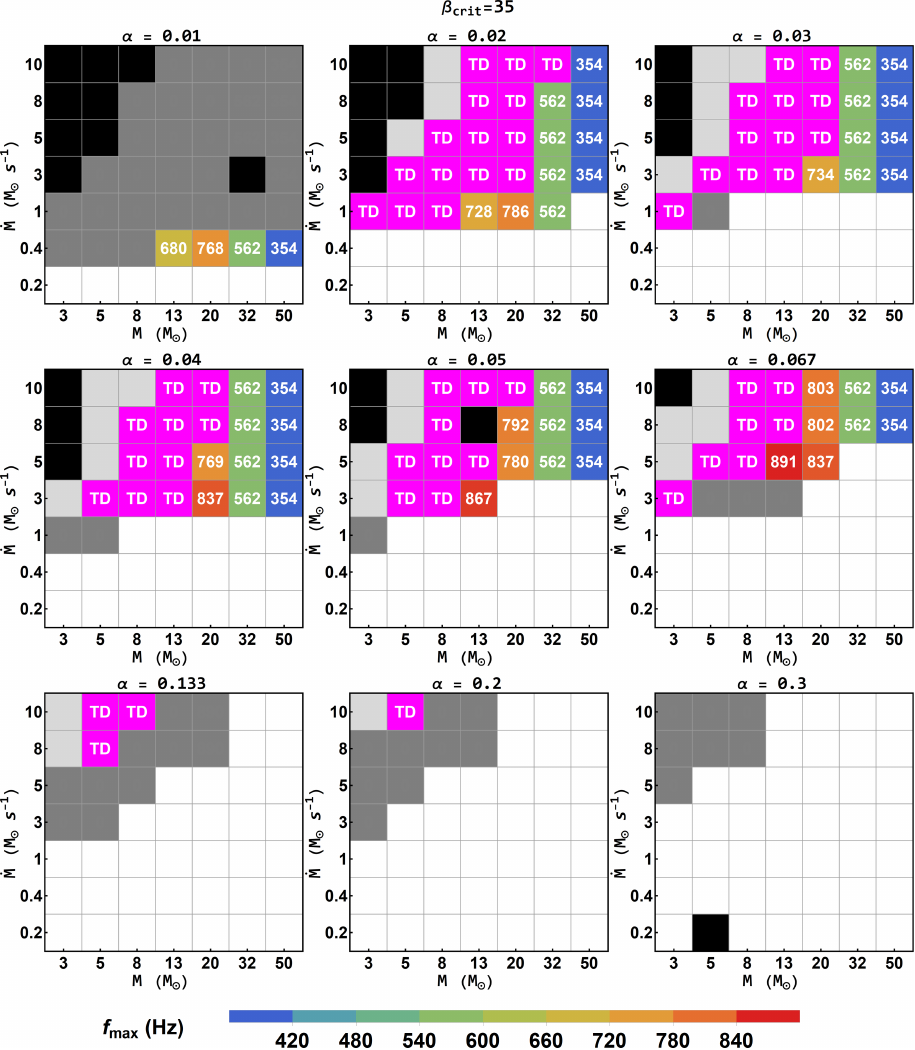}
\caption{An extended version of Fig. \ref{fig:AngularFrequancy_alpha002M20_PaperRas} in the main text, plotting the peak orbital frequency $f_{{\rm max}}$ of the GWs emitted by an inspiraling fragment. We compute this by evaluating the orbital frequency ($f_{\rm GW}$) at the maximum between the tidal disruption radius ($r_{\rm t}$) and the ISCO. Panels represent the same $\{M, \dot{M}, \alpha\}$ parameter choices as in Fig. \ref{app: BiggestInitialMass_9Plots}.  The color scheme is the same as in Fig. \ref{fig:AngularFrequancy_alpha002M20_PaperRas}. The rainbow color system labels the frequency, as represented in the bar legend below the figure. The frequency generally increases as the central mass decreases, or for a few examples, as the viscosity increases, while it does not significantly depend on the accretion rate.}
\label{app: AngularFrequency_9Plots}
\end{figure*}

\begin{figure*}
\includegraphics[width=0.96\textwidth]{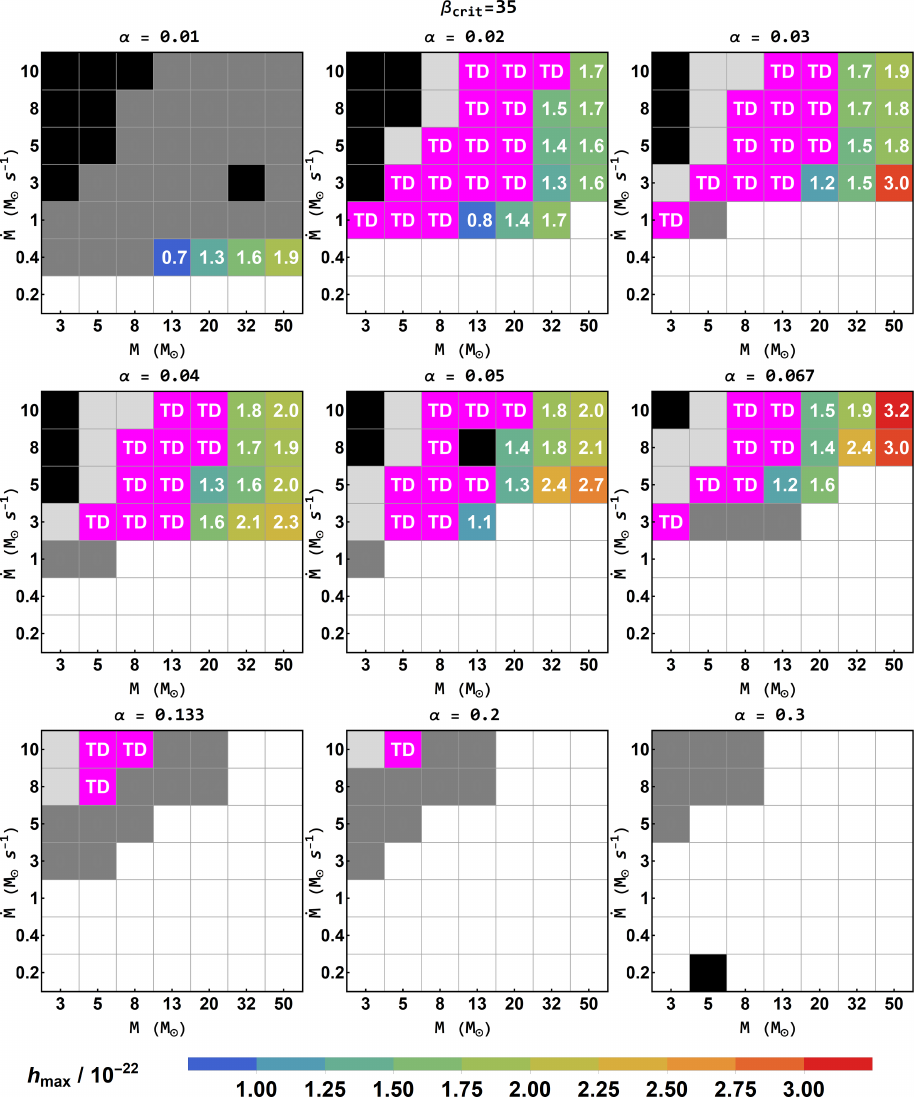}
\caption{An extended version of Fig. \ref{fig:Strain_alpha002M20_PaperRas} in the main text, plotting the peak strain $h_{{\rm max}}$ of the GWs emitted by an inspiraling fragment normalized by $10^{-22}$.  As previously we compute this by evaluating the strain ($h_{\rm GW}$) at the maximum between the tidal disruption radius ($r_{\rm t}$) and the ISCO. Panels represent the same $\{M, \dot{M}, \alpha\}$ parameter choices as in Fig. \ref{app: BiggestInitialMass_9Plots}. The color scheme is the same as in Fig. \ref{fig:Strain_alpha002M20_PaperRas}. The rainbow color system represents the strain, as represented in the bar legend below the figure. The peak GW strain is generally insensitive to $\dot{M}$, while it grows for increasing $M$ or $\alpha$.}
\label{app: Strain_9Plots}
\end{figure*}

Finally, in the extended survey of GW properties, we find little change in $f_{{\rm max}}$ (Fig. \ref{app: AngularFrequency_9Plots}).  
Likewise, the peak strain $h_{{\rm max}}$ (Fig. \ref{app: Strain_9Plots}) is qualitatively similar to the behavior seen in the main text, though the dependence on both $M$ and $\alpha$is now more clear: the peak GW strain grows modestly for increasing either $M$ or $\alpha$.

\end{document}